\documentclass[aps,prx,superscriptaddress,floatfix,showpacs,notitlepage,twocolumn, 10pt]{revtex4-2}

\usepackage{graphicx}
\usepackage{dcolumn}
\usepackage{bm}
\usepackage{amsmath,amssymb}
\usepackage{amsthm}
\usepackage{mathrsfs}
\usepackage{indentfirst}
\usepackage{float}
\usepackage{braket}
\usepackage[utf8]{inputenc}
\usepackage{bm}
\usepackage{graphicx}
\usepackage{physics}
\usepackage{mathtools}
\usepackage{color,graphicx} 
\usepackage{bbold}

\theoremstyle{definition}

\usepackage{relsize}

\usepackage{hyperref}
\usepackage{hypernat}
\hypersetup{
colorlinks=true,
citecolor=black,
linkcolor=black,
urlcolor=blue,
pdfpagemode=UseNone,
pdfstartview=FitH}

\begin{document}

\preprint{APS/123-QED}
\title{Resource-State Quantum RAM for Fast and Error-Correctable Queries}
\author{Francesco Cesa}\email{francesco.cesa@uibk.ac.at}
\affiliation{Institute for Quantum Optics and Quantum Information of the Austrian Academy of Sciences, 6020 Innsbruck, Austria}
\affiliation{Institute for Theoretical Physics, University of Innsbruck, 6020 Innsbruck, Austria}
\affiliation{Department of Physics, University of Trieste, Strada Costiera 11, 34151 Trieste, Italy}
\affiliation{Istituto Nazionale di Fisica Nucleare, Trieste Section, Via Valerio 2, 34127 Trieste, Italy}
\author{Hannes Bernien}
\affiliation{Institute for Quantum Optics and Quantum Information of the Austrian Academy of Sciences, 6020 Innsbruck, Austria}
\affiliation{Institute for Experimental Physics, University of Innsbruck, 6020 Innsbruck, Austria}
\affiliation{Pritzker School of Molecular Engineering, University of Chicago, Chicago, IL 60637, USA}
\author{Hannes Pichler}\email{hannes.pichler@uibk.ac.at}
\affiliation{Institute for Quantum Optics and Quantum Information of the Austrian Academy of Sciences, 6020 Innsbruck, Austria}
\affiliation{Institute for Theoretical Physics, University of Innsbruck, 6020 Innsbruck, Austria}

\begin{abstract}
\noindent Quantum devices can process data in a fundamentally different way than classical computers. To leverage this potential, many algorithms require the aid of a quantum Random Access Memory (QRAM), i.e. a module capable of efficiently loading datasets onto the quantum processor. However, a realisation of this building block is still outstanding due to its formidable resource requirements, which become even more demanding in quantum error-correction schemes. Here we show that the challenge of implementing QRAM can be entirely reduced to a state-preparation problem: since such resource-state is independent on the memory, our approach allows one to prepare it offline, opening the door to new design strategies. As an example, we introduce a heralded `QRAM factory' which enables improved fidelities with high acceptance rate. More broadly, our results introduce the concept of resource-state QRAM: we study its performance in noisy settings, showing that it preserves the noise-resilience of standard QRAM, and discuss how it can be efficiently combined with quantum error-correction. Finally, we propose an implementation with neutral-atom hardware, where our analysis suggests that high-fidelity and low-latency queries can be implemented.
\end{abstract}

\maketitle

\begin{figure*}[t!]
    \begin{minipage}{1.0\textwidth}
        \includegraphics[width=\textwidth]{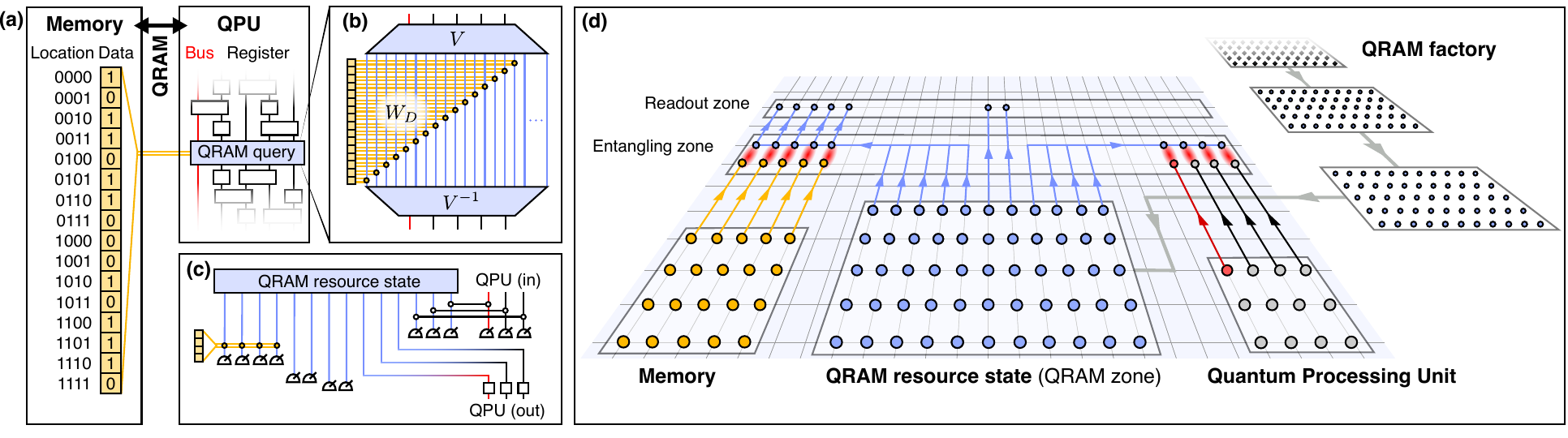}
    \end{minipage}%
\caption{\textbf{Quantum RAM.} 
(a) A memory stores $N$ data bits $D_{\bm{x}}\in\left\{0,1\right\}$ in locations $\bm{x}=\left(x_0,\dots,x_{\log N-1}\right)$. The QPU imports the data coherently according to \emph{quantum} queries to the memory: an input state $\ket{\psi}=\sum_{\bm{x}}\psi_{\bm{x}}\ket{\bm{x}}$ plays the role of a quantum \emph{address}, querying a superposition of data; the QRAM responds with the output state $\sum_{\bm{x}} \psi_{\bm{x}}\ket{\bm{x}}\otimes\ket{D_{\bm{x}}}$, with the data loaded on the bus qubit initialised in $\ket{0}$. (b) Unified sketch of fast and noise-resilient approaches according to Eq.~\eqref{basic_query}. The operation $V$ takes as input the address $\ket\psi$, and processes it together with $\mathcal{O}(N)$ ancillas, to provide a `one-hot encoded' state which is interfaced with the memory by a \emph{loading} operation $W_D=\otimes_{\bm{x}}Z_{\bm{x}}^{D_{\bm{x}}}$; finally, $V^{-1}$ compresses back the state. (c) Our QRAM design answers to arbitrary queries with the aid of a resource-state (RS) $\ket{\Phi}$ on a set of ancillary qubits. We start by performing Bell measurements; then we load the data via single-qubit Pauli operations and measurements, retrieving the output on the unmeasured qubits. (d) We blueprint end-to-end implementations with neutral-atom processors, where physical qubits are encoded in single atoms trapped in optical tweezers.
Coherent atom transport allows to inject the QRAM RS, and to interface the QPU with it. A parallel QRAM factory guarantees continuous supply of the RS via a fast and efficient generation protocol based on dynamical rearrangement. A classical memory can be stored either on classical hardware, or on atoms as in the figure (the latter configuration also allowing for quantum memories).}
\label{fig1}
\end{figure*}

\noindent Quantum computation (QC) promises significant algorithmic speedups for specific tasks~\cite{nielsen2010quantum, dalzell2023quantum}, fueling innovation across numerous modern technologies. Remarkable progress has been achieved in developing quantum processing units (QPUs), witnessed by increasing system sizes, improved performances, and the incorporation of quantum error-correction (QEC)~\cite{Ma_2023, Finkelstein_2024, Cao_2024, Bluvstein_2023, reichardt2024logicalcomputationdemonstratedneutral, radnaev2025universalneutralatomquantumcomputer, PhysRevX.13.041052, Marques_2021, Krinner_2022, google2023suppressing, acharya2024quantumerrorcorrectionsurface, egan2021fault, postler2022demonstration, paetznick2024demonstrationlogicalqubitsrepeated, aghaee2025scaling}. However, critical challenges remain: among these is the development of a quantum Random Access Memory (QRAM)~\cite{giovannetti2008quantum}. This module is analogous to classical RAM, which enables fast import of  data from a memory to the central processing unit (CPU). Specifically, when querying a memory location labeled by the address $\bm{x}$, the RAM responds by furnishing the corresponding databit $D_{\bm{x}}$. Analogously, a QRAM provides a  fast quantum-mechanical interface between the QPU and datasets as in Fig.~\ref{fig1}(a); a \emph{quantum query} is therefore defined as
\begin{equation}\label{qRAM_generic}
    U_{\text{QRAM}}\left[\; \sum_{\bm{x}} \psi_{\bm{x}} \ket{\bm{x}}\otimes\ket{0}\; \right] = \sum_{\bm{x}} \psi_{\bm{x}} \ket{\bm{x}} \otimes \ket{ D_{\bm{x}}},
\end{equation}
where $\psi_{\bm{x}}$ labels the amplitude of the  computational basis state $\ket{\bm{x}}$ of the QPU register. The queried data $D_{\bm{x}}$ is written coherently onto a \emph{bus} qubit that here is initialised in $\ket{0}$. Minimal modifications extend Eq.~\eqref{qRAM_generic} to quantum memories~\footnote{See Supplemental Material.}. Central to many foundational algorithms, QRAM queries are ubiquitous in QC, e.g. implementing `black-box oracles', aiding quantum search and quantum walk, and realising multi-controlled gates~\cite{deutsch1992rapid, bernstein1993quantum, grover1996fast, ivanov2006engineering, low2024trading}. Applications include quantum chemistry~\cite{Cao_2019, berry2015simulating, babbush2018encoding}, machine learning~\cite{harrow2009quantum, biamonte2017quantum} and data centers~\cite{PhysRevA.108.032610}. 
More generally, QRAM allows coherent access to pre-calculated lookup-tables, speeding up, e.g., modern variations of Shor's factoring algorithm~\cite{Gidney2021howtofactorbit}.\\
\indent A number of QRAM designs have been proposed, typically envisioning specific hardware~\cite{giovannetti2008quantum,giovannetti2008architectures, hann2019hardware, chen2021scalable, weiss2024quantum, wang2024quantum}, and more recently small-scale prototypes were demonstrated~\cite{pq3x-cmw9, shen2025experimentalrealizationbucketbrigadequantum, zhang2025demonstratingcoherentquantumrouters}. However, realizing Eq.~\eqref{qRAM_generic} fundamentally requires a space-time circuit volume scaling with $N$, the memory size, which poses formidable practical challenges. Furthermore, while QRAM benefits of a remarkable resilience against errors~\cite{giovannetti2008quantum, hann2021resilience}, nevertheless QEC is expected to be necessary for large-scale implementations. Under this perspective, the challenge is exacerbated by the demand for $\mathcal{O}(N)$ non-Clifford gates, which are particularly costly in standard QEC paradigms. Altogether, such practical and theoretical complexity hindered large-scale experimental demonstrations and raised debates on the feasibility of QRAMs, pinning bold question marks on the potential of quantum computers~\cite{aaronson2015read, jaques2023qramsurveycritique}.  \\
\indent Here, we develop a new framework for QRAM design. Specifically, we show that the complexity of implementing QRAM can be recast into a \emph{state-preparation} problem, thereby introducing the concept of resource-state QRAM. In our approach, fast and robust mid-circuit queries are implemented through simple Clifford operations and single-qubit Pauli measurements (SQPMs): as illustrated in Fig.~\ref{fig1}(c), this is achieved by consuming a previously assembled data-independent resource-state (RS) $\ket{\Phi}$, which encompasses all the non-Cliffordness required by Eq.~\eqref{qRAM_generic}. Importantly, such protocol has no additional cost, and inherits the exponentially enhanced fidelity of the celebrated `bucket-brigade' protocol~\cite{giovannetti2008quantum, hann2021resilience}.\\ 
\indent Thus, the resource-state QRAM allows one to rethink at data-lookup problems in terms of the high-fidelity assembly of $\ket{\Phi}$ in a QRAM factory, where offline operation can be exploited through state-preparation techniques. Indeed, typically preparing a specific \emph{known} state is simpler than applying a circuit to an unknown state~\cite{gottesman1999demonstrating}: this is a largely exploited concept in QC, with paradigmatic examples being the fault-tolerant injection of magic states~\cite{gottesman1999demonstrating, PhysRevA.71.022316}, measurement-based quantum computation~\cite{PhysRevLett.86.5188, briegel2009measurement} or quantum repeater schemes~\cite{RevModPhys.95.045006}. As an example, here we develop a high-rate `try-until-succeed' scheme for preparing $\ket{\Phi}$: this allows us to boost the fidelity of queries in a way that would not be possible in the standard paradigm, showcasing how resource-state QRAM opens to new opportunities and strategies.\\
\indent Finally, we propose and analyse an implementation of resource-state QRAM with neutral-atom arrays~\cite{Bluvstein_2023, reichardt2024logicalcomputationdemonstratedneutral, radnaev2025universalneutralatomquantumcomputer}. We find that dynamical atom rearrangement~\cite{Bluvstein_2022} naturally combines with the QRAM factory, and also we note that, in a QEC setting, it would exploit optimally our Clifford query via transversal gates. This motivates our design in Fig.~\ref{fig1}(d), which complements the `zoned' architecture of modern processors~\cite{Bluvstein_2023, reichardt2024logicalcomputationdemonstratedneutral, radnaev2025universalneutralatomquantumcomputer} with a dedicated QRAM zone. The latter is continuously replenished by the parallel QRAM factory, and supplies continuous queries. On this example, we show that the query and the factory rate match the clock-time of fault-tolerant operation, indicating that resource-state QRAM could be integrated with negligible latency.

\section*{RESULTS}
\begin{figure}[t!]\centering\includegraphics[width=0.47 
\textwidth]{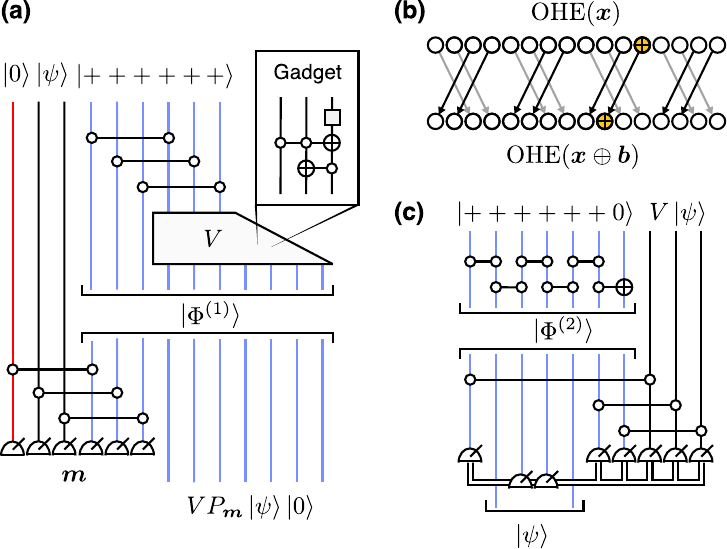}
\caption{(a) Gate-teleportation for the random OHE $VP_{\bm{m}}$: we first prepare the resource-state (RS) $|\Phi^{(1)}\rangle$, and then apply Bell measurements (BMs) with the input $\ket\psi$. The box highlights the fundamental gadget used for $V$.
(b) The OHE of a bit-string $\bm{x}$; empty and yellow-crossed circles circles represent $\ket
0$ and $\ket{+}$ respectively. For generalised OHEs $VP_{\bm{m}}$, the locations are permuted via $\pi_{\bm{m}}$. Here, $N=16$, $\bm{x}=(1,1,0,1)$ and $\bm{b}=(0,1,0,0)$. (c) Clifford implementation of $V^{-1}$ (for $N=2$). We pre-assemble a stabiliser RS $\smash{|\Phi^{(2)}\rangle}$ via a 1D Clifford circuit. The input is then entangled with $\smash{|\Phi^{(2)}\rangle}$; then, SQPMs enable the inversion. The first SQPMs are in the $X$ basis; in the second layer of SQPMs we choose either the $Z$ or the $X$ basis depending on the previous outcomes. 
}
\label{fig3}
\end{figure}

\noindent \textbf{Framework.} Here, we highlight the key points of fast and noise-resilient (bucket-brigade-type) QRAMs~\cite{giovannetti2008quantum}. We consider memories of size $N$, and QPU registers featuring $\log N$ qubits. In the main paradigm~\cite{giovannetti2008quantum},  Eq.~\eqref{qRAM_generic} is executed in three steps as in Fig.~\ref{fig1}(b),
\begin{equation}\label{basic_query}
    U_\text{QRAM} = \;V^{-1} W_D \;V. 
\end{equation} 
This can be read as follows: (i)~A `one-hot encoding' (OHE)~\cite{harris2012digital, bengio2017deep}, discussed below, is applied via $V$; (ii)~The dataset $D$ is loaded with $W_D$; (iii)~The OHE is inverted. \\
\indent More precisely, $V$ is a linear map that `expands' the $\log N+1$ register and bus qubits onto $2N-1$ qubits as 
\begin{equation}\label{main_OHE}
   V\;\; : \;\; \ket{\bm{x}}\ket{0} \xrightarrow[]{\text{one-hot}} \ket{\text{OHE}(\bm{x})}\ket{\text{NOHE}(\bm{x})}.
\end{equation}
The first term, $\ket{\text{OHE}(\bm{x})}$, is a product state on $N$ qubits: they are all in $\ket{0}$, except for a `pointer-qubit', identified by $\bm{x}$, which is in the state $\ket{+}=(\ket{0}+\ket{1})/\sqrt{2}$. In Fig.~\ref{fig3}(b) we give an explicit example. The second factor, $\ket{\text{NOHE}(\bm{x})}$, is a $N-1$ qubit state representing a `nested' OHE of $\bm{x}$, ensuring reversibility in a noise-resilient way (Methods). For $V\ket{\bm{x}}\ket{1}$ the output is similiar, but the pointer qubit is flipped to $\ket{-}=(\ket{0}-\ket{1})/\sqrt{2}$.\\
\indent The loading acts on the first  $N$ qubits as $W_D=\bigotimes_{\bm{x}}Z_{\bm{x}}^{D_{\bm{x}}}$, leaving the last $N-1$ qubits invariant; thus, during step (ii) the pointer is flipped as $\ket{+}\!\rightarrow\!\ket{-}$ if $D_{\bm{x}}=1$ (since $Z\ket{\pm}=\ket{\mp}$). \\
\indent Finally, $V^{-1}$ inverts the OHE: 
\begin{equation}
    V^{-1} W_D\ket{\text{OHE}(\bm{x})}\ket{\text{NOHE}(\bm{x})} = \ket{\bm{x}}\ket{D_{\bm{x}}},
\end{equation}
retrieving the data on the QPU.\\
\indent Following this recipe, Eq.~\eqref{qRAM_generic} can have $\mathcal{O}(\log N)$ circuit-depth, and a remarkable $\mathcal{O}\left(\text{polylog} N\right)$ infidelity scaling~\cite{giovannetti2008quantum, hann2021resilience}. However, $V$ requires $\mathcal{O}(N)$ long-range non-Clifford gates on $\mathcal{O}(N)$ ancillary qubits. It therefore incurs significant costs~\cite{di2020fault}, both in terms of physical implementation and of non-Clifford factory demands, obstructing efficiency and speed of the main algorithm. \\
\indent In contrast our approach implements Eq.~\eqref{basic_query} via simple Clifford operations (mostly SQPMs) on $\ket{\psi}\otimes\ket{\Phi}$, where $\ket{\psi}$ is the input state. The complexity of $V$, outsourced to the \emph{offline} preparation of $\ket{\Phi}$, can therefore be tackled with state-preparation tools, which can be advantageous over directly implementing $V$ on $\ket{\psi}$.  \\

\begin{figure*}[t!]
    \begin{minipage}{1.0\textwidth}
        \includegraphics[width=\textwidth]{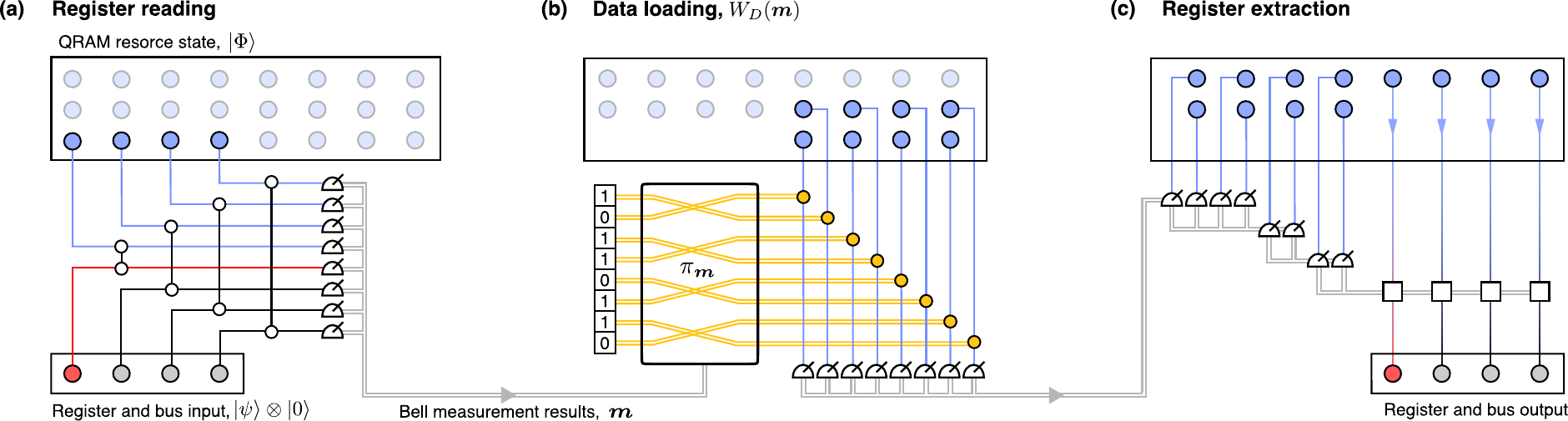}
    \end{minipage}%
\caption{\textbf{Clifford query.} (a) The QPU register is initially in the state $\ket{\psi}$ on $\log N$ qubits (grey), while the RS $\ket{\Phi}$ is stored on $\mathcal{O}(N)$ ancillas (blue); here we initialise the bus in $\ket{0}$ (red). In the first step, of depth $\mathcal{O}(1)$, we perform BMs between the QPU register and bus, and a subset of the RS qubits. The outcome of these measurements is denoted by $\bm{m}$ and used in the next step. (b) In the second step, of depth $\mathcal{O}(1)$, the data (yellow) is loaded via conditional Clifford gates, on another subset of the QRAM ancillas. Depending on $\bm{m}$ a permutation of the data according to $\pi_{\bm{m}}$ (see main text) is performed prior to the gate application to achieve the correct data loading operation, $W_D(\bm{m})$. The involved qubits are then measured in the $X$ basis (bottom). (c) In the last step, of depth $\mathcal{O}\left(\log N\right)$, almost all the ancillary qubits, except $\log N +1$, are measured in adaptive Pauli bases ($X$ or $Z$); the answer to the query is retrieved on the remaining unmeasured qubits, after a final Pauli correction operation. Overall, the query is executed as Eq.~\eqref{modified_query} in $\mathcal{O}(\log N)$ depth, including $\log N$ BMs and $\mathcal{O}(N)$ SQPMs.  }
\label{fig2}
\end{figure*}

\begin{figure}[t!]
\centering\includegraphics[width=0.45 \textwidth]{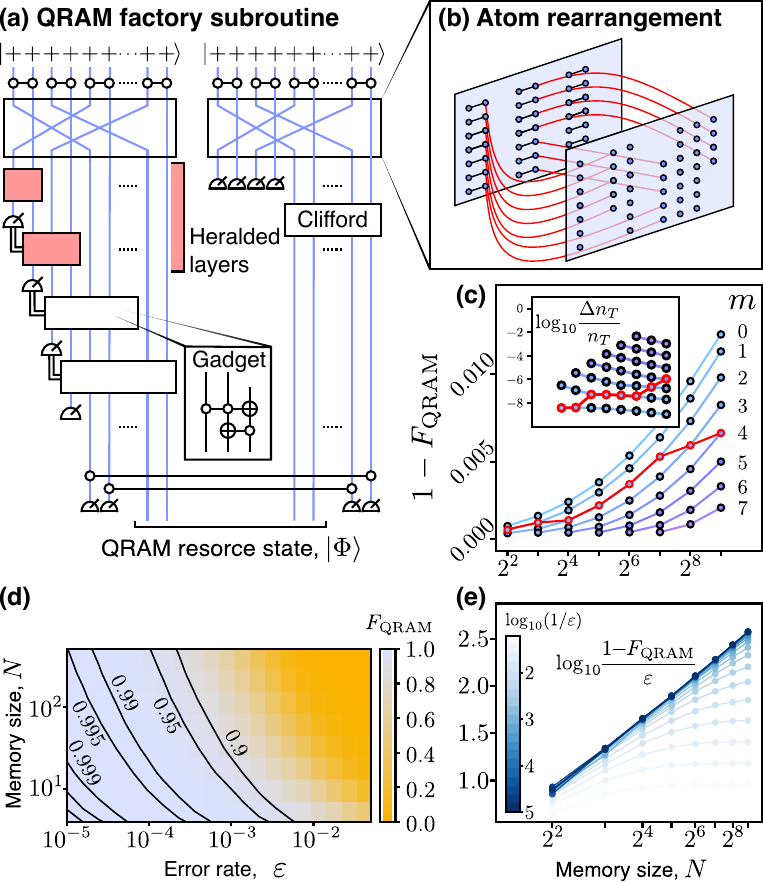}
\caption{(a,b) High-rate protocol for preparing $\ket{\Phi}$ in the QRAM factory. We first prepare $\mathcal{O}(N)$ Bell pairs, and then reconfigure the system layout via $\mathcal{O}(1)$ global atom motions. Next, we prepare the RSs $|\Phi^{(1)}\rangle$ (iterating the inset gadget) and $|\Phi^{(2)}\rangle$ (via Clifford gates). Both these subroutines also include adaptive Pauli gates and measurements. Finally, we merge the two in $|\Phi\rangle$ via BMs, which is also achieved via local operations. (c) By heralding the first $m$ factory layers, we can extensively boost the fidelity. The inset shows the relative waste of Toffoli gates $\Delta n_T=\mathcal{O}(\log N)$, in units of the naive Toffoli count $n_T=\mathcal{O}(N)$. In red we highlight a potential strategy for choosing $m$. (d) Query fidelity under continous depolarising noise, where at every step each qubit (also if idle) depolarises with probability $\varepsilon$. (e) We extract the fidelity scaling from the collapse to a straight line in a log-loglog plot of $(1-F_\text{QRAM})/\varepsilon=\mathcal{O}(\log^\alpha N)$ for small $\varepsilon$; a linear fit returns $\alpha \simeq 2.2$. Analogous results persist across all tested error models (Supplemental Material).}
\label{fig4}
\end{figure}

\noindent \textbf{Clifford query.} We now show how the complexity of QRAM can be outsourced to the preparation of a resource-state $\ket{\Phi}$, which we define on the way. \\
\indent Our query construction is inspired by the concept of gate-teleportation (GT), which plays a foundational role in QEC~\cite{gottesman1999demonstrating}. While GT is typically only efficient for minimally non-Clifford tasks, we show that it can aid a highly non-Clifford subroutine such as QRAM. This is based on the observation that  Eq.~\eqref{basic_query} can be rewritten as 
\begin{equation}\label{modified_query}
    U_\text{QRAM} = P^{-1}_{\bm{m}} V^{-1}W_D(\bm{m}) V P_{\bm{m}},
\end{equation}
where $P_{\bm{m}}$ is an arbitrary Pauli operator (labeled by $\bm{m}$) acting on the QPU register. Here,  $W_D(\bm{m})=\pi_{\bm{m}}W_D \pi^{-1}_{\bm{m}}$, and  $\pi_{\bm{m}}$ is a simple permutation of the OHE qubits. Importantly, this holds true for \emph{any} Pauli operator $P_{\bm{m}}$. In our construction, $\bm{m}$ will be intrinsically random. This rewriting allows us to design a modified query protocol comprising the following three Clifford steps. \\
\indent Our step (i) is shown in Fig.~\ref{fig3}(a): a protocol similar to GT~\cite{Nielsen_1997,  gottesman1999demonstrating} applies $VP_{\bm{m}}$ in Eq.~\eqref{modified_query} to the input QPU state $\ket{\psi}$. As in standard GT, we employ the preassembled RS $\smash{\ket{\Phi^{(1)}}}\equiv (\mathbb{1}\otimes V)\ket{\Psi}^{\otimes 1+\log N}$ where $\ket{\Psi}=\tfrac{1}{\sqrt{2}}(\ket{0+}+\ket{1-})$. Performing $\log N$ Bell measurements (BMs) between the RS and the input, results in a state $VP_{\bm{m}}\ket{\psi}$, where $\bm{m}$ is random and determined by the BMs outcome. Crucially, this is now Clifford~\footnote{We note that this contrasts with typical GT, where typically highly non-Clifford byproducts must be applied to make the protocol deterministic (see Methods). }. 
\indent Step (ii) implements $W_D({\bm{m}})$, which depends on the BM outcomes in step (i). This is achieved efficiently by a simple adaptation of $W_D$ in classical processing. Indeed, as shown in Fig.~\ref{fig3}(b), $\pi_{\bm{m}}$ is a highly structured permutation of the pointers, which we implement as $W_D(\bm{m})=\bigotimes_{\bm{x}}Z_{\bm{x}}^{D_{\bm{x}\oplus \bm{b}}}$, with $\bm{b}$ determined by $\bm{m}$. Importantly, this does not require to process the memory. \\
\indent Step (iii) realizes $P^{-1}_{\bm{m}}V^{-1}$ in Eq.~\eqref{modified_query} with Clifford operations. For this, first note that $V$ can be constructed by iterating the circuit gadget in Fig.~\ref{fig3}(a). Even though this is non-Clifford, it can be \emph{inverted} with Clifford operations (Methods): this is possible because we only require inversion on the image of $V$, which is an exponentially smaller subspace [see Eq.~\eqref{main_OHE}]. In particular, we can design a pattern of \emph{adaptive} SQPMs and Clifford gates that deterministically maps any state of the form $W_D(\bm{m})VP_{\bm{m}}\ket{\psi}$ to the desired output $U_\text{QRAM}\ket{\psi}$.\\
\indent This completes our proof, as one could implement the query through Clifford operations on $\ket{\psi}\ket{\Phi^{(1)}}$. In this case, step (iii) requires adaptive two-qubit Clifford gates. We can eliminate such two-qubit gates via a measurement-based scheme~\cite{PhysRevLett.86.5188, briegel2009measurement}. Specifically, we construct a \emph{stabiliser} RS $\smash{\ket{\Phi^{(2)}}}$ with the following property: after entangling any state $\ket{\Xi}=V\ket{\psi}$ with $\smash{\ket{\Phi^{(2)}}}$ via BMs, $\mathcal{O}(\log N)$ layers of adaptive SQPMs extract the desired output $V^{-1}\ket{\Xi}$. \\
\indent In summary, given $\ket{\psi}\ket{\Phi^{(1)}}\ket{\Phi^{2}}$, we can implement Eq.~\eqref{modified_query} on $\ket{\psi}$ via one layer of BMs followed by $\mathcal{O}(\log N)$ layers of adaptive SQPMs. Finally, we can compress $\ket{\Phi^{(1)}}\ket{\Phi^{(2)}}$ in a unique RS $\ket{\Phi}$ using standard Clifford techniques (Methods). This RS can be assembled prior to the query in the QRAM factory without using any information on $\ket{\psi}$, nor on the memory content.\\
\indent Fig.~\ref{fig2} details the resulting compact query protocol. This is very efficient especially for QEC, as it only requires Clifford operations during the online computation. The query runtime $T_\text{query} \simeq 2\tau\log\!N$, corresponding to half the duration in previous proposals~\cite{hann2021resilience}, is dominated by the mid-circuit readout time $\tau$; within fault-tolerant operation, this also sets the computational clock-time. \\
\\
\noindent \textbf{Noise-resilience and QEC.} In presence of errors, it is known that `bucket-brigade' QRAM schemes feature a remarkable fidelity scaling~\cite{giovannetti2008quantum, hann2021resilience}, which only decays as a function of $\mathcal{O}(\log^\alpha N)$ - as opposed to the naively expected $\mathcal{O}(\text{poly} N)$. Resource-state QRAM can benefit of the same mechanism, by incorporating its principles in the preparation of $\ket{\Phi}$ to constrain the error propagation. We certify this numerically for various settings; Figs.~\ref{fig4}(d,e) report results for continuously depolarising qubits. To estimate the near-term potential, we note that with error-rates $\varepsilon\!\sim\! 10^{-4}$, QRAMs of size $N \gtrsim 512$ have fidelity above $95\%$.\\
\indent In a QEC scenario, as a consequence of the noise-resilience above, in our design the code distance needed for achieving a given target $F_\text{QRAM}$ scales as
\begin{equation}\label{code_distance}
    d \simeq \log\Big( (1-F_\text{QRAM})^{-1}\text{polylog}N) \Big)\Big/\log(1/\varepsilon),
\end{equation}
featuring an exponential reduction compared to alternative schemes~\cite{babbush2018encoding}, where $d\sim \log N$. While the QRAM space-time volume is fundamentally $\mathcal{O}(N)$, Eq.~\eqref{code_distance} implies that the QEC overhead \emph{factor} is the same as for a circuit of volume $\mathcal{O}(\text{polylog} N)$. Similarly, while the $T-$count is the same as for any lookup subroutine~\cite{babbush2018encoding}, the number of noisy $T-$states consumed (including distillation~\cite{PhysRevA.71.022316}) is reduced by a factor $\mathcal{O}(\text{polylog}N)$. \\
\indent For fault-tolerant (FT) queries, in our construction all operations are fully \emph{transversal} in many paradigmatic settings, including the surface and CSS codes~\cite{nielsen2010quantum}. It follows that a FT query 
proceeds via SQPMs only, by leveraging measurement-based QEC~\cite{Raussendorf_2006} e.g. within foliated~\cite{PhysRevLett.117.070501} or fusion-based~\cite{PhysRevLett.131.120604, PhysRevX.13.041013} schemes. Since our protocol is based on destructive SQPMs, we simply measure the physical qubits individually: similarly to Knill QEC~\cite{knill2005quantum}, the results reveal both logical measurement outcomes and syndrome snapshots. Altogether, the query parallelises all the operations we would apply to physical qubits, to logical blocks. As recently demonstrated~\cite{Bluvstein_2023, reichardt2024logicalcomputationdemonstratedneutral}, this is achieved elegantly with neutral atoms by interfacing the blocks via atom transport.\\

\noindent \textbf{QRAM factory.} We now elaborate on the cost of preparing $\ket{\Phi}$. Fig.~\ref{fig4}(a) schematises how a QRAM factory can operate in $2\log N$ layers, which progressively incorporate $\mathcal{O}(N)$ Toffoli gates (Methods). Thanks to the scaling of $F_\text{QRAM}$ above, these $T-$states need only be distilled to precision $1/\text{polylog}N$; thus, each consumes $\text{polylog}(\log N)$ noisy $T-$states through $\mathcal{O}(\log\log\log N)$ distillation rounds. This is the fundamental $T-$cost required by any QRAM implementation. \\
\indent In our approach such cost is recast into offline state-preparation. One advantage is that this can be combined with heralding procedures. In Methods we present one such factory protocol, wherein we `retry-until-succeed' the first layers of the assembly. Fig.~\ref{fig4}(c) reports numerical results, showing an extensive fidelity boost $\Delta F_\text{QRAM} \sim \mathcal{O}(\varepsilon \text{poly}( m\log N))$, where $m$ is a parameter quantifying the amount of heralding. Crucially, one can choose $m$ such that the rejection rate and the overhead of `wasted' resources remains small (figure inset). This is beneficial both within FT operation, where the syndrome information heralds errors, but also without QEC, e.g. for neutral atoms or superconductors, where leakeges can be physically detected~\cite{Ma_2023, Scholl_2023, PhysRevX.12.021049, Wu_2022, PhysRevX.13.041013, PhysRevX.14.011051}. \\ 
\\
\noindent \textbf{Latency}. While in principle the query runtime is $T_\text{query}\simeq 2\tau\log N$, the actual latency of resource-state QRAM is also influenced by the interplay between $T_   \text{query}$ and the factory rate $T_\Phi$. Such interplay, in turn, depends on the design of the QRAM factory; in our proposal [Fig.~\ref{fig4}(b)], the most delicate step (in terms of runtime) is a system reconfiguration, which enables long-range gates. In Methods we focus on neutral-atom processors~\cite{Bluvstein_2023, reichardt2024logicalcomputationdemonstratedneutral, radnaev2025universalneutralatomquantumcomputer}, where we show that coherent rearrangement~\cite{Bluvstein_2022} allows to implement such reconfiguration in $\mathcal{O}(1)$ long-range moves. Particularly, this yields 
\begin{equation}\label{main_T_phi}
     T_\Phi\lesssim 3\sqrt{6}T
     N^{1/4} + 2\tau\log\!N,
\end{equation}
where the first term is related to rearrangement, and the second to SQPMs. Specifically, $T$ is the time required for transporting an atom across the minimal trap distance. Considering current experiments~\cite{Bluvstein_2022}, we set $\tau\simeq 500\mu\mathrm{s}$ and $T\simeq33\mu\mathrm{s}$. With these parameters, $T_\Phi$ is comparable to $T_\text{query}$, as the rearrangement time remains negligible up to $N\simeq 2^{24}$ (corresponding to $\simeq 2\mathrm{MB}$ of memory), which is much larger than current system sizes of single processors~\cite{manetsch2024tweezerarray6100highly}. For $1\mathrm{kB}$ of memory we find $T_\text{query}\simeq 13\mathrm{ms}$ and $ 1/T_\Phi\simeq 0.1\mathrm{kHz}$, wherein only $\simeq 13\%$ of the time is spent on atom transport. Since $T_\Phi\simeq T_\text{query}$, we can parallelise a query with the preparation of a new copy of $\ket{\Phi}$ for the next query, resulting in an efficient cycle, not adding substantial runtime. Remarkably, despite the overheads for syndrome extraction and distillation, we show that this also holds in the QEC setting (Methods). 

\section*{Discussion}
\noindent We now discuss the implications of our results, specially focusing on future directions to be addressed.\\
\indent We introduced a QRAM design where all the query complexity is reduced to a state-preparation problem. The cost of QRAM can now be tackled independently in an `offline factory' which does not affect the quantum state processed in the QPU, and is memory-independent. The fact that resource-state QRAM offers new optimisation opportunities is exemplified by our heralding procedure, but future directions can now further exploit heralding strategies, and potentially design specialised QRAM purification protocols. The remaining online task, which is fast and has zero non-Clifford cost, is specially simplified in the QEC setting, where it can be implemented via transversal operations only. We also note that, while such considerations assume a resource theory where non-Clifford operations represent the dominant operational cost, the benefits of resource-state QRAM would likely persist in other cost models.  \\
\indent At a fundamental level, memory lookups will arguably be among the most costly subroutines of QC. Previous works pointed out that the overhead associated with QRAM requires careful evaluations of its role in quantum advantage~\cite{aaronson2015read} and of its `opportunity-cost'~\cite{jaques2023qramsurveycritique}. In this respect, we note that resource-state QRAM motivates re-evaluations of opportunity costs in an algorithm-specific way. For instance,  with resource-state QRAM the cost can be invested even before the memory is specified, changing the the cost allocation options with respect to previous approaches.\\
\indent Beyond such theoretical aspects, developing QRAM is a formidable technological challenge. By analogy with classical RAM, a sound approach would be to consider QRAM as a separate module, featuring specialised design and hardware components, with high-fidelity interfaces with the QPU. Resource-state QRAM is well matched with such `divide-and-conquer' strategy, natively incorporating a distinction between the QRAM module (which operates as a factory) and the processor, and allows even for different hardware solutions for QRAM and QPU. As a physical QRAM realization, we have specifically proposed an implementation with neutral-atoms, where simulations suggest that our architecture could be efficiently integrated with low latency. This may be adapted to trapped-ion processors, which have also developed qubit shuttling~\cite{Pino_2021}, but the general ideas behind our proposal can find application also in other platforms, including superconducting~\cite{acharya2024quantumerrorcorrectionsurface} or photonic~\cite{aghaee2025scaling} systems.\\

\section*{Acknowledgements}
\noindent We thank Dolev Bluvstein, Hans Briegel,  Vittorio Giovanetti, Mikhail Lukin and Peter Zoller for discussions, and Gabriele Calliari for support with the numerical simulations. This work is supported by the ERC Starting grant QARA (Grant No.~101041435), the Horizon
Europe programme HORIZON-CL4-2022-QUANTUM02-SGA via the project 101113690 (PASQuanS2.1) and
by the Austrian Science Fund (FWF) (Grant No.~DOI~10.55776/COE1). HB gratefully acknowledges funding from the Air Force Office of Scientific Research (Grant No. FA9550-21-1-0209), the Office of Naval Research (Grant No. N00014-23-1-2540), and the Army Research Office (Grant no. W911NF2410388).

\bibliography{biblio.bib}

\section*{Methods}
\noindent Our proposed QRAM module is based on several concepts, which are detailed in the Supplemental Material (SM). Here, we highlight the key points, outlining how they lead to the final QRAM design.   

\subsection*{Nested one-hot encoding}

\begin{figure*}[t!]
    \begin{minipage}{1.0\textwidth}
        \includegraphics[width=\textwidth]{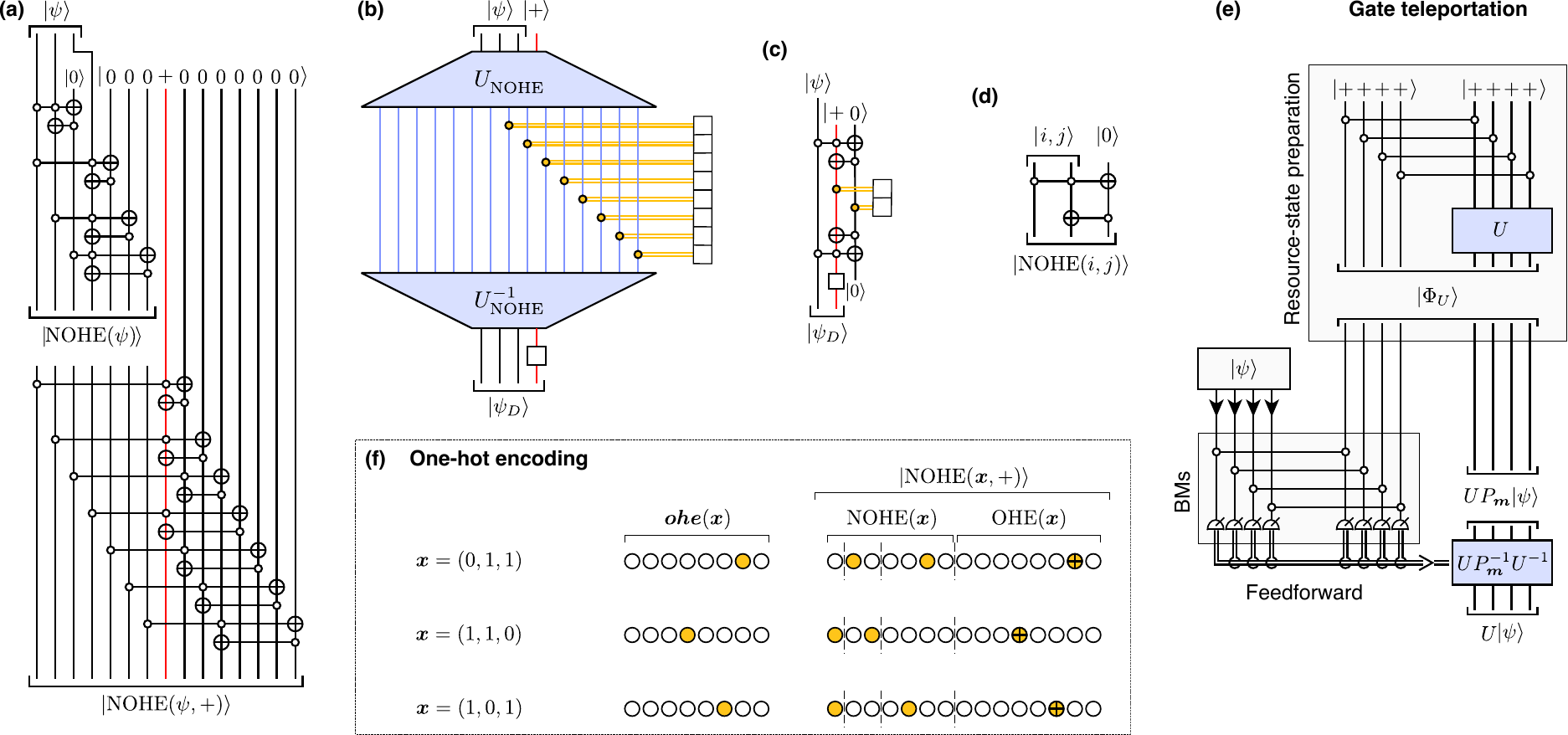}
    \end{minipage}%
\caption{\textbf{Nested one-hot encoding, quantum RAM and gate teleportation.} (a) Circuit decomposition for realising a NOHE (here we display $U_\text{NOHE}^{(2N)}$ for $N=8$). The state $\ket{\text{NOHE}(\psi)}$ is obtained by applying the circuit in the top-left part; by adding a \emph{bus} qubit in the $\ket{+}$ state and iterating the same procedure, we obtain $\ket{\text{NOHE}(\psi,+)}=V\ket{\psi,0}$. Note that the circuit consists of the iteration of a native gadget, featuring a Toffoli gate followed by a controlled-not. (b) An example of QRAM model based on the NOHE; the red lines are classically-controlled $\mathrm{Z}$ gates, conditioned on the memory elements being $D_{\bm{x}}=1$. Abstractly this is similar, e.g., to the well-known bucket-brigade protocol~\cite{giovannetti2008quantum}. (c) Explicit QRAM model for $N=2$, where the address is a single qubit. (d) Action of the minimal gadget for the circuit decomposition of $U_\text{NOHE}$, when applied to an arbitrary two-qubit computational basis state. (e) Depiction of standard gate-teleportation (GT) for realising an arbitrary gate $U$ to an input defined on $\log N$ qubits. First, a RS $|\Phi_U\rangle=(\mathbb{1}\otimes U)\ket{\Psi_{00}}^{\otimes\log N}$ is prepared, by applying $U$ to one half of $\log N$ Bell pairs (here, we set $\ket{\Psi_{00}}=(\ket{00}+\ket{01}+\ket{10}-\ket{11})/2$, see the SM). Second, Bell measurements (BMs) are applied between the input and part of the RS qubits; here, we implemet BMs by first applying a controlled-$Z$ gate, $CZ=\mathbb{1}-2\ket{11}\bra{11}$, and then measuring the qubits individually in the $X$ basis. Third, feedforward of the BM outcomes $\bm{m}$ regulates the application of a byproduct operation $UP_{\bm{m}}U^{-1}$, where $P_{\bm{m}}$ is a Pauli operator. (f) Explicit depiction of the various variations of one-hot encoding, on the example of $N=8$. Here empty or red-colored circles represent the $\ket{0}$ or the $\ket 1$ state respectively; red-colored circles with a black cross represent the $\ket{+}$ state. }
\label{fig_M1}
\end{figure*}

\noindent Let $\bm{x}=(x_0,x_1,\dots,x_{\log N -1})\in\left\{0,1\right\}^{\log N}$ be a bitstring of $\log N$ bits. A one-hot encoding (OHE) encodes $\bm{x}$ in $N$ bits by setting all the encoding bits to $0$, except a `hot' bit, which is set to $1$ and acts as a \emph{pointer}. Defining $\mu(\bm{x})=\sum_{K=0}^{\log N -1}x_K2^K$, the encoding bitstring $\bm{ohe}(\bm{x})$ has components $\bm{ohe}_\alpha(\bm{x})=\delta_{\alpha, \mu(\bm{x})}$, with $\alpha = 0,1,\dots, N-1$. We introduce a similar concept in the quantum domain, but using the $\ket{+}$ state as a pointer:
\begin{equation}
    \ket{\text{OHE}(\bm{x})} = \frac{\ket{0}^{\otimes N} + \ket{\bm{ohe}(\bm{x})}}{\sqrt{2}} = \ket{0,0,\dots,0,+,0,\dots},
\end{equation}
with the $\ket{+}$ state at position $\mu(\bm{x})$. Note that the map $\ket{\bm{x}}\rightarrow\ket{\text{OHE}(\bm{x})}$ is not unitary.\\
\indent To apply the OHE, Fig.~\ref{fig_M1}(a) shows a circuit implementing the unitary $U_\text{NOHE}^{(N)}$, which we refer to as `nested' OHE (NOHE, see the SM). This takes as input $\log N$ register qubits in an arbitrary state $\ket{\psi}$, plus $N-\log N - 1$ ancillary qubits initialised in $\ket{0}$; the output $\ket{\text{NOHE}(\psi)}$ is the NOHE of $\ket{\psi}$, and it is such that if we add a \emph{bus} qubit initialised in $\ket{+}$, then $|\text{NOHE}(\bm{x},+)\rangle = |\text{NOHE}(\bm{x})\rangle |\text{OHE}(\bm{x})\rangle$. Connecting to the main text, we have $V=U_\text{NOHE}^{(2N)} H_\text{bus}$, where $H=(X\!+\!Z)/\!\sqrt{2}$, such that 
\begin{equation}
    V\ket{\psi,0}= \sum_{\bm{x}} \psi_{\bm{x}} \ket{\text{NOHE}(\bm{x})} \ket{\text{OHE}(\bm{x})}
\end{equation}
for any state $\ket{\psi}=\sum_{\bm{x}} \psi_{\bm{x}}\ket{\bm{x}}$. Here we exchanged the order of $\text{NOHE}(\bm{x})$ and $\text{OHE}(\bm{x})$ with respect to Eq.~\eqref{main_OHE}. A depiction of these concepts is in Fig.~\ref{fig_M1}(f).\\
\indent In practice, for a register featuring $\log N$ qubits (plus the bus qubit), $V$ can be understood as an isometry,
\begin{equation}\label{V_spaces}
    V \; : \; \mathscr{H}^{(\log N +1)} \xrightarrow[]{\text{nested one-hot encoding}} \mathscr{H}^{(N-1)}\otimes\mathscr{H}^{(N)},
\end{equation}
where $\mathscr{H}^{(n)}\sim\mathbb{C}^{2^n}$ is the $n$ qubit Hilbert space. We refer to the last $N$ qubits as the `OHE qubits'. 

\subsection*{Rewriting the QRAM operation}
\noindent Eq.~\eqref{modified_query} builds upon three main observations. Writing $P_{\bm{m}}=Z^{\otimes \bm{a}}X^{\otimes\bm{b}}$, where e.g. $Z^{\otimes \bm{a}}=\bigotimes_iZ_i^{a_i}$ the first observation is that for any Pauli operator $P_{\bm{m}}$ acting on the QPU register, one has 
\begin{equation}
    VX^{\otimes \bm{b}}V^{-1} = U_{\bm{m}} \otimes \pi_{\bm{m}}.
\end{equation}
Therein, $U_{\bm{m}}$ is a deeply non-Clifford byproduct, but it only acts on the first $N-1$ output qubits in Eq.~\eqref{V_spaces}. Differently $\pi_{\bm{m}}$ is a very simple and highly structured \emph{permutation} of the \emph{physical} OHE qubits. The second (straightforward) observation is that $U_\text{QRAM}Z^{\otimes \bm{a}}=Z^{\otimes \bm{a}}U_\text{QRAM}$. The third is that, trivially, $W_DU_{\bm{m}} = U_{\bm{m}} W_D$.\\
\indent Combining these concepts, $U_{\bm{m}}$ cancels out, and Eq.~\eqref{modified_query} follows:
\begin{equation}\label{U_QRAM_blocks}
    U_\text{QRAM} = \Big[P_{\bm{m}}^{-1}V^{-1}\Big]\Big[\pi_{\bm{m}}W_D\pi_{\bm{m}}^{-1}\Big]\Big[V P_{\bm{m}}\Big].
\end{equation}
Crucially, we can account for $\pi_{\bm{m}}$, which substitutes $W_D$ with $W_D(\bm{m})$, in \emph{classical} processing: we reabsorbe it in a reordering of the dataset, which does not require any reading of the memory. The special structure of $\pi_{\bm{m}}$ allows this to be achieved efficiently (SM).  

\begin{figure*}[t!]
    \begin{minipage}{1.0\textwidth}
        \includegraphics[width=\textwidth]{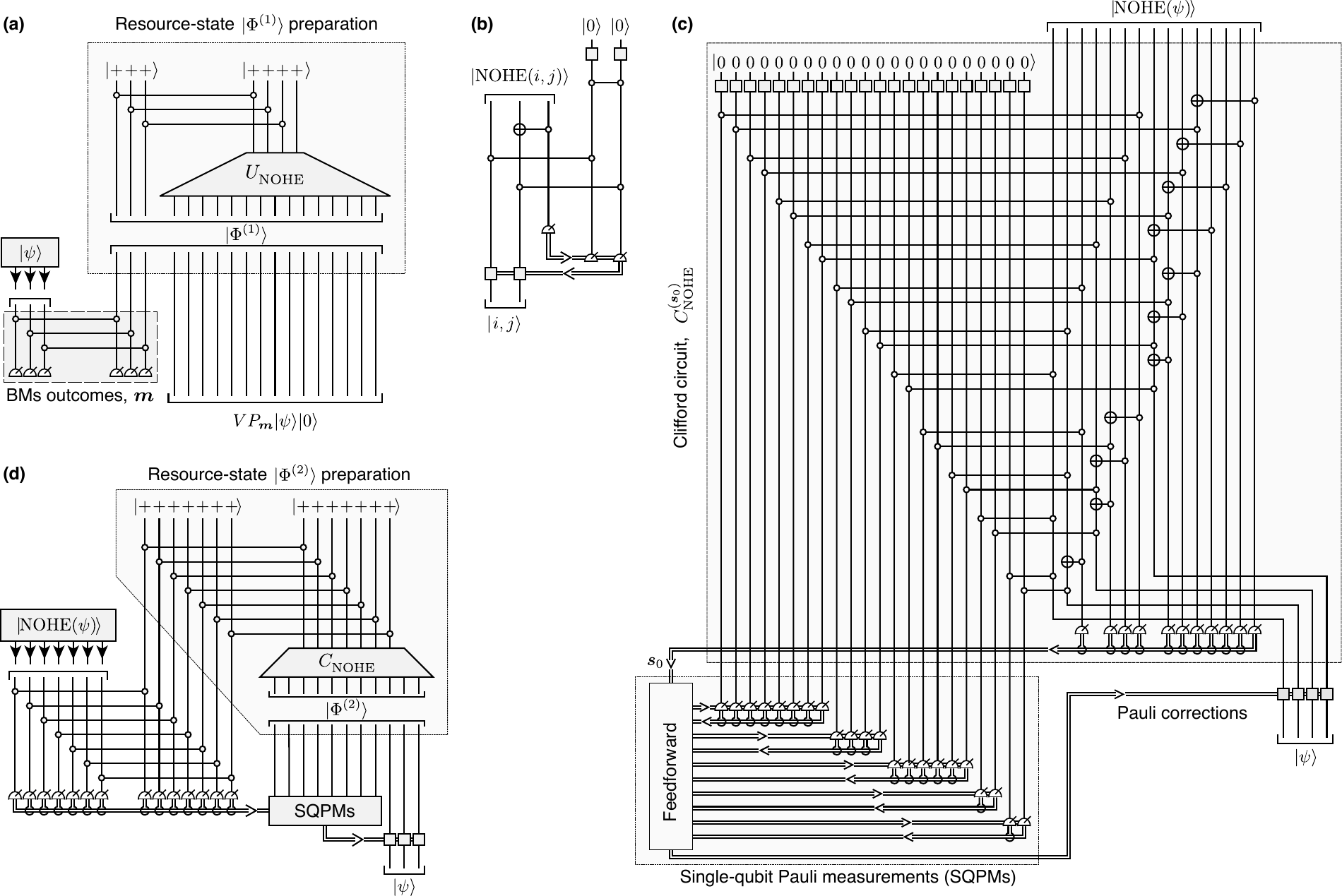}
    \end{minipage}%
\caption{\textbf{Clifford implementation of NOHE, and its Clifford inversion.} (a) Gate-teleportation (GT) inspired implementation of $VP_{\bm{m}}$ in $\mathcal{O}(1)$ online depth, where $P_{\bm{m}}$ is a random Pauli operator revealed by the outcomes of the Bell measurements (BMs). The first part prepares the RS $|\Phi^{(1)}\rangle$, which is independent on the input: and can be executed at any time prior to the query and contains all the non-Clifford complexity. The second part implements BMs between the input $\ket{\psi}$ and the RS. In contrast with GT, here we do \emph{not} correct for the byproduct, and keep any obtained output $VP_{\bm{m}}\ket{\psi}$. (b) Although the basic operation $\ket{i,j}\rightarrow\ket{\text{NOHE}(i,j)}$ is non-Clifford, it can be inverted with a Clifford circuit - with the aid of the \emph{stabiliser} ancillary state $\ket{\Psi_{00}}=CZ\ket{+,+}$. Therein, the first measurement is in the $X$ basis, while the second two are either in the $X$ or in the $Z$ basis, depending on the previous outcome; the last layer of gates single-qubit gates is of the form $Z^\alpha\otimes Z^\beta$, with $\alpha,\beta$ determined by the previous measurement outcomes. (c) Generic circuit for the Clifford inversion of the NOHE. This is composed by two parts: the first is a circuit $C_\text{NOHE}$ featuring deterministic operations a series of controlled-not gates (which are Clifford) followed by a set of measurements in the $X$ basis. The second part is a sequence of \emph{adaptive} single-qubit Pauli measurements, where each qubit is measured either in the $X$ or in the $Z$ basis depending on previous measurement outcomes. (d) Resource-state protocol for inverting $U_\text{NOHE}$. The first part prepares a stabiliser RS $|\Phi^{(2)}\rangle$ by applying $C_\text{NOHE}$ to one partition of a set of Bell pairs. The second part implements BMs between the RS and the input $\ket{\text{NOHE}(\Psi)}=U_\text{NOHE}\ket{\psi}$. The third part implements a sequence of adaptive SQPMs analogous as the one above, eventually yielding the output $\ket{\psi}$.  
 }
\label{fig_M2}
\end{figure*}

\subsection*{Clifford generalised NOHE via gate-teleportation}

\noindent In Eq.~\eqref{U_QRAM_blocks}, step (i) requires to apply $VP_{\bm{m}}$, where $P_{\bm{m}}$ is a Pauli operator. We realize this with the Clifford procedure in Fig.~\ref{fig_M2}(a), during which $P_{\bm{m}}$ is determined randomly. This is inspired by the gate-teleportation (GT) paradigm~\cite{Nielsen_1997,gottesman1999demonstrating}, recalled in Fig.~\ref{fig_M1}(e). Specifically, we use the pre-assembled resource-state (RS) 
\begin{equation}\label{Phi_1}
    \ket{\Phi^{(1)}} = \frac{1}{N}\sum_{\bm{y},\bm{z}} (-)^{\bm{y}\cdot\bm{z}} \ket{y}_\mathcal{I}\otimes\ket{\text{NOHE}(\bm{z},+)}_\mathcal{D},
\end{equation}
where $\mathcal{I}$ and $\mathcal{D}$ feature $\log N$ and $2N\!-\!1$ qubits. Initially the state of the register qubits $\mathcal{R}$ is $\ket{\psi}$. Then, we implement BMs between $\mathcal{R}$ and $\mathcal{I}$, with outcomes $\bm{m}$, which leaves $VP_{\bm{m}}\ket{\psi}\ket{0}$ on the unmeasured qubits $\mathcal{D}$. \\
\\
\noindent \textbf{Comments.} This protocol only features Clifford operations - the $\mathcal{O}(\log N)$ parallel BMs. However, the non-stabiliser RS $|\Phi^{(1)}\rangle$ must be given. The main complexity of $V$, specially for QEC (see discussion below), is therefore outsourced to the RS preparation. Crucially, this inherits the benefits of RS-based approaches in quantum information, e.g. `T-gate injection'~\cite{gottesman1999demonstrating}: we no longer need to apply $V$ to an \emph{arbitrary} input $\ket{\psi}$ during the algorithm (i.e., online). In contrast, we simply need to prepare a RS, which can be done at any time before the query (offline) in a dedicated `factory' - allowing for efficiently exploiting, e.g., state purification and post-selected protocols.\\
\indent Typically, GT is limited to relatively simple operations due to its intrinsic randomness: if we wanted to apply \emph{exactly} $V$, to cope with $P_{\bm{m}}$ we would need to apply the byproduct correction $VP_{\bm{m}}V^{-1}$, which is in general almost as complex as $V$ itself. Our crucial insight in the rewriting above, is that when it comes to QRAM we do \emph{not} need to correct for the byproduct: to all effects, the random OHE $VP_{\bm{m}}$ is completely sufficient.

\begin{figure*}[t!]
    \begin{minipage}{1.0\textwidth}
        \includegraphics[width=\textwidth]{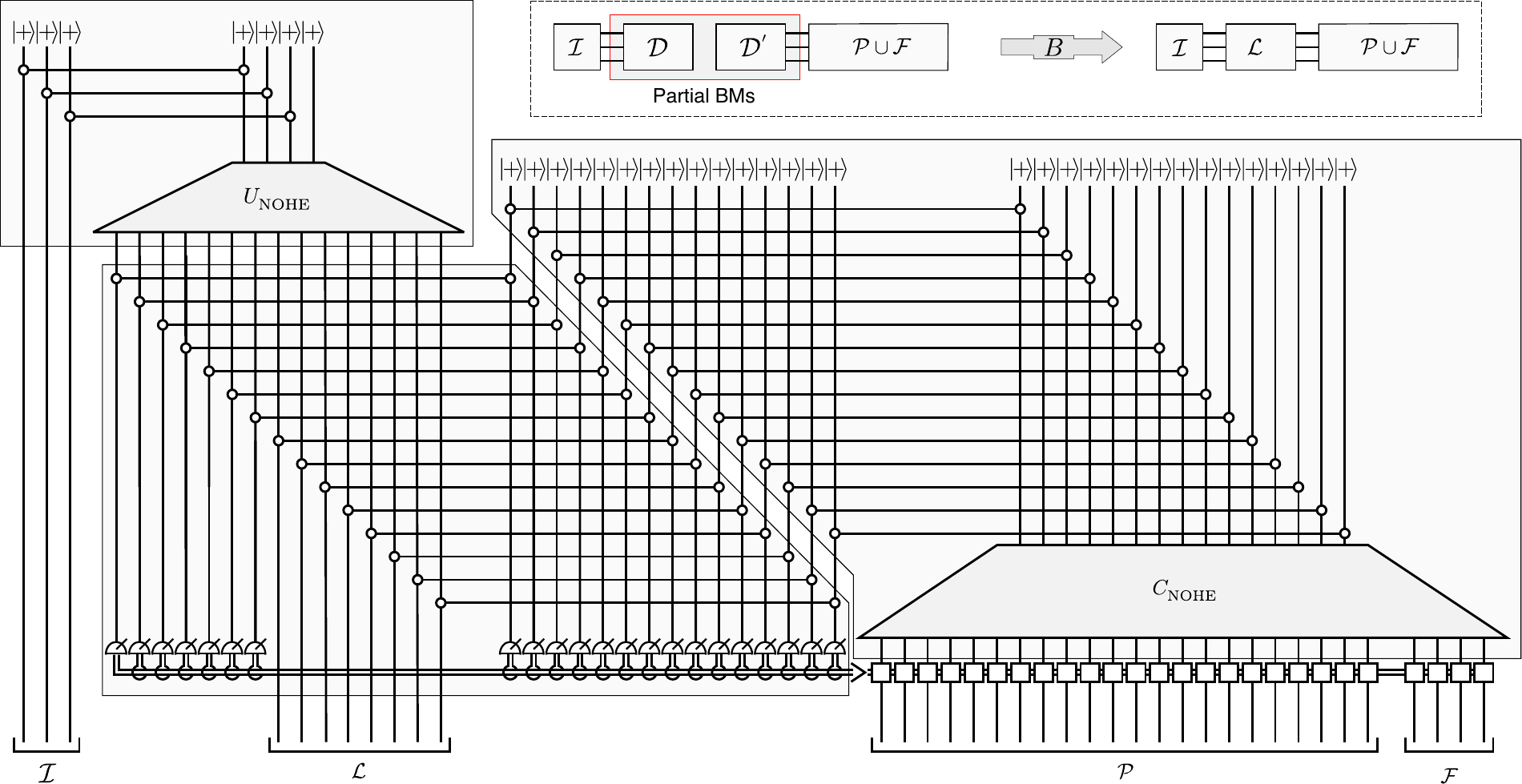}
    \end{minipage}%
\caption{\textbf{Circuit for the resource-state.} The resource-state $\ket\Phi$ can be understood as emerging from a `synthesis' of the previously introduced states $|\Phi^{(1)}\rangle$ and $|\Phi^{(2)}\rangle$. Here, we display a circuit which defines it formally. In the main figure, the top-left and right panels highlights the preparation of $|\Phi^{(1)}\rangle$ and $|\Phi^{(2)}\rangle$ respectively; the bottom-left panel contains a `partial' Bell measurement, where not all the qubits are projected. This is also summarised in the scheme above, which highlights the action of the contraction $B$; therein, horizontal lines informally represent the presence of entanglement between sets of qubits. }  
\label{fig_M3}
\end{figure*}

\subsection*{Clifford NOHE inversion}
\noindent Step (iii) of Eq.~\eqref{U_QRAM_blocks} requires to implement $P_{\bm{m}}^{-1}V^{-1}$. Recalling that $V^{-1}=H_\text{bus}U_\text{NOHE}^{-1}$, the only complex part is $U_\text{NOHE}^{-1}$, which is deeply non-Clifford. However, regarding $U_\text{NOHE}$ as an isometry from $\log N$ to $N-1$ qubits, and only considering its inversion on the image $U^{(N)}_\text{NOHE}(\mathscr{H}^{\log N})$, we invert it via Clifford operations only. Below we summarise the key insights achieving this.\\
\\
\noindent \textbf{Minimal case.} Fig.~\ref{fig_M2}(b) displays our protocol for inverting $U_\text{NOHE}^{(4)}$: given any three-qubit input state of the form $\ket{\chi}=U_\text{NOHE}^{(4)}\ket{\psi,0}$, it returns $\ket{\psi}$. We note that this is similar to the method of Ref.~\cite{Gidney2018halvingcostof} for inverting the AND operation, with the key difference that our protocol only requires SQPMs, and no adaptive entangling operation. This is crucial for efficiently integrating this minimal building-block in our generic measurement-based construction for arbitrary $N$.\\
\\ 
\noindent \textbf{General case.} Fig.~\ref{fig_M2}(c) illustrates the NOHE inversion for arbitrary $N$. This builds on the fact that $U^{(N)}_\text{NOHE}$ is implemented by iterating $U_\text{NOHE}^{(4)}$: hence, we invert it by iterating the procedure above. \\
\indent The resulting inversion protocol therefore combines simple Clifford gates and SQPMs. Importantly, in Fig.~\ref{fig_M2}(c) we \emph{first} execute the gates, and \emph{then} the SQPMs. This is allowed by the Cliffordness of the procedure: Pauli byproducts are propagated up to a SQPM, where they are accounted for by interpreting the outcome. This allows to divide the circuit in two parts: the first, $C_\text{NOHE}$, contains the Clifford operations and SQPMs that are \emph{pre-determined}; the second contains \emph{adaptive} SQPMs. Since the outcomes $\bm{s}_0$ of the measurements in $C_\text{NOHE}$ are probabilistic, the first part of the circuit includes $\bm{s}_0$ as a classical parameter, 
\begin{equation}
    C_\text{NOHE}^{(\bm{s}_0)} \;\; : \;\; \mathscr{H}_\mathcal{D} \longrightarrow \mathscr{H}_\mathcal{P}\otimes\mathscr{H}_\mathcal{F}.
\end{equation}
Here, $\mathcal{D}$, of size $|\mathcal{D}| = N-1$, is the set of qubits supporting the input, while $\mathcal{P}$, with $|\mathcal{P}|=2(N-\log N - 1)$, contains those which will undergo the adaptive SQPMs; finally, $\mathcal{F}$ collects the $|\mathcal{F}|=\log N$ output qubits. \\
\indent The final adaptive SQPMs are performed similarly to the second round of measurements for $N=4$. For general $N$, our result is as follows: given any arbitrary state $\ket\Xi\in U_{\text{NOHE}}\left( \mathscr{H}^{(\log N)} \right)$, then we implement
\begin{equation}\label{SQPM_inversion}
    C_\text{NOHE}^{(\bm{s}_0)} \ket{\Xi} \xrightarrow[\;\;\;\;\;\text{SQPMs}\;\;\;\;\;]{} U^{-1}_{\text{NOHE}} \ket{\Xi};
\end{equation}
moreover, the protocol is always completed in $T_f=2\log N-3$ measurement rounds. \\
\\
\noindent \textbf{RS based implementation.} In Fig~\ref{fig_M2}(d) we show how this inversion is executed with the aid of the RS
\begin{equation}\label{Phi_2}
    \ket{\Phi^{(2)}} = \frac{1}{2^{N-1}}\sum_{\bm{Y}, \bm{Z}} (-)^{\bm{Y}\cdot\bm{Z}} \ket{\bm{Y}}_{\mathcal{D}'}\otimes C_\text{NOHE}\ket{\bm{Z}}_{\mathcal{P}\cup\mathcal{F}},
\end{equation}
where the indexes run over $\bm{Y}, \bm{Z}\in\left\{0,1\right\}^{N-1}$ and $|\mathcal{D}'|=N-1$. Importantly, since $C_\text{NOHE}$ is Clifford, $|\Phi^{(2)}\rangle$ is a \emph{stabiliser} state; this is crucial for the inversion scheme to succeed via SQPMs only. Moreover, this allows to leverage standard stabiliser formalism techniques, to drastically optimise $C_\text{NOHE}$, eliminating the non-adaptive measurements; remarkably, after these optimisations we end up with a compact RS $|\Phi^{(2)}\rangle$, which can be prepared efficiently from an initial set of \emph{linear} cluster states~\cite{PhysRevLett.86.5188}.

\subsection*{QRAM resource-state}
\noindent Fig.~\ref{fig_M3} provides an operational definition of $\ket{\Phi}$. This can be understood as a Clifford `merging' of $|\Phi^{(1)}\rangle$ and $|\Phi^{(2)}\rangle$,
\begin{equation}\label{Phi_P}
    \ket{\Phi} = B\Bigg[\ket{\Phi^{(1)}}\otimes\ket{\Phi^{(2)}}\Bigg],
\end{equation}
via a linear contractive map $B:\mathscr{H}^{(3N-2)}\rightarrow\mathscr{H}^{(N)}$ acting on a subset of the qubits. This describes `partial' BMs, and thus is intrinsically random. However, since $|\Phi^{(2)}\rangle$ is a stabiliser state, we cope with this via final Pauli corrections, making the output deterministic.

\subsection*{Query protocol}
\noindent The query protocol in Fig.~\ref{fig2} follows directly by combining the concepts above. Since this requires several additional technical details, we provide an extended account for it in the SM, where we explicitly analyse all the steps. 

\subsection*{Noise-resilience}

\noindent Let the (potentially faulty) output of the QRAM be the density matrix $\rho_D(\psi)$. Then, the \emph{query fidelity} reads
\begin{equation}
    F_\text{QRAM} = \frac{1}{2^N}\sum_D \int \mathrm{d}\psi \;  \Big\langle \psi_D \Big|\rho_D(\psi) \Big|\psi_D \Big\rangle,
\end{equation}
where $2^{-N}\sum_D$ and $\int\mathrm{d}\psi$ average over all possible datasets and inputs. Typically, it scales as $F_\text{QRAM}\sim \exp\left\{-\text{poly} N\right\}$, as $\mathcal{O}(N\log N)$ space-time events are potentially faulty. This makes near-term implementations unfeasible, and also obstructs QEC, as the needed QEC resources would scale super-exponentially in the processor size. In contrast, our NOHE method results in 
\begin{equation}
    1-F_\text{QRAM} \; \lesssim \; \text{polylog} N;
\end{equation}
this is inherited from the \emph{bucket-brigade} (BB) protocol introduced by Giovannetti and colleagues~\cite{giovannetti2008quantum}, whose noise-resilience was rigorously proved~\cite{hann2021resilience}.\\
\\
\noindent \textbf{Mechanism.} At its core, this built-in noise-resilience derives from the strongly constrained error propagation within the circuit decomposition of $V$ in Fig.~\ref{fig_M1}(a), where - analogously to the BB - the large majority of faults propagate in such a way to eventually disentangle from the final output. To understand how this occurs in our protocol, recall that our procedure is based on the destructive consumption of the RS $\ket{\Phi}$: all qubits are measured individually and disentangled from the rest, except for the $\log N+1$ output qubits. Crucially, by design most of the entropy generated by faults is left in the measured qubits, and thus removed from the output. While we use feedforward of the (potentially faulty) measurement outcomes to decide the future measurements bases, standard concepts in measurement-based QC~\cite{PhysRevLett.86.5188, PhysRevA.68.022312, briegel2009measurement} show that this is equivalent to the non-suppressed errors in the BB (see the SM). \\
\indent Similarly to Ref~\cite{hann2021resilience}, we prove a bound of the form
\begin{equation}
    F_\text{QRAM} \geq 4\Bigg[\frac{1}{N}\sum_{j}p_j |\mathcal{G}_j| - \frac{1}{2}\Bigg]^2,
\end{equation}
where $\mathcal{G}_j\subseteq \left\{0,1\right\}^{\log N}$ is a family of error-free classical addresses defined in the SM, and $\left\{p_j\right\}$ is a probability distribution specified by the error model. Crucially, the NOHE inherits from the BB the fact that
\begin{equation}\label{expected_good_locations}
    \sum_jp_j|\mathcal{G}_j| \simeq N -C \varepsilon N\log^\alpha N - \mathcal{O}(\varepsilon^2),
\end{equation}
where $\varepsilon$ is the error-rate, $C$ is a combinatorial term and $\alpha$ is the dominating exponent. This eventually results in 
\begin{equation}\label{logarithmic_error}
    1-F_\text{QRAM} \leq  8C\varepsilon \log^\alpha N + \mathcal{O}(\varepsilon^2),
\end{equation}
demonstrating the polylogarithmic error scaling.\\
\\
\noindent \textbf{Numerical analysis.} Beyond the analytical discussion above, we certify the scaling of $F_\text{QRAM}$ via numerical simulations, demonstrating the expected behaviour and extracting the dominating exponents $\alpha$. Despite the dimension of the Hilbert space scaling as $2^{\mathcal{O}(N)}$, specialised techniques~\cite{hann2021resilience} allow simulating queries up to $N\simeq 10^4$. Our simulations are based on standard quantum Monte Carlo methods, wherein for all parameters we converge to a precision $\sigma\lesssim 10^{-5}$. \\
\indent We consider several relevant error-models, including standard depolarisation, biased Pauli errors (up to full dephasing), qubit losses, atom heating and phenomenological noise; we confirm Eq.~\eqref{logarithmic_error} across all the tested settings, with $\alpha$ showing (weak) dependence on the model. Detailed results and methods are reported in the SM.\\
\indent Fig.~\ref{fig4}(e) shows our analysis for the strongest noise model considered, where all qubits (both if idle and active) continuously depolarise with rate $\varepsilon$ per circuit step; since we are interested in the high-fidelity regime $\varepsilon\ll 1$, we aim at extracting the first coefficient in the expansion for small $\varepsilon$, which is defined as $A_1(N)=\lim_{\varepsilon\rightarrow 0}(1-F_\text{QRAM})/\varepsilon$. By decreasing $\varepsilon$ up to $10^{-5}$, collapse to a common curve probes that we reached the limit $\varepsilon\rightarrow 0$ for the considered $N$. Moreover, in the plot the vertical axis is logarithmic, while the horizontal axis is log-log (i.e., it linearly displays $\log\log N$), so that collapse to a straight line demonstrates the logarithmic scaling in Eq.~\eqref{logarithmic_error}; by linearly fitting and extracting the slope, we can calculate $\alpha \simeq 2.2$.

\begin{figure*}[t!]
    \begin{minipage}{1.0\textwidth}
        \includegraphics[width=\textwidth]{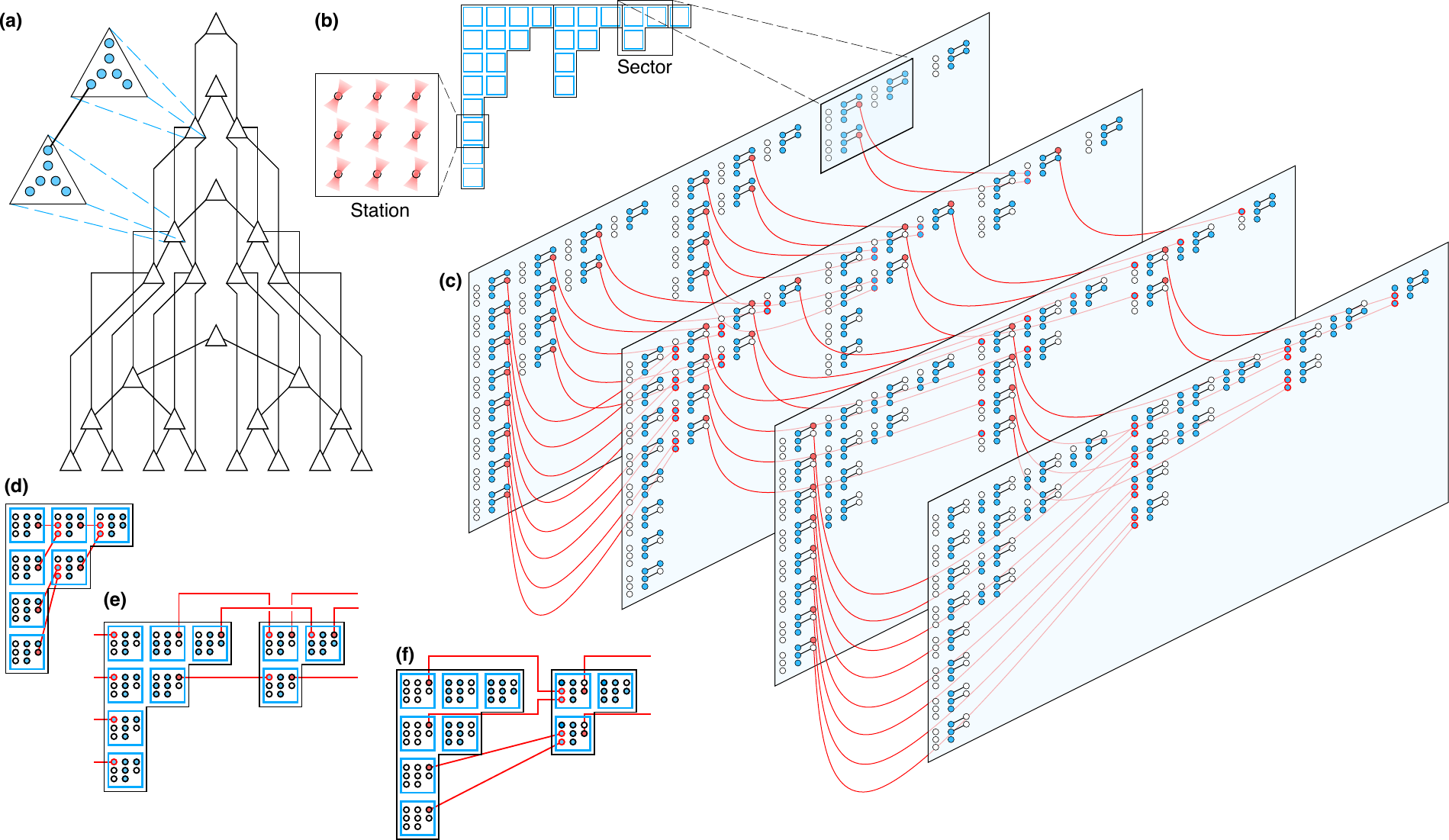}
    \end{minipage}%
\caption{\textbf{Example of rearrangement subroutine for the QRAM factory.} (a) Graphical representation of the space-time connectivity needed for implementing the circuit $U_\text{NOHE}$, realising a nested bifurcation-graph (NBG, see the SM). As highlighted in the figure, triangules are related to groups of six qubits; bonds between such triangles represent the presence of an entanglement bond between them, in the sense that a Bell pair is shared. The NBG is the spatial connectivity that we aim at building by sharing Bell pairs via dynamical atom rearrangement: after this is achieved, the circuit for $U_\text{NOHE}$ can be implemented via local operations only. (b) The spatial configuration of static tweezer traps (generated, e.g., via a spatial light modulator, SLM) needed for the rearrangement. Specifically, we consider 
$\log N$ vertical \emph{sectors}, each comprising a growing number of \emph{stations} organized in columns of growing size. Each station contains $9$ SLM tweezer traps. (c) Depiction of the rearrangement subroutine, which achieves the NBG connectivity via $3$ parallel rearrangements. The rearrangements are performed with an additional tweezer array, generated by an acousto-optical deflector (AOD), which allows to transport the atoms. At the beginning of the protocol, each station stores $5$ atoms, wherein $4$ compose $2$ separate Bell pairs (represented by the bonds), while the fifth is disentangled. Then, the three parallel rearrangements reconfigure the system layout, eventually realising the needed NBG connectivity. (d,e,f) Details of the rearrangements, which are explained extensively in the SM.   
 }
\label{fig_M5}
\end{figure*}

\subsection*{Quantum error-corrected QRAM}

\noindent Our proposed QRAM module overcomes several QEC challenges, which have long fueled skepticism on the possibility of error-correcting (and thus scaling-up) a QRAM - and QC more in general~\cite{aaronson2015read}. The first challenge is the conjectured exponential error scaling. The second is the non-Cliffordness of $V$, demanding unbearable resources for FT implementations, as $\mathcal{O}(N)$ non-Clifford gates need to be applied \emph{during} the query.\\
\indent In our QRAM design, errors scale polylogarithmically, and the query is \emph{entirely} Clifford. Most importantly, we show below that, with standard QEC schemes, any operation can be implemented fault-tolerantly via \emph{transversal gates} (TGs) only. Since our protocol is entirely based on SQPMs, measurement-based approaches~\cite{Raussendorf_2006} to QEC are beneficial. A paradigmatic recipe is to pre-assemble a 3D stabiliser state which foliates the desired QEC code~\cite{PhysRevLett.117.070501}. Modern approaches grow this RS during the computation, realising a fusion-based scheme~\cite{PhysRevLett.131.120604, PhysRevX.13.041013}. For neutral atom implementations, this can be specially beneficial for correcting atom losses, e.g. from heating. \\
\\
\noindent \textbf{Fault-tolerant query.} 
\noindent The query is entirely transversal on codes where e.g. the following operations are transversal: (i) Either the CNOT or the $CZ$ gate; (ii) Single-qubit \emph{destructive} measurements in the $X$ and $Z$ basis. This is the case e.g. of CSS codes, and of the surface code. When these requirements are satisfied, for the query we only need $\log N$ parallel destructive BMs (implemented e.g. through $CZ$s and $X$ measurements), and $\mathcal{O}(\log N)$ rounds of SQPMs on $\mathcal{O}(N)$ logical qubits. Thus, there is no $T$ cost associated with the query. \\
\indent To further understand the advantage of our query scheme, recall that typically one would continuously extract the syndrome via expensive protocols, featuring many ancillas, several rounds of state-verification and repeated measurements~\cite{nielsen2010quantum, shor1996fault, knill2005quantum, PhysRevA.57.127}. Importantly, our procedure only contains \emph{destructive measurements} of the logical qubits in the Pauli basis; combining the physical SQPMs outcomes also provides important syndrome information. This can potentially be leveraged to eliminate the need for redundant syndrome extraction, e.g. through Knill-type teleportation-based QEC~\cite{knill2005quantum}, or by exploiting concepts of `algorithmic fault-tolerance'~\cite{Zhou_2025}.

\subsection*{QRAM factory details}
\noindent Preparing $\ket\Phi$ is the most demanding part, both with and without QEC: any remarkable complexity of QRAM is now outsourced here. Our approach has several advantages: (1) The preparation does not directly impact the online query: once the main algorithm is running, any highly complex protocol is already executed, and $\ket\Phi$ is a resource furnished to the processor. (2) We no longer need to apply $U_\text{NOHE}$ to an \emph{arbitrary} input: rather, we prepare a \emph{specific}, known RS, independent on algorithm and memory. This is a much simpler task, as discussed above regarding the GT implementation of step (i). (3) Heralding procedures based on error-detection and `try-until-succeed' can be leveraged, as discussed below. \\
\\
\noindent \textbf{Cost evaluation} Fig.~\ref{fig_M3} employs $2N-\log N - 2$ Toffoli gates. Since this is the only necessary non-Clifford gate, a sound strategy is to distill ad-hoc RSs~\cite{PhysRevA.63.052314, PhysRevA.87.032321, Haah_2018}, minimising the `magic-state factories' (MSFs). Alternatively, one can distill $T-$states~\cite{PhysRevA.71.022316, PhysRevA.86.052329}. Since the preparation of $\ket\Phi$ does not impact the runtime of the principal algorithm, it is convenient to minimise the number of needed MSFs, and thus use minimal decompositions of the Toffoli gate in four $T$ gates~\cite{PhysRevA.87.022328}. Crucially, the times $T_\text{query}$ and $T_\Phi$ needed for a query and for preparing $\ket{\Phi}$ satisfy $T_\Phi\simeq T_\text{query}$: thus, the QRAM factory can operate \emph{in parallel} with the query, assembling a new copy of $\ket{\Phi}$ while an old copy is consumed. Since the query has no $T-$cost, the QRAM factory can employ all the available MSFs. \\
\\
\noindent \textbf{Heralded preparation.} We elaborate in detail on the heralded QRAM factory in the SM. The key idea is as follows. We choose an integer $m\ll\log N$, such that also $2^m\equiv M \ll 1/\varepsilon$, where $\varepsilon$ is the operational error. Then, we implement error-detection on all the gates where the first $m$ bits of the address act as controls; whenever an error is detected, we abort the protocol and start again. In total, this therefore requires heralding the first $\mathcal{O}(M\log N)$ gates, which is a negligible fraction of the $\mathcal{O}(N)$ total gates. The average Toffoli waste $\sim \varepsilon M^2\ll N$ is small by construction, compared with the total cost. However, owing to the strongly constrained error-propagation of our protocol, which results in the observed polylogarithmic infidelity scaling of Eq.~\eqref{logarithmic_error}, these first gates provide a major, extensive contribution to the final fidelity. In the SM, we show that the bound in Eq.~\eqref{expected_good_locations} is modified as
\begin{equation}
    \sum_jp_j|\mathcal{G'}_j| \simeq N -C \varepsilon N\log^\alpha N + C\varepsilon N m^\alpha - \mathcal{O}(\varepsilon^2) .
\end{equation}
The fidelity improvement then follows directly, as also observed numerically in Fig.~\ref{fig4}(c). We note that simulations suggest a scaling considerably more favorable than the bound above, approximately as $1-F_\text{QRAM} = \mathcal{O}(\varepsilon\log N \log (N/M))$, therefore underlying an extensive improvement of the form $\Delta F_\text{QRAM} \sim \varepsilon m \log N$.

\subsection*{Neutral atom blueprint}

\noindent We now detail our implementation with modern neutral-atom processors, which are described e.g. in Refs.~\cite{Bluvstein_2023, reichardt2024logicalcomputationdemonstratedneutral, radnaev2025universalneutralatomquantumcomputer}. Here, the physical qubits are single atoms trapped in optical tweezers. Key to this QC platform is the possibility to coherently rearrange the atoms via the tweezers~\cite{Bluvstein_2022}, allowing for mid-circuit reconfiguration of the system connectivity. Recently, this allowed to probe logical error-corrected QC~\cite{Bluvstein_2023, reichardt2024logicalcomputationdemonstratedneutral}, by harnessing atom transport for native transversal gates. This strongly motivates our focus on neutral-atoms, as it naturally exploits our Clifford query. Moreover, atom shuttling combines surprisingly well with the QRAM factory, allowing for efficient, high-rate preparation of $\ket{\Phi}$.\\
\\
\noindent \textbf{Neutral atom QRAM.} Fig.~\ref{fig1}(d) illustrates our implementation. The qubits $\mathcal{A}$ supporting the RS $\ket\Phi$ are stored statically, realising a `QRAM storage zone'. Between queries, low-rate syndrome extraction is performed to detect errors on $\ket{\Phi}$ due to decoherence.\\
\indent For the query, we interface the $\log N$ qubits in $\mathcal{I}\subset\mathcal{A}$ with the $\log N$ QPU qubits: we can employ coherent atom transport, or leverage pre-distilled Bell pairs for entanglement swapping, and perform BMs. As previously explained, this directly provides syndrome information useful to correct errors during the gates and the transport (via `Knill QEC'~\cite{knill2005quantum}). The rest of the query is implemented via local, highly parallelised operations on the remaining \emph{physical single atoms} in $\mathcal{A}$. For the measurements, (at least) two approaches are possible: (i) As shown in recent experiments with alkaline-earth atoms, specific atomic structures allow single-qubit non-destructive measurements~\cite{Ma_2023, PhysRevX.13.041034, PhysRevX.13.041035, PhysRevX.13.041051}; (ii) Otherwise, coherent transport to a `readout zone' can be used, as demonstrated with alkali atoms~\cite{Bluvstein_2023, radnaev2025universalneutralatomquantumcomputer}.\\
\indent Finally, we extract the $\log N +1$ qubits in $\mathcal{F}\subset\mathcal{A}$, and feed them back to the QPU; for this, again we can use coherent transport, or pre-assembled Bell pairs. This completes an entirely transversal QRAM query. \\
\\
\noindent \textbf{Rearrangement subroutine.} Central to our QRAM design is the preparation of $\ket\Phi$, which is carried out independently with a scheme detailed in the SM. One crucial part is a subroutine that essentially implements $U_\text{NOHE}$ in Fig.~\ref{fig_M1}. Similarly to any QRAM method (e.g., the BB~\cite{giovannetti2008quantum}), for this the major challenge is the non-local connectivity. Here, we leverage coherent atom transport to artificially build the needed connectivity while preserving the error-resilience. However, at the hardware level there are several stringent requirements arising from the physical tools (e.g., the acousto-optical deflector, AOD employed to move the tweezers), which strongly constrain the possibility of parallelising atom motions~\cite{Bluvstein_2022, Tan_2024}; moreover, atom transport is typically slow, as to minimise atom losses. Surprisingly, we find that for the specific task of implementing the QRAM factory, the needed atom rearrangements combine remarkably well with these requirements: as we show in the SM, $\mathcal{O}(1)$ system reconfigurations are sufficient. \\
\indent Specifically, Fig.~\ref{fig_M5} details a dynamical rearrangement scheme, which is compatible with all the AOD constraints, and is sufficient for the QRAM factory to then proceed via local operations only. This comprises $3$ parallel rearrangements, wherein each atom is moved for a distance bounded by $\mathcal{O}(N)$. A more sophisticated scheme, discussed in the SM, makes an optimal use of the 2D geometry, compactifying the storage positions. While this now requires $6$ parallel rearrangements, it reduces quadratically the maximal motion distance to $\mathcal{O}(\sqrt{N})$, resulting in a significantly faster protocol.\\
\\
\noindent \textbf{Runtime tradeoff.} In Eq.~\eqref{main_T_phi}, three terms contribute to the factory runtime,
\begin{equation}
    T_\Phi = T_g + T_m + T_r,
\end{equation}
representing the time employed for performing the gates ($T_g$), the measurements ($T_m$) and the atom rearrangement ($T_r$). The first two both scale as $\mathcal{O}(\log N)$, but $T_g$ is negligible as in neutral-atom experiments~\cite{Bluvstein_2022} the typical gate time ($\tau_g\sim 100\mathrm{ns} $) is much smaller than the typical readout time ($\tau\sim 500\mu\mathrm{s}$). We find that $T_m=2\tau\log N$, matching exactly the query runtime $T_\text{query}$. \\
\indent The most crucial calculation concerns $T_r$. Following Refs.~\cite{Bluvstein_2022, Tan_2024}, the time needed for moving an atom across a distance $d$ scales as $T(d)=T_0\sqrt{d/d_0}$, where $T_0\simeq 200\mu\mathrm{s}$ is the measured time for moving an atom across a distance of $d_0 = 110 \mu\mathrm{m}$ with negligible loss and decoherence. This formula assumes tweezer motion with constant acceleration, optimized so that the atom loss rate (which is the dominating source of errors) is independent on the transport distance. Then, the fact that we only employ $\mathcal{O}(1)$ total rearrangements, none exceeding a distance $\mathcal{O}(\sqrt{N})$, yields $T_r\lesssim \mathcal{O}(TN^{1/4})$, where $T=T_0\sqrt{l/d_0}$ and $l$ is the minimal distance between tweezers in the processor. In the SM, we derive the prefactor to be bounded by $3\sqrt{6}$, resulting in Eq.~\eqref{main_T_phi}; our quantitative estimates assume $l\simeq 3\mu\mathrm{m}$, which is aligned with current experimental parameters~\cite{Bluvstein_2022}. \\
\indent In the SM we show how this estimation is modified in a QEC setting, finding that a similar tradeoff holds. Indeed, due to the repeated cycles of syndrome extraction, the FT query runtime is multiplied by a factor $\mathcal{O}(d)$, leading $T'_\text{query}\simeq dT_\text{query}$. Similar overheads impact $T_g$ and $T_m$. The rearrangement time, however, is only multiplied by a factor $\mathcal{O}(\sqrt{d})$: this is because (assuming a surface code implementation), distances incur a $\mathcal{O}(d)$ factor, but the atom motion time only scales with the square-root of the distance. It follows that the rearrangement impacts progressively less the time tradeoff between query and factory. Finally, $T_\Phi$ now is also affected by a time overhead due to $T-$state factories; in the SM we show that this only grows as $\sim \text{polylog}( \log (\log N))$, and remains small with respect to all the other contributions. Explicit estimations, accounting for the prefactors, are presented in our SM, showing that $T_\Phi/T_\text{query}=\mathcal{O}(1)$ holds in the FT setting for any system size relevant to single-module processors.

\subsection*{Data availability}
\noindent Data and codes are available on upon reasonable request~\cite{zenodo}.

\end{document}


\preprint{APS/123-QED}

\title{Supplemental material for\\
Resource-State Quantum RAM for Fast and Error-Correctable Queries
}
\author{Francesco Cesa}\email{francesco.cesa@uibk.ac.at}
\affiliation{Institute for Quantum Optics and Quantum Information of the Austrian Academy of Sciences, 6020 Innsbruck, Austria}
\affiliation{Institute for Theoretical Physics, University of Innsbruck, 6020 Innsbruck, Austria}
\affiliation{Department of Physics, University of Trieste, Strada Costiera 11, 34151 Trieste, Italy}
\affiliation{Istituto Nazionale di Fisica Nucleare, Trieste Section, Via Valerio 2, 34127 Trieste, Italy}
\author{Hannes Bernien}
\affiliation{Institute for Quantum Optics and Quantum Information of the Austrian Academy of Sciences, 6020 Innsbruck, Austria}
\affiliation{Institute for Experimental Physics, University of Innsbruck, 6020 Innsbruck, Austria}
\affiliation{Pritzker School of Molecular Engineering, University of Chicago, Chicago, IL 60637, USA}
\author{Hannes Pichler}\email{hannes.pichler@uibk.ac.at}
\affiliation{Institute for Quantum Optics and Quantum Information of the Austrian Academy of Sciences, 6020 Innsbruck, Austria}
\affiliation{Institute for Theoretical Physics, University of Innsbruck, 6020 Innsbruck, Austria}


\maketitle

\tableofcontents

\newpage

\section{Non-Cliffordness of the QRAM operation}\label{section_non_cliffordness}
\noindent Here, we investigate the non-Cliffordness of the main building blocks of QRAM. For this, we first recall the concept of Clifford hierarchy, and then show that the QRAM operation, in general, requires to implement circuits which are at deep layers in the hierarchy.

\subsection{Clifford hierarchy: basics}
\noindent Recall that the Clifford hierarchy 
\begin{equation}
    \mathcal{C}_0\subset\mathcal{C}_1\subset\mathcal{C}_2\subset\mathcal{C}_3\subset\dots
\end{equation}
is defined as follows. First, $\mathcal{C}_0$ is set to be the Pauli group - that is, the group of all operators that can be written as a tensor product of Pauli operators $X, Y, Z$, possibly up to phases $\pm 1$ and $\pm i$. Next, the $K-$th layer of the Clifford hierarchy is defined such that it maps Pauli operators to the Clifford layer below. That is, 
\begin{equation}
    \mathcal{C}_K = \left\{U \;\;\; \text{such that} \;\;\;  UPU^\dag\in\mathcal{C}_{K-1} \;\;\; \forall \;\;\; P\in\mathcal{C}_0\right\}.
\end{equation}
We note that, formally, $U\in\mathcal{C}_K$ for some $K$ implies that $U\in\mathcal{C}_{K'}$ for any $K'\geq K$. Typically, we are interested in the \emph{lowest} layer an operation $U$ belongs to. For instance, if $U\in\mathcal{C}_K$ but $U\notin\mathcal{C}_{K-1}$, we say that $U$ belongs to the layer $K$ of the Clifford hierarchy.\\ 
\indent With this definition, $\mathcal{C}_1$ is said to be the Clifford group; it contains important operations such as the Hadamard gate $H$, and the controlled-not $CX$ and controlled-phase $CZ$ gates. This can be easily proved to have all the features of a group. However, for $K\geq 2$, $\mathcal{C}_K$ is not closed anymore: for instance, $\mathcal{C}_2$ is already universal for quantum computation~\cite{nielsen2010quantum}, implying that in principle any operation $U\in\mathcal{C}_K$ belonging to any Clifford layer can be implemented to arbitrary precision by combining elements from $\mathcal{C}_2$. For this reason, gates from $\mathcal{C}_2$, such as the $T-$gate $T=e^{-iZ\pi/8}$ or the Toffoli gate $CCX= \mathbb{1}-2\ket{11-}\bra{11-}$, are often referred to as `magic gates', where the phrasing `quantum magic' is frequently used for the notion of `non-Cliffordness'.\\
\\
\noindent \textbf{Examples.} In general, given an operator $U$, it might not be evident what Clifford layer it falls in. However, there are specific classes of operators which are simple to analyze. One first instance are multi-controlled phase gates, of the form $C_nZ = \mathbb{1}-2\ket{1}\bra{1}^{\otimes n+1}$. This is a gate acting on $n+1$ qubits, where $n$ `control' qubits trigger a conditional $Z$ gate on the last `target' qubit. It can be easily shown that with this definition, one has
\begin{equation}
    C_nZ \notin\mathcal{C}_{n-1} \;\;\; \text{and} \;\;\;  C_nZ \in\mathcal{C}_n,
\end{equation}
meaning that $C_nZ$ belongs to layer $n$ of the Clifford hierarchy. This can be seen very easily by induction, by noting that $(C_nZ) X (C_n) = X (C_{n-1}Z)$, where $X$ acts on any of the $n+1$ involved qubits, and that $C_0Z=Z$. We note that $C_2Z$ is equivalent to the Toffoli gate up to a Hadamard rotation, and in general $C_nZ$ is equivalent to the multi-controlled-not gate with $n$ controls (up to a $H$ on the target). \\
\indent A second class of operators whose Clifford layer can be easily understood is given by the single-qubit operators of the form $T_n=e^{-iZ\pi/2^{n+1}}$, for whicih one can easily see that
\begin{equation}
    T_n \notin\mathcal{C}_{n-1} \;\;\; \text{and} \;\;\;  T_n \in\mathcal{C}_n.
\end{equation}
Again, this can seen by induction, as $T_nXT_n^\dag = T_{n-1}X$, and $T_0=-iZ\in\mathcal{C}_0$. Note that $T_2=T$ is the $T$ gate, one of the most important `magic gates' for fault-tolerant quantum computation. 

\subsection{Proof of deep non-Cliffordness of QRAM}
\noindent We now discuss the generic non-Cliffordness of $U_\text{QRAM}$ for a memory of size $N$. We note that in general the Clifford layer of $U_\text{QRAM}$ depends on how the memory is composed. For instance, if we had $D_{\bm{x}}=0$ for all memory locations, then $W_D=\mathbb{1}$, implying that also $U_\text{QRAM} = VW_D V^{-1}=\mathbb{1}$ is the identity. However, the composition of the memory only enters the loading operation, and it is simple to see that
\begin{equation}
    W_D \in\mathcal{C}_0 \;\;\;\forall \;\;\; D,
\end{equation}
as it is always defined as a tensor product of $Z$ operations and identities. Note that this also holds for the QRAM construction proposed in this work, where $W_D(\bm{m})\in\mathcal{C}_0$ for any $\bm{m}$. Thus, in general all the non-Cliffordness of QRAM is captured by $V$, which does not depend on the dataset $D$. We now prove that for a memory of size $N$ one must necessarily have 
\begin{equation}\label{V_non_Clifford}
    V \in \mathcal{C}_{\log N +1}; 
\end{equation}
intuitively, this can be understood by noting that indeed for any 
$D$, $U_\text{QRAM}$ is a large multi-controlled gate, where the $\log N$ QPU register qubits trigger a conditional $X$ gate on the bus qubit. The proof proceeds as below.\\
\\
\noindent \textbf{Proof.} To prove Eq.~\eqref{V_non_Clifford}, recall that we defined the QRAM operation such that 
\begin{equation}
    U_\text{QRAM} \ket{\bm{x}}\ket{0} = \ket{\bm{x}}\ket{D_{\bm{x}}},
\end{equation}
and that we consider implementations of the form $U_\text{QRAM} = VW_DV^{-1}$. We moreover assume that any ancillas that are used during the implementation of $U_\text{QRAM}$ must necessarily be disentangled at the end, for any input state. Now, from the fact that, manifestly, one has $U_\text{QRAM}^2=\mathbb{1}$~\footnote{Simply note that $W_D^2=\mathbb{1}$ since $W_D$ is a tensor product $Z$ gates and identities, and therefore, trivially,  $U_\text{QRAM}^2=VW_DV^{-1}VW_DV^{-1}=VW_D^2V^{-1}=\mathbb{1}$.}, we immediately see that if $D_{\bm{x}}=1$ for a specific $\bm{x}$, then one also has $U_\text{QRAM} \ket{\bm{x}}\ket{1} = \ket{\bm{x}}\ket{0}$. That is, $D_{\bm{x}}=1\implies U_\text{QRAM}\ket{\bm{x},1}=\ket{\bm{x},0}$. Now, let us choose the dataset as follows: \begin{align}
    D_{\bm{x}} = \left\{\begin{matrix}
        0 & \text{if} & \bm{x} = \bm{1}\equiv (1,1,1,\dots,1) \\
        1 & \text{otherwise}
    \end{matrix} \right.
\end{align}
It follows now that $U_\text{QRAM}\ket{\bm{x}\neq\bm{1}}\ket{1}=\ket{\bm{x}\neq\bm{1}}\ket{0}$ for any $\bm{x}\neq \bm{1}$. For unitarity, we then also must have $U_\text{QRAM}\ket{\bm{1}}\ket{1}=\pm\ket{\bm{1}}\ket{1}$, where at most one sign is undecided (here we are also using the assumption that possible ancillas always end up disentangled at the end). More generally, with this choice we can write
\begin{equation}
    U_\text{QRAM} \ket{\bm{x}}\ket{B} = (\pm)^{B\delta_{\bm{x}, \bm{1}}}  \ket{\bm{x}}\ket{B\oplus \delta_{\bm{x}, \bm{1}} \oplus 1 }. 
\end{equation}
We now consider the following protocol: (i)~First, we apply $U_\text{QRAM}$; (ii)~Second, we apply $-Z$ to the bus qubit; (iii)~We apply $U_\text{QRAM}$ again. On the computational basis, this results in
\begin{equation}
    -U_\text{QRAM}Z_\text{bus}U_\text{QRAM} \ket{\bm{x}}\ket{B} = (-)^{B+\delta_{\bm{x}, \bm{1}}} \ket{\bm{x}}\ket{B}.
\end{equation}
Also recalling that $U_\text{QRAM} = U_\text{QRAM}^\dag$, this directly implies 
\begin{equation}
    U_\text{QRAM}Z_\text{bus}U_\text{QRAM}^{-1} = - Z_\text{bus} C_{\log N-1}Z.
\end{equation}
Now, from the section above, it is simple to recognize that the operator to the right-hand side belongs to layer $\log N -1$ of the Clifford hierarchy. This implies that, \emph{with this choice of the memory} $D$, one has
\begin{equation}
    U_\text{QRAM} \in\mathcal{C}_{\log N} \;\;\;\; \text{(specific case of memory)}.
\end{equation}
However, as we noted above, while the entire operation $U_\text{QRAM}$ depends on $D$, and specifically its Cliffordness depends on it, on the other hand $V$ is independent on $D$. Thus, since by definition $U_\text{QRAM}=VW_DV^{-1}$ where $W_D\in\mathcal{C}_0$ always, from $U_\text{QRAM} \in\mathcal{C}_{\log N}$ it follows directly that $V\in\mathcal{C}_{\log N +1}$, as we wanted to prove.

\newpage

\section{Nested one-hot encoding}\label{SM_section_NOHE}

\noindent In this section, we give a detailed account of the notion of nested one-hot encoding (NOHE). As we show in Section~\ref{equivalence_BB}, the following concepts are fundamentally equivalent to the famous bucket-brigade (BB) QRAM approach~\cite{giovannetti2008quantum}; our construction has the additional crucial feature that we could find an efficient implementation of the circuit decomposition with reconfigurable neutral-atom arrays. Moreover, our implementation only employs half the qubits with respect to the BB; we attribute to this a slightly more favorable error scaling, as reported in Section~\ref{noise_resilience} - though the qualitative behaviour is essentially the same.

\subsection{Definitions}
\noindent Let $\bm{z}=(z_0,\dots,z_{K-1})\in\left\{0,1\right\}^K$ be any string of $K$ bits, which can be thought of as representing the integer $\mu(\bm{z})=\sum_Jx_J2^J$. The standard `one-hot encoding' (OHE) $\bm{ohe}(\bm{z})$ is a string of $2^K$ bits, such that $\bm{ohe}(\bm{z})_j=\delta_{j,\mu(\bm{z})}$; that is, all bits are set to $0$ except one `hot' bit, in position $\mu(\bm{z})$, which is set to $1$. This acts as a \emph{pointer}. For instance, if $\bm{z}=(0,1,0)$, then $\bm{ohe}(\bm{z})=(0,0,1,0,0,0,0,0)$. In practice, this allows to represent a computational-basis state with a space location. \\
\indent Let now $\bm{x}=\left(x_0,\dots,x_{\log N -1}\right)$ be a bitstring addressing a memory location; we define the \emph{nested} one-hot encoding (NOHE) as 
\begin{equation}
    \text{NOHE}(\bm{x}) = \Big(\bm{ohe}^{(0)}(\bm{x}), ..., \bm{ohe}^{(\log N-1)}(\bm{x})\Big), 
\end{equation}
where we set $\bm{ohe}^{(K)}(\bm{x}) \equiv x_K \bm{ohe}\left(x_0,\dots,x_{K-1}\right)$ and $\bm{ohe}^{(0)}=1$ by default. In practice, $\bm{ohe}^{(K)}(\bm{x})$ is a string of $2^{K}$ bits, representing a variant of the OHE of the marginal string $(x_0,\dots,x_{K-1})$ containing the first $K$ bits of $\bm{x}$: specifically, now the hot qubit is not necessarily set to $1$, but instead it is equal to the $K+1$ bit in $\bm{x}$. For instance, if $\bm{x}=(0,1,1,x_3,x_4,\dots)$, then $\bm{ohe}^{(3)}(\bm{x})=(0,0,0,0,0,0,x_3,0)$. Thus, altogether $\text{NOHE}(\bm{x})$ features $N-1$ bits $\alpha=0,1,\dots,N-2$. In the following, it will be useful to parameterize $\alpha = 2^K -1 + k$, where $K=0,\dots, \log N - 1$ and $k=0,\dots, 2^K-1$. Note that the parameters $(K,k)$ are uniquely defined. In summary, with this notation we get the $\alpha-$th component of the NOHE as
\begin{equation}
    \text{NOHE}_\alpha(\bm{x}) = x_K\delta_{k,\mu(x_0,\dots,x_{K-1})}
\end{equation}
for $\alpha >0$, and $\text{NOHE}_{\alpha=0}(\bm{x})=x_0$.
\\
\indent We generalise this to the quantum domain by linearity: for any $\log N$ qubit state $\ket{\psi}=\sum_{\bm{x}}\psi_{\bm{x}}\ket{\bm{x}}$ we set $\ket{\text{NOHE}(\psi)} \equiv \sum_{\bm{x}}\psi_{\bm{x}} \ket{\text{NOHE}(\bm{x})}$. We identify the NOHE with a map defined (informally) such that 
\begin{equation}\label{U_isometrical_action}
    U^{(N)}_\text{NOHE}\ket{\psi}=\ket{\text{NOHE}(\psi)}
\end{equation}
for any arbitrary state $\ket\psi$ on $\log N$ qubits. We note that this does not uniquely define the unitary $U_\text{NOHE}$, which will be entirely specified by a circuit decomposition.

\subsection{Bus qubit}
\noindent In the main text, for a central building-block of QRAM we introduce a `bus qubit' and define the operation
\begin{equation}\label{V_HU}
    V = U^{(2N)}_\text{NOHE}H_\text{bus}.
\end{equation}
In practice, to the $\log N$ address qubits we add a new qubit in the $\ket{0}$ state; then, we implement the NOHE on this $\log N +1$ qubit address. On a computational basis input we therefore have
\begin{equation}\label{V_NOHE_SM}
    V\ket{\bm{x}}\ket{0}=\ket{\text{NOHE}(\bm{x},+)}=\ket{\text{NOHE}(\bm{x})}\ket{\text{OHE}(\bm{x})},
\end{equation}
where we defined the following state on $N$ qubits:
\begin{equation}
    \ket{\text{OHE}(\bm{x})} = \frac{\ket{0}^{\otimes N} + \ket{\bm{ohe}(\bm{x})}}{\sqrt{2}}.
\end{equation}
This state is factorised, and represents a computational state in the original Hilbert space via the physical location of the pointer $\ket{+}$: if e.g. $\bm{x}=(0,1,0)$, then $\ket{\text{OHE}(\bm{x})}=\ket{0,0,+,0,0,0,0,0}$. The term $\ket{\text{NOHE}(\bm{x})}$ can be understood as a byproduct of the circuit, which also ensures reversibility (note that $\langle \text{OHE}(\bm{x})|\text{OHE}(\bm{y})\rangle = (1+\delta_{\bm{x},\bm{y}})/2$). Fig.~5(f) in Methods illustrates this concept. With respect to the notation in the main text, in Eq.~\eqref{V_HU} we have exchanged the order of $\text{NOHE}(\bm{x})$ and $\text{OHE}(\bm{x})$ for simplicity; here we follow the order above.

\begin{figure*}[t!]
    \begin{minipage}{1.0\textwidth}
        \includegraphics[width=\textwidth]{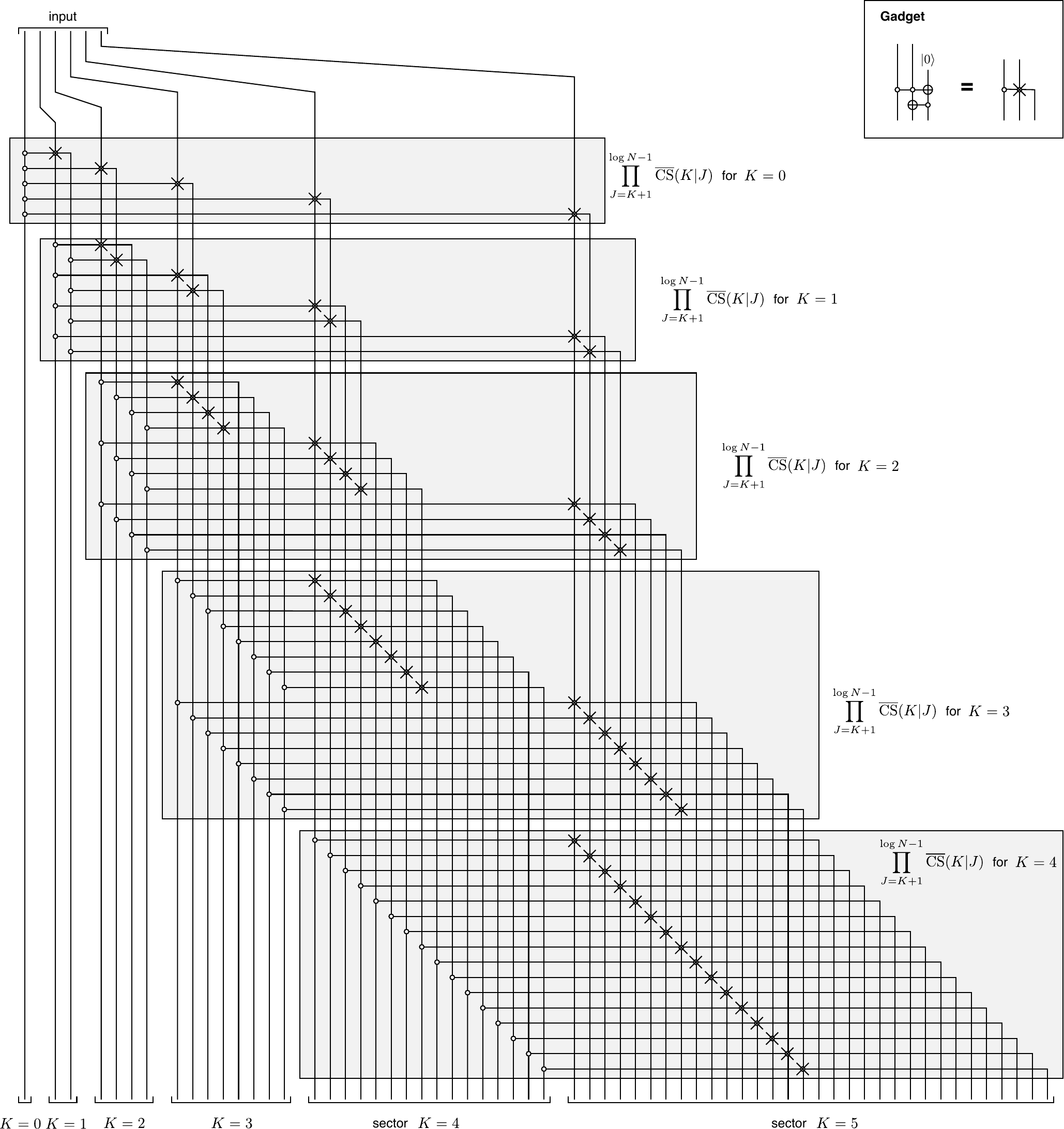}
    \end{minipage}%
\caption{\textbf{Non parallelised NOHE circuit decoposition}. Circuit diagram for the NOHE operation in the non parallel decomposition of Eq.~\eqref{U_OHE_decomposition}. Here, for graphical convenience we use the gadget displayed in the inset to introduce the ancillas; as a consequence, we also do not display the initial swaps of Eq.~\eqref{U_OHE_decomposition}. Note that, in this circuit, we show $U_\text{NOHE}^{(N)}$ for $N=64$; regarding the rightmost input qubit as the bus, this therefore aids queries to a memory of size $N=32$. For generic $N$, this circuit features $\log N -1$ main layers (highlighted by the gray boxes), but note that the operations within each layer do not commute, and are therefore sequential - resulting in a total time depth $ (\log^2 N -\log N)/2$. }
\label{non_parallel_NOHE}
\end{figure*}

\subsection{Circuit decomposition}
\noindent In Eq.~\eqref{U_isometrical_action} we define $U_\text{NOHE}$ as an isometry from $\log N$ to $N-1$ qubits, which can be embedded in a unitary circuit by introducing $N-\log N - 1$ ancillary qubits in $\ket{0}$. That is, we subtend $U^{(N)}_{\text{NOHE}}\Big[\ket{\psi}\ket{0}^{\otimes N-\log N -1}\Big]=\ket{\text{NOHE}(\psi)}$. This can be decomposed in the iteration of a simple primitive: 
\begin{equation}\label{U_OHE_decomposition}
    U^{(N)}_{\text{NOHE}} = \Bigg[\prod_{K=0}^{\log N -2} \prod_{J=K+1}^{\log N-1} \overline{CS}(K|J)\Bigg]\Bigg[\prod_{K=2}^{\log N - 1}S_{K,2^K-1}\Bigg].
\end{equation}
Therein, $S_{b,c}$ swaps the qubits $b\leftrightarrow c$, i.e. $S_{b,c}\ket{\psi_b}\otimes\ket{\phi_c} = \ket{\phi_b}\otimes\ket{\psi_c}$, and $CS(a|b,c) = \ket{0}_a\!\bra{0} \otimes\mathbb{1}_{b,c} + \ket{1}_a\!\bra{1} \otimes S_{b,c}$ is the Friedkin gate (controlled-swap); moreover, for $K<J$ we set 
\begin{equation}
    \overline{CS}(K|J) = \prod_{\alpha = 2^K-1}^{2(2^K-1)} CS(\alpha | \alpha + 2^J-2^K, \; \alpha + 2^J).
\end{equation}
Here, we label the string of $N-1$ qubits via $\alpha = 0,1,\dots,N-2$; moreover, we use the ordering convention that $\prod_{k=k_\text{min}}^{k_\text{max}}U_k= U_{k_\text{max}}U_{k_\text{max}-1}\dots U_{k_\text{min}}$. \\
\begin{figure*}[t!]
    \begin{minipage}{1.0\textwidth}
        \includegraphics[width=\textwidth]{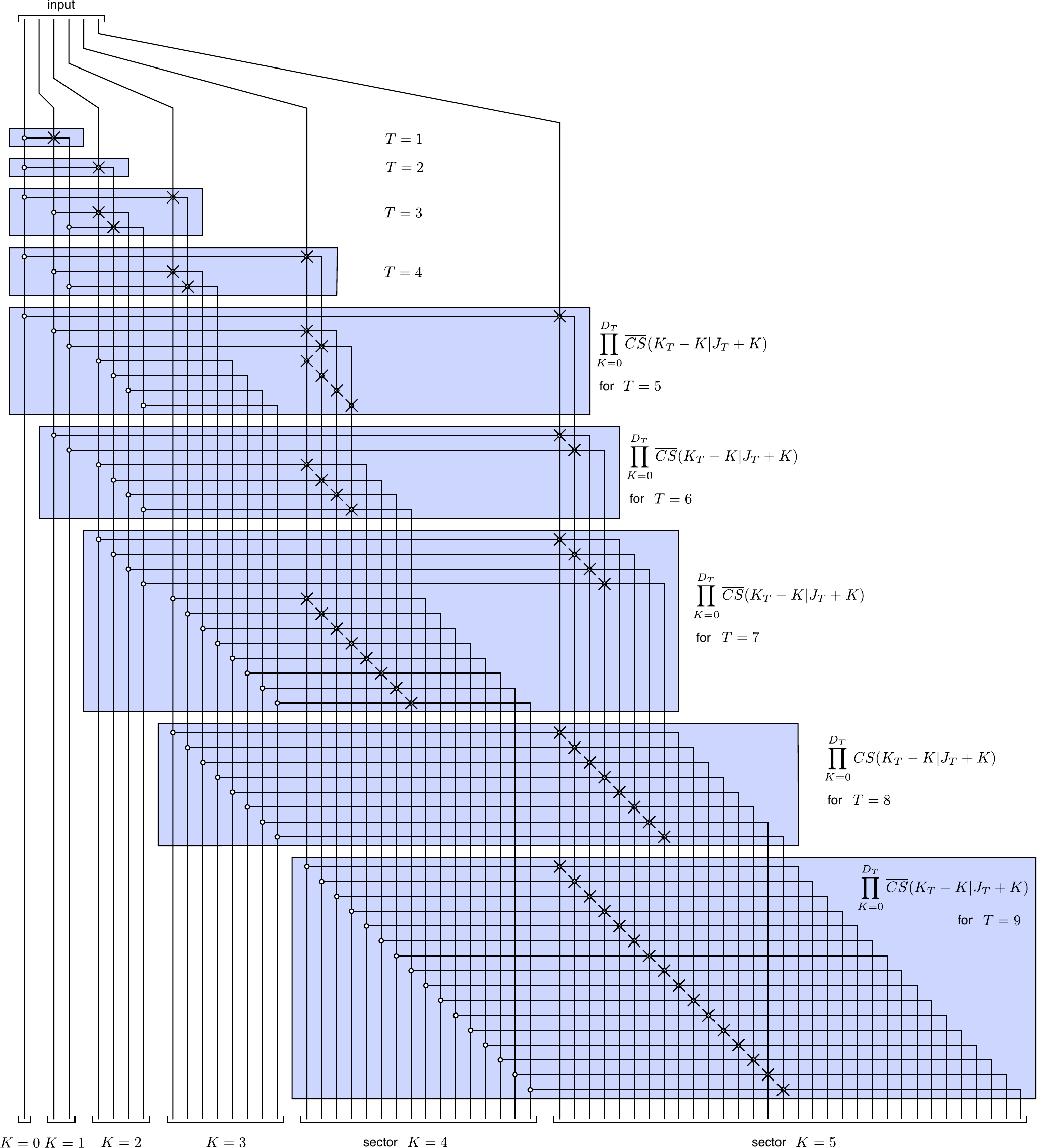}
    \end{minipage}%
\caption{\textbf{Parallelised NOHE circuit decoposition}. Circuit diagram for the NOHE operation in the parallelised decomposition of Eq.~\eqref{fast decomposition_main}. This circuit is obtained by straightforward re-ordering and parallelising the decomposition shown in Fig.~\ref{non_parallel_NOHE}; also here, we display $U^{(N)}_\text{NOHE}$ with $N=64$. This decomposition features $2\log N -3$ sequential layers (highlighted by the blue boxes); within each layer, all the gates commute, such that each layer can be executed in parallel in one single step - whence the total time-depth of the circuit is exactly $T_f=2\log N  -3$.}
\label{parallel_NOHE}
\end{figure*}
\indent In addition, Eq.~\eqref{U_OHE_decomposition} can be further simplified by realising that the Friedkin gate is always applied with one of the targets in the $\ket{0}$ state; this allows to substitute it with the `gadget' displayed in Fig.~5(d) of the Methods, consisting of a Toffoli and a controlled-not gate. This saves one controlled-not per Friedkin gate. Altogether, this results in the circuit decomposition shown in Fig.~5(a) of the Methods, for the explicit example of $N=8$.\\
\indent  Finally, while Eq.\eqref{U_OHE_decomposition} apparently suggests a $\mathcal{O}(\log^2 N)$ circuit depth, it can be parallelised to exactly $T_f\equiv 2\log N-3$ layers. More precisely, in Section~\ref{Circuit decmoposition of NOHE} we rewrite Eq.~\eqref{U_OHE_decomposition} in the parallelised form
\begin{equation}\label{fast decomposition_main}
    U^{(N)}_{\text{NOHE}} = \Bigg[\prod_{T=1}^{T_f}\prod_{K=0}^{D_T} \overline{CS}(K_T - K|J_T + K)\Bigg] \Bigg[ \prod_{K=2}^{\log N-1} S_{K,2^K-1} \Bigg],
\end{equation}
wherein all the terms in the inner productory over $K$ commute, and can therefore executed in parallel. Here, we defined $K_T = \left \lfloor{\frac{T-1}{2}}\right\rfloor$, $J_T = \text{min}\left\{T, \log N-1\right\}$ and $D_T = K_T - \text{max}\left\{0, T-\log N+1\right\}$. Thus, the NOHE can be implemented in time $T_f=\mathcal{O}(\log N)$, i.e., with circuit-depth linear in the QPU register size. Note that for implementing QRAM on a memory of size $N$ we need to apply $U_\text{NOHE}^{(2N)}$; thus, for the purpose of QRAM the exact duration of the one-hot encoding procedure is $2\log N - 1$.

\subsection{Equivalence to the Bucket-brigade QRAM}\label{equivalence_BB}
\begin{figure*}[t!]
    \begin{minipage}{1.0\textwidth}
        \includegraphics[width=\textwidth]{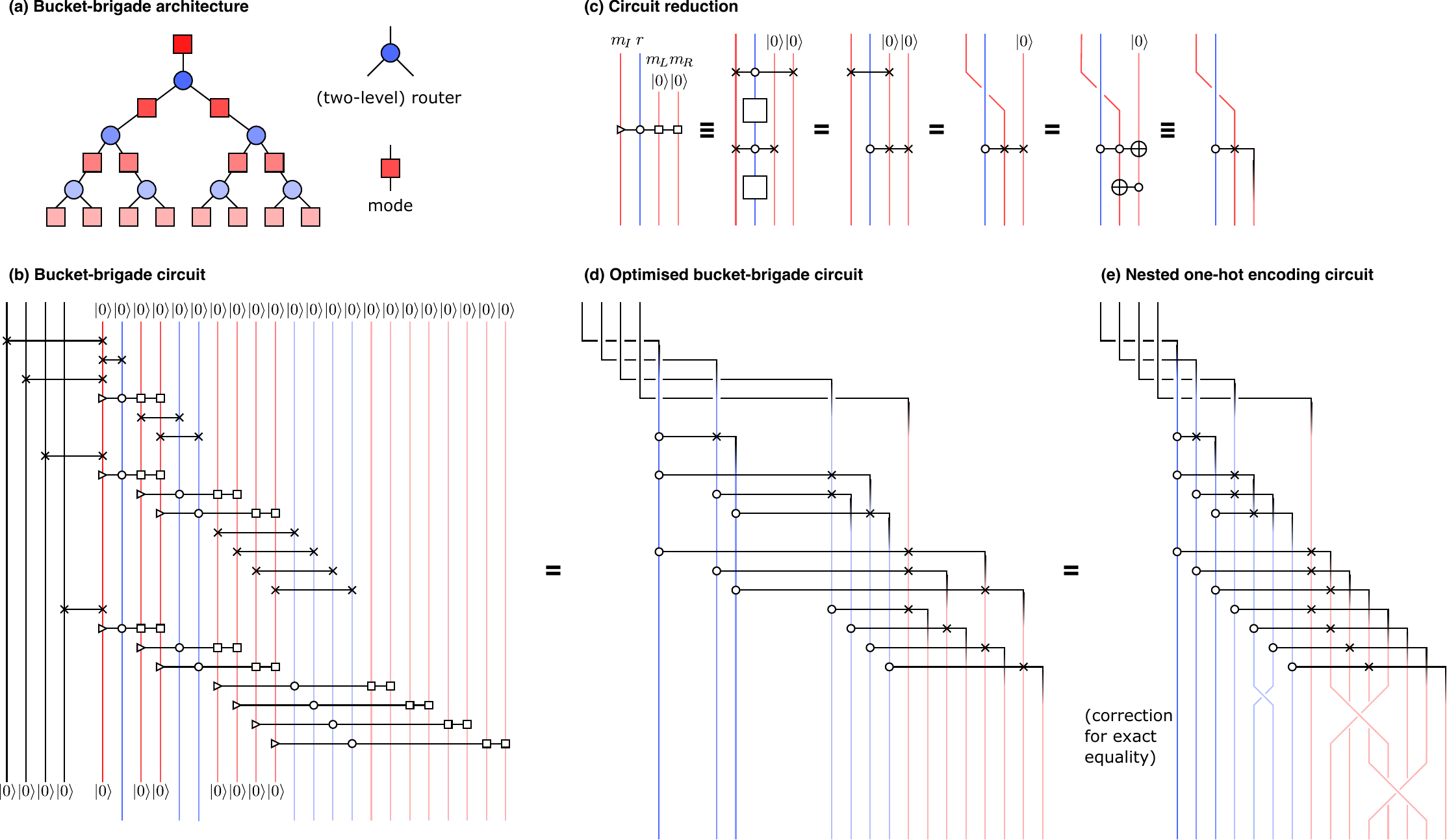}
    \end{minipage}%
\caption{\textbf{Circuit equivalence of bucket-brigade and nested one-hot encoding.} (a,b) Standard architecture and circuit layout for the `two-level router' variation of the BB protocol~\cite{giovannetti2008quantum, hann2021resilience}. (c) Chain of circuit equalities for our formal reduction from the BB to the NOHE. Note that the first and the last implicitly defines two gadgets used for ease of notaion in the diagrams below. (d) The chain of equalities above is used to simplify the BB circuit and deform it into a form almost equal to our NOHE circuit. (e) Exact equivalence can be achieved by relabeling (virtually swapping) certain qubits.  }
\label{fig_BB_to_NOHE}
\end{figure*}

\noindent On a mathematical level, the circuit decomposition of our NOHE is equivalent to the bucket-brigade protocol, which was analytically proved to be noise-resilient in Ref.~\cite{hann2021resilience}. The major differences in our circuit is that formally (i) It employs roughly half the number of ancillas, (ii) It uses half the number of $T$ gates and (iii) We could find an efficient implementation with reconfigurable neutral-atom processors. However, as we detail below, the two circuits are equivalent in terms of information flow (and thus of error propagation), and one thus expects the very same bounds to apply~\footnote{Note that the fact that the same \emph{scaling bounds} apply derives from the observation that errors propagate in an equivalent way. However, the circuits are not equal, implying that the \emph{exact scaling} can differ, as indeed we observe in our numerics.}. \\
\indent More precisely, the equivalence can be shown as in Fig.~\ref{fig_BB_to_NOHE}. This builds upon two main observations, which involve the two main operations within the bucket-brigade routing. For completeness, we stress that here we are considering the two-level router variant of the bucket-brigade~\cite{hann2021resilience}. The following discussion builds on the core methods of~\cite{giovannetti2008quantum}, whose circuit decompositions can be found e.g. in Ref.~\cite{hann2021resilience}. We will show that the bucket-brigade can be deformed in our NOHE circuit, in such a way that all the circuit features, including error propagation, are preserved.
\begin{enumerate}
    \item \textbf{Routing.} The core operation of the bucket brigade is the routing operation, whose fundamental building block is schematised and defined in the leftmost gadget of Fig.~\ref{fig_BB_to_NOHE}(c). Precisely, this primitive involves four qubits: an incident mode $m_I$, a router $r$ and two output modes $(m_L, m_R)$ (left and right, respectively); the incident mode and the router are initially in any qubit state, while the output modes are initialised in $\ket{0}$. In Fig.~\ref{fig_BB_to_NOHE}(c) we show how this primitive can be rewritten by instead implementing a swap $m_I\leftrightarrow m_L$, followed by a Friedkin gate which does not involve $m_I$. Crucially, since $m_L$ is initialised in $\ket{0}$, we can further substitute the swap by simply virtually relabeling $m_I\rightarrow m_L$. This has the effect of sparing one ancillary qubit per application of the routing primitive. 
    \item \textbf{Router initialisation.} After the routing, the last layer of modes is used to initialise a new layer of routers, by simply swapping in the state of the incident modes. Again, this can clearly be substituted by simply relabeling the modes $m_I\rightarrow r$, and identifying the last layer of modes for an address as the new layer of routers.  
\end{enumerate}
By iterating this concept to its natural limit within the bucket-brigade protocol, we can deform the original circuit in Fig.~\ref{fig_BB_to_NOHE}(b) to the optimised version in Fig.~\ref{fig_BB_to_NOHE}(d), where we show that only the last layer of modes is actually necessary, while all the other modes can be avoided. Fig.~\ref{fig_BB_to_NOHE}(d) is derived by incorporating two further details with the concepts above: First, we substitute the Friedkin gate with a Toffoli followed by a CNOT as explained in Methods, which simply spares a CNOT; this is the last step in Fig.~\ref{fig_BB_to_NOHE}(c). Second, we relabel the qubits in a specific way, such that almost all the qubit connections are local, and the residual nonlocal connectivity can be efficiently implemented via dynamical atom rearrangement.\\
\indent In summary, we have therefore shown that the conventional bucket-brigade protocol can be deformed into our proposed NOHE, in such a way that all the circuit features are manifestly preserved; as such, the noise resilience arguments of Ref.~\cite{hann2021resilience} apply directly to our proposal.  \\

\newpage

\section{Rewriting the QRAM operation}
\noindent \noindent In the main text, one key result is the rewriting of the QRAM operation as
\begin{equation}\label{modified_query}
    U_\text{QRAM} = P_{\bm{m}}^{-1} V^{-1} W_D(\bm{m}) V P_{\bm{m}};
\end{equation}
here $P_{\bm{m}}\in\mathcal{C}_0$ is a Pauli operator acting on the QPU register, and $W_D(\bm{m})=\pi_{\bm{m}}W_D\pi_{\bm{m}}^{-1}$, where $\pi_{\bm{m}}$ is a simple permutation of the OHE encoding qubits. This is also illustrated in Fig.~\ref{fig_rewriting}(a). We now show how this result is derived. \\
\\
\noindent \textbf{Proof.} Our rewriting builds upon three main observations. Before proceeding, let $P_{\bm{m}}=X^{\otimes \bm{b}}Z^{\otimes\bm{a}}$ be a generic Pauli operator acting on the QPU register; here, $\bm{m}=(\bm{a},\bm{b})\in\left\{0,1\right\}^{2\log N}$ and we set e.g. $Z^{\otimes\bm{a}}\equiv\bigotimes_{i}Z_i^{a_i}$. We emphasize that $P_{\bm{m}}$ only acts on the QPU register, and \emph{not} on the bus qubit.\\
\indent The first observation is that
\begin{equation}
    VX^{\otimes \bm{b}} = U_{\bm{m}}\otimes\pi_{\bm{m}}\; V.
\end{equation}
Therein, $U_{\bm{m}}$ and $\pi_{\bm{m}}$ act respectively on the first $N-1$ and on the last $N$ output qubits: $U_{\bm{m}}\otimes\pi_{\bm{m}}V\ket{\bm{x},0}=U_{\bm{m}}\ket{\text{NOHE}(\bm{x})}\otimes \pi_{\bm{m}}\ket{\text{OHE}(\bm{x})}$. More precisely,
\begin{equation}
    U_{\bm{m}} = U_\text{NOHE}^{(N)} P_{\bm{m}} U_\text{NOHE}^{(N)\dag} 
\end{equation}
is a complicated, deeply non-Clifford byproduct operator; however, $\pi_{\bm{m}}$ instead is a very simple permutation of the $N$ OHE output qubits. This follows directly from the concept of OHE: since Pauli $X$ operators permute the computational basis states, if then a OHE is applied, this permutation in the Hilbert space results in a physical permutation of the pointer qubits. For concreteness, let $\bm{x}=(0,1,1)$, and $\bm{b}=(0,0,1)$; then, $V\ket{\bm{x},0} = \ket{0,0,0,0,0,0,+,0}$, while $VX^{\otimes\bm{b}}\ket{\bm{x},0} = \ket{0,0,+,0,0,0,0,0}=\pi_{\bm{m}}\ket{0,0,0,0,0,0,+,0}$. \\
\indent The second observation is that
\begin{equation}
    U_\text{QRAM}Z^{\otimes \bm{a}} = Z^{\otimes \bm{a}} U_\text{QRAM}.
\end{equation}
For this, note that $U_\text{QRAM}$ is a multiple-controlled gate, where the $\log N$ register qubits simply act as controls; thus, $U_\text{QRAM}$ commutes with any gate which is diagonal in the QPU computational basis - such as $Z^{\otimes \bm{a}}$.\\
\indent The third observation is that since by definition $W_D$ only acts on the $N$ OHE qubits, then $W_DU_{\bm{m}}=U_{\bm{m}}W_D$.\\
\indent Finally, Eq.~\eqref{modified_query} follows straightforwardly by merging these three facts, and by defining $W_D(\bm{m})=\pi_{\bm{m}}W_D\pi_{\bm{m}}^{-1}$. In summary, the QRAM operation is rewritten as 
\begin{equation}\label{U_QRAM_blocks}
    U_\text{QRAM}=\Big[P_{\bm{m}}V\Big]\Big[\pi_{\bm{m}}W_D\pi_{\bm{m}}^{-1}\Big]\Big[VP_{\bm{m}}\Big],
\end{equation}
where we highlighted the modified versions of steps (i-iii). Importantly, this holds true for \emph{any} Pauli operator $P_{\bm{m}}$ acting on the QPU register. In our construction, this is intrinsically random, as we discuss below.

\newpage 

\section{Efficient adaptive loading}
\begin{figure*}[h!]
    \begin{minipage}{0.9\textwidth}
        \includegraphics[width=\textwidth]{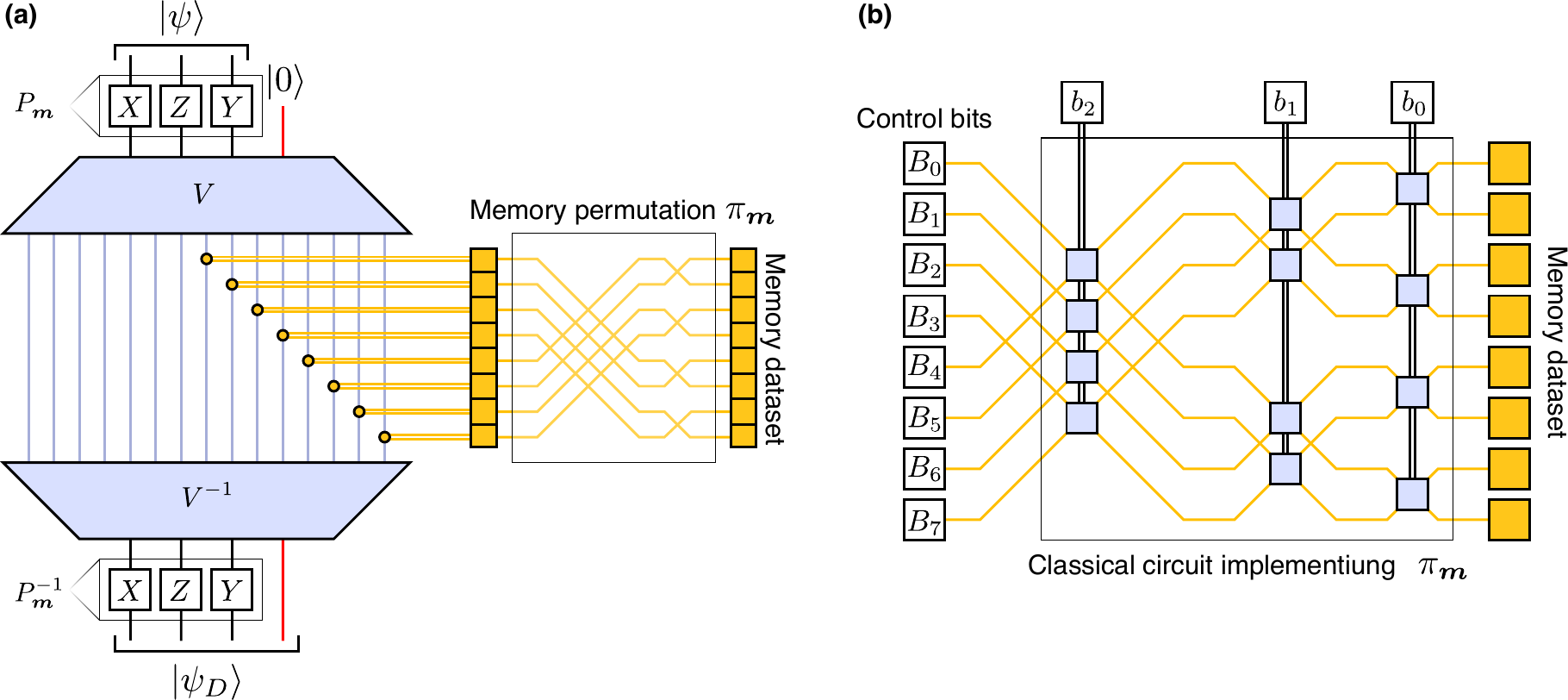}
    \end{minipage}%
\caption{(a) Representation of our rewriting of $U_\text{QRAM}=V^{-1}W_DV$ as $U_\text{QRAM}=P^{-1}_{\bm{m}}V^{-1}W_D(\bm{m})VP_{\bm{m}}$, where $P_{\bm{m}}$ is any Pauli operator acting on the QPU register and $W_D(\bm{m}) = \pi_{\bm{m}}W_D\pi_{\bm{m}}^{-1}$ is the same loading operation, but with a highly structured permutation $\pi_{\bm{m}}$ of the memory locations. Here, we have $\bm{b}=(1,0,1)$, and correspondingly the permutation is implemented via two parallel swaps. (b) Classical circuit implementing the loading. Specifically, we use a circuit of swaps, controlled by the $\log N$ bits $b_K$, to activate the $N$ control bits $B_l$ in such a way to implement $W_D(\bm{m})$. The viola squares therein switch the input signals (in yellow) if the corresponding bit $b_K$ (entering with the double black line) is activated, $b_K=1$. } 
\label{fig_rewriting}
\end{figure*}

\noindent Step (ii) for implementing $U_\text{QRAM}$ according to Eq.~\eqref{U_QRAM_blocks} requires to efficiently implement the loading operation $W_D(\bm{m})$. Importantly, in our protocol $\bm{m}$ is not known prior to the start of the query, as it is only determined by the random BM outcomes during step (i). Here, we discuss how the loading can always be implemented efficiently for any obtained $\bm{m}$. \\
\indent Let now $l\in\mathcal{L}=\left\{0,1,\dots,N-1\right\}$ label the qubits $W_D(\bm{m})$ has to be applied to. At an abstract level it is sufficient to proceed as follows. Given $\bm{m}$, for each qubit $l$ we compute the string $\bm{x}=\bm{b}\oplus\mu^{-1}(l)$; then, we apply a $Z$ gate to $l$ if $D_{\bm{x}}=1$. This is executed in parallel on all qubits $\mathcal{L}$, allowing for a $\mathcal{O}(1)$ implementation. Crucially, as we show below, we do not need the classical control software to `lookup' the memory.  \\
\indent We model the controls on $\mathcal{L}$ as follows. To each $l$, we associate one `control bit' $B_l\in\left\{0,1\right\}$, and one `control device' $\mathscr{C}_l$ pointing at $l$; when $\mathscr{C}_l$ is triggered, it only looks up $B_l$, and applies a $Z$ gate to $l$ if $B_l=1$. All the control devices can be triggered in parallel. In addition, the classical memory is coupled to the circuit displayed in Fig.~\ref{fig_rewriting}(b): this is a network of swaps, which are controlled by only $\log N$ external control bits $\mathscr{B}_0,\mathscr{B}_1\dots, \mathscr{B}_{\log N -1}$; note that each $\mathscr{B}_K$ controls $N/2$ swaps simultaneously. When $W_D(\bm{m})$ is called, we have already registered $\bm{m}=\left(\bm{a},\bm{b}\right)$ on classical memory; we thus straightforwardly set $\mathscr{B}_K=b_K$. Next, we activate the circuit of swaps; this sets the control bits to exactly $B_l = D_{\bm{b}\oplus \mu^{-1}(l)}$. Finally, we trigger all the controls $\mathscr{C}_l$ in parallel, resulting in the implementation of $W_D(\bm{m})$. Thus, the desired loading is executed without any `reading' of the memory by classical hardware.

\newpage

\section{Clifford generalised one-hot encoding via gate-teleportation}
\noindent Here, we discuss more in detail our protocol for implementing the step (i) of Eq.~\eqref{U_QRAM_blocks} as a Clifford subroutine of depth $\mathcal{O}(1)$. We specifically discuss its relation with the paradigmatic gate-teleportation (GT) primitive~\cite{gottesman1999demonstrating}.

\subsection{Gate teleportation and Bell measurements: conventions and principles}

\noindent We first review the main concepts needed for understanding our protocol for step (i). It is useful to first set the ground by stating our conventions for Bell measurements (note that we use a cluster-state inspired definition of the Bell basis, corresponding to a $\pi/2$ rotation of one of the involved qubits). Then, we detail the standard paradigm of GT. 

\subsubsection{Bell measurements}
\noindent In this work, we adopt the following conventions. We define the two-qubit Bell basis as 
\begin{equation}
    \ket{\Psi_{a,b}} = \frac{1}{2}\sum_{i,j}(-)^{ai+bj+ij} \ket{i,j},
\end{equation}
where $(a,b)\in\left\{0,1\right\}^2$. For simplicity, we will commonly refer to the first element as $\Psi\equiv \Psi_{0,0}$. It can be easily verified that $\left\{\Psi_{a,b}\right\}$ is indeed a complete orthonormal basis, i.e. that $\langle\Psi_{a,b}|\Psi_{a',b'}\rangle = \delta_{a,a'}\delta_{b,b'}$. We note that this is related to another common choice for the Bell basis $\ket{\overline{\Psi}_{a,b}}\equiv 2^{-1/2}\sum_k(-)^{ak}\ket{k,k\oplus b}$ by the simple transformation $|\overline{\Psi}_{a,b}\rangle = \left(\mathbb{1}\otimes H\right)|\Psi_{a,b}\rangle$. \\
\indent We note that our Bell basis is related to the conventional $Z$ basis as $\ket{\Psi_{a,b}}=CZ\;H^{\otimes 2}\ket{a,b}$. It follows that we can perform BMs, i.e. measurements in the two-qubit Bell basis, by (i) First, applying a controlled-phase gate $CZ$, and (ii) Second, measuring the $X$ basis of both qubits. We note that for perfect equivalence we should as well (iii) Third, apply a final $CZ$ gate; however, we will typically not be interested in the measured qubits after the measurement, so we will always skip this unnecessary step. Then, measuring $X=\pm 1$ on the first (second) qubit is equivalent to measuring $a=(1\pm 1)/2$ (and $b=(1\pm 1)/2$ respectively). In summary, a BM with outcomes $(a,b)$ is described by the `projector' 
\begin{equation}
    \mathbb{P}^{(a,b)}_\text{BM} = \frac{1+(-)^aX}{2} \otimes \frac{1+(-)^b X}{2} CZ.
\end{equation}
Note that this is not formally a projector, as it is not idempotent (that is, $\mathbb{P}^{(a,b)}\mathbb{P}^{(a,b)}\neq\mathbb{P}^{(a,b)}$); in contrast, with this definition $CZ\mathbb{P}_\text{BM}^{(a,b)}$ would technically be a projector.

\subsubsection{Gate teleportation: basic details}
\noindent We now discuss the conventional paradigm of GT~\cite{Nielsen_1997}. This is also detailed in Fig.~5(e) of Methods. Let $U:\mathscr{H}^{(n)}\rightarrow \mathscr{H}^{(m)}$ be any quantum operation, e.g. a unitary gate, whose input is the Hilbert space of $n$ qubits. Note that for what follows the codomain can be of any size. For instance, in our case we are interested in the GT of $V$, so that $n=\log N$ (where $N$ is, as usual, the size of the memory) and $m=2N-1$. Then, the goal of GT is to apply $U$ to any arbitrary input state $\ket{\psi}\in\mathscr{H}^{(n)}$. For this, it leverages a pre-assembled resource-state (RS) of the form $\ket{\Phi_U}=(\mathbb{1}\otimes U)\ket{\Psi}^{\otimes n}$. That is ,we apply $U$ on one partition of $n$ Bell pairs. Explicitly, this therefore reads
\begin{equation}
    \ket{\Phi_U} = \frac{1}{2^n} \sum_{\bm{y},\bm{z}} (-)^{\bm{y}\cdot\bm{z}} \ket{\bm{y}}_\mathcal{I}\otimes U\ket{\bm{z}}_\mathcal{F},
\end{equation}
where we labeled by $\mathcal{I}$ the first partition of the Bell pairs. Note that $|\mathcal{F}|=m$; thus, according to the discussion above, we might have $|\mathcal{F}|\neq|\mathcal{I}|$ in general. \\
\indent Let the input state $\ket{\psi}$ be supported on $\mathcal{R}$, with $|\mathcal{R}|=|\mathcal{I}|=n$. Then, GT proceeds by performing Bell measurements between corresponding qubits in $\mathcal{R}$ and $\mathcal{I}$. Let $\bm{m}=(\bm{a},\bm{b})$ be the measurement outcomes. It is simple to show that all the measurement outcomes are equiprobable, thus every $\bm{m}$ has probability $2^{-2n}$. Moreover, the resulting (normalized) conditional state on the unmeasured qubits $\mathcal{F}$ is exactly
\begin{equation}
    2^{2n}\mathbb{P}^{(a,b)}_\text{BM} \Big[\ket{\psi}_\mathcal{R}\ket{\Phi_U}_{\mathcal{I}\cup\mathcal{F}}\Big] = UP_{\bm{m}}\ket{\psi},
\end{equation}
where the Pauli operator $P_{\bm{m}}\in\mathcal{C}_0$ takes the form
\begin{equation}
    P_{\bm{m}} = X^{\otimes\bm{b}}Z^{\otimes\bm{a}}. 
\end{equation}
Up to this point, GT only needed Clifford operations (namely, $n$ parallel Bell measurements) to be implemented - once the RS $\ket{\Phi_U}$ is given.\\ 
\indent Crucially, note that the Pauli correction $P_{\bm{m}}$ acts on $\ket{\psi}$ \emph{prior} to the application of the desired operation $U$. Thus, to complete our protocol, in the standard GT paradigm we need to apply the \emph{byproduct} correction $UP_{\bm{m}}U^{-1}$: it is easy to see that after this, the obtained state is exactly the desired $U\ket{\psi}$.

\subsubsection{Comment: the efficiency of GT for general purposes}
\noindent Unfortunately, in general this strategy is only efficient when $U$ is relatively simple. This can be understood in terms of the Clifford hierarchy: note indeed that if $U\in\mathcal{C}_K$ belongs to a certain level of the Clifford hierarchy, then the byproduct generally is just one layer below. More precisely,
\begin{equation}
    U\in\mathcal{C}_K \;\;\;\; \implies \;\;\;\; UP_{\bm{m}}U^{-1}\in\mathcal{C}_{K-1} \;\;\;\; \text{for at least some } \bm{m}.
\end{equation}
Indeed, GT has proved to be a very successful paradigm in fault-tolerant QC, where it allows to implement simple non-Clifford gates from $\mathcal{C}_2$, e.g. the $T$ gate, via Clifford operations only. Specifically, this is a crucial building-block of many QEC schemes, which employ GT to inject $T$ gates. However, when $U$ is very deep in the Clifford hierarchy, GT is not advantageous: in practice, even though the Clifford complexity is lowered by one layer via GT, the fact that the byproduct correction is random compensates this achievement, as it imposes the application of an unpredictable, deeply non-Clifford gate. 
\indent In principle, QRAM seems to clearly fall into the category of operations which do not benefit from GT. Indeed, as we showed $V\in\mathcal{C}_{\log N +1}$ is very deep in the Clifford hierarchy, and in general correcting for any byproduct is extremely hard, specially on a fault-tolerant processor. One of our most important results in this work is realising that the byproduct does \emph{not} need to be corrected: instead, we can accept \emph{any} random outcome $VP_{\bm{m}}$ from the GT, and still continue with the query. 

\subsection{Gate-teleportation for generalised NOHE}
\noindent For realizing $U_\text{QRAM}$ according to the recipe in Eq.~\eqref{U_QRAM_blocks}, step (i) requires to apply $VP_{\bm{m}}$ to the input state $\ket{\psi}\ket{0}$, where $\ket{\psi}$ is the QPU register state and $P_{\bm{m}}$ is a (arbitrary, not yet defined) Pauli operator acting on it. We realize $VP_{\bm{m}}$ with a Clifford procedure, during which $P_{\bm{m}}$ is determined randomly. Similarly to GT, we use the pre-assembled resource-state (RS) 
\begin{equation}\label{Phi_1}
    \ket{\Phi^{(1)}} = \frac{1}{N}\sum_{\bm{y},\bm{z}} (-)^{\bm{y}\cdot\bm{z}} \ket{y}_\mathcal{I}\otimes\ket{\text{NOHE}(\bm{z},+)}_\mathcal{D};
\end{equation}
where $\mathcal{I}$ and $\mathcal{D}$ feature $\log N$ and $2N\!-\!1$ qubits. \\
\indent Initially the state of the QPU register is $\ket{\psi}$, supported on qubits $\mathcal{R}$; the bus qubit, in principle initialized in $\ket{0}$, does not need to be provided as input. Then, we simply implement BMs between $\mathcal{R}$ and $\mathcal{I}$, collecting the outcomes $\bm{m}$; this directly leaves us with the state $VP_{\bm{m}}\ket{\psi}\ket{0}$ on the unmeasured qubits $\mathcal{D}$. We explicit in the SM the simple relation between $\bm{m}$ and $P_{\bm{m}}$. 

\newpage

\section{Clifford nested one-hot encoding inversion}
\noindent Step (iii) of the recipe in Eq.~\eqref{U_QRAM_blocks} requires to implement $P_{\bm{m}}^{-1}V^{-1}$. The last term, $P_{\bm{m}}^{-1}$, is a Pauli operation that can be implemented straightforwardly at the end. Recalling that $V^{-1}=H_\text{bus}U_\text{NOHE}^{-1}$ and noting that also $H_\text{bus}$ is Clifford, the only complex part remains the inversion $U_\text{NOHE}^{-1}$. In principle, this is deeply non-Clifford, as it is the inverse of the non-Clifford circuit in Eq.~\eqref{U_OHE_decomposition}. However, below we show that if we regard instead $U_\text{NOHE}$ as an isometry from $\log N$ to $N-1$ qubits, and only consider its inversion on the image $U^{(N)}_\text{NOHE}(\mathscr{H}^{\log N})$ in the codomain, where $\mathscr{H}^{(n)}\sim\mathbb{C}^{2^n}$ is the Hilber space of $n$ qubits, then we can invert it via Clifford operations only. 

\subsection{Inverting a minimal NOHE}
\noindent To illustrate the main idea, we first elaborate on the example of $N=4$; according to Eq.~\eqref{V_HU}, this aids $U_\text{QRAM}$ on a memory featuring $2$ databits. Then, from Eq~\eqref{U_OHE_decomposition} the NOHE can be written as the isometry $U^{(4)}_{\text{NOHE}}\equiv\sum_{i,j}\ket{i,j\oplus ij,ij}\bra{i,j}$. For inverting it, we need to implement the right-inverse $G \equiv \sum_{i,j,k}\ket{i,j\oplus k}\bra{i,j,k}$, such that $GU_{\text{NOHE}}=\mathbb{1}$ is the identity on $\mathscr{H}^{(2)}$. Due to the manifest non-reversibility of $G$, it cannot be implemented via  unitary operations; however, we now outline a protocol to execute $G$ on the image of $U_{\text{NOHE}}$ by introducing measurements. Crucially, this will only involve Clifford resources: specifically, simple Clifford gates, SQPMs, Pauli operations and a stabilizer RS. To this end, let us first consider a generic three-qubit input state $\ket{\chi}=\sum_{i,j,k} \chi_{i,j,k}\ket{i,j,k}$. Our goal is to construct the state $G\ket{\chi}=\sum_{i,j,k}\chi_{i,j,k}\ket{i,j\oplus k}$, under the assumption that $\ket\chi\in U_{\text{NOHE}}\left(\mathscr{H}^{(2)}\right)$. For this, we proceed as in Fig.~6(b) of Methods: (a) We implement a controlled-not gate between the third and the second qubit; (b) We entangle the first and second qubit with an ancillary Bell pair via controlled-phase gates; (c) We measure the third input qubit in the $X$ basis, collecting the outcome $s_0\in\left\{0,1\right\}$; (d) We measure the ancillas in a single-qubit Pauli basis, choosing either the $Z$ (if $s_0=0$) or the $X$ basis (if $s_0=1$), and we collect the outcomes $\bm{s}_1=\left(\alpha, \beta\right)\in\left\{0,1\right\}^2$; (e) We finally apply single-qubit Pauli gates on the original qubits, depending on $\left(s_0,\bm{s}_1\right)$.\\
\indent Explicitly, the conditional state after the $X$ basis measurement in step (c) reads
\begin{equation}
   \ket{\chi|s_0}\!= \!\!\sum_{a,b}(-)^{a,b}\!\ket{a,b}\otimes\sum_{i,j,k}(-)^{ai+b(j+k)+s_0k}\chi_{i,j\oplus k, k} \!\ket{i,\!j};
\end{equation}
here the ancillary qubits are written to the left of the tensor product. Next, we implement step (d). If we measured $s_0=0$, we measure in the $Z$ basis; the system is then projected to
\begin{equation}\label{chi_0}
   \ket{\chi|s_0=0,\bm{s}_1} =Z^\alpha\otimes Z^\beta \sum_{i,j,k} \chi_{i,j,k} \ket{i,j\oplus k}.
\end{equation}
Thus, in step (v) we simply need to apply the Pauli correction $Z^\alpha\otimes Z^\beta$ to cancel the byproduct and retrieve our target output $G\ket{\chi}$. Differently, if $s_0=1$ we measure in the $X$ basis; now the conditional state iss
\begin{equation}\label{chi_1}
   \ket{\chi|s_0=1,\bm{s}_1} = Z^\beta\otimes Z^\alpha \sum_{i,j,k}(-)^{i(j+ k) + k} \chi_{i,j,k} \ket{i,j\oplus k},
\end{equation}
which for general inputs $\ket\chi$ is not trivially connected to the target output. However, recall now that $\ket\chi\in U_{\text{NOHE}}\left(\mathscr{H}^{(2)}\right)$. This subspace is characterised by the projector
\begin{equation}
    \mathbb{P} = \sum_{i,j,k} \delta_{ij,0}\delta_{k,ik} \ket{i,j,k}\bra{i,j,k}.
\end{equation}
Crucially, if $\ket\chi$ is stabilised by this projector, i.e. $\mathbb{P}\ket{\chi}=\ket{\chi}$, then on all its non-vanishing components we get $\left(ij+ik+k\right)\equiv 0\;\mathrm{mod}2$, meaning that the phases in the equation above are all equal. Thus, if the input is from the subspace of interest, again we can retrieve the desired output by simply applying the Pauli correction $Z^\beta\otimes Z^\alpha$. In both cases, we deterministically apply $G$, under the assumption that $\ket{\chi}=\ket{\text{NOHE}(\psi)}$ for some $\psi\in\mathscr{H}^{(2)}$. 

\subsection{General case}
\noindent For larger $N$, recall that $U_\text{NOHE}$ can be implemented by iterating the same isometry $G$ introduced above as in Fig.~5(a) of Methods, eventually resulting in a larger isometry from $\log N$ qubits to $N-1$ qubits. Since the basic building-block is the same, it can be inverted by iterating the inversion procedure outlined for $N=4$. This is shown explicitly for $N=16$ in Fig.~6(c) of Methods. The circuit therein can be understood by comparison with the circuit for realising $U_\text{NOHE}$ in Fig.~5(a) of Methods: we invert it by applying the destructive protocol in Fig.~6(b) in place of the direct (non-Clifford) unitary inversion of the gadget in Fig.~5(d). \\ 
\indent Note that in Fig.~6(c) of Methods we \emph{first} execute all the unitary gates, and \emph{then} all the SQPMs. This is allowed since only Clifford operations are involved, implying that we do not need to physically implement the adaptive Pauli corrections - that is, the byproducts of the form $Z^\alpha\otimes Z^\beta$ in Eqs.~\eqref{chi_0} and~\eqref{chi_1}. Instead, byproducts can be propagated through the circuit up to the point of a SQPM, where they are accounted for by adequately interpreting the measurement outcomes.\\
\indent This eliminates the need for mid-circuit adaptive gates, thereby separating the circuit in two parts, as highlighted in Fig.~6(c) of Methods: the first part $C_\text{NOHE}$ contains the Clifford operations and Pauli measurements that are \emph{pre-determined}; the second part contains \emph{adaptive} SQPMs on the ancillas, where the measurement bases are chosen depending on previous measurement outcomes. Since these outcomes are probabilistic, $C_\text{NOHE}$ also has a \emph{classical} output $\bm{s}_0\in\left\{0,1\right\}^{N-\log N -1}$, which collects them; formally, when during the application of $C_\text{NOHE}$ on the input $\ket{\Xi}$ we obtain the measurement outcomes $\bm{s}_0$, we write the output as $C_\text{NOHE}^{(\bm{s}_0)}\ket{\Xi}$. In summary, this operation implements a map of the form
\begin{equation}
    C_\text{NOHE}^{(\bm{s}_0)} \;\; : \;\; \mathscr{H}_\mathcal{D} \longrightarrow \mathscr{H}_\mathcal{P}\otimes\mathscr{H}_\mathcal{F}.
\end{equation}
Here, $\mathcal{D}$ is the set of quits which supports the input, and has size $|\mathcal{D}| = N-1$; the qubits in $\mathcal{P}$, with $|\mathcal{P}|=2(N-\log N - 1)$, are those which will then be subjected to the adaptive SQPMs; finally, $\mathcal{F}$ collects the $|\mathcal{F}|=\log N$ qubits which will support the output.\\ 
\indent The final sequence of adaptive SQPMs is performed by leveraging standard techniques of the stabiliser formalism: measurement outcomes different from $+1$ can always be accounted for by propagating virtual Pauli byproducts through the Clifford circuit, and thereby updating subsequent SQPM bases. The number of rounds of parallel SQPMs equals exactly the number of circuit-layers of $U_\text{NOHE}$: measurements clearly correspond to gadgets, and the measurement outcomes corresponding to each gadget only influence the measurement basis of subsequent gadgets. This can therefore be summarised as follows: given any arbitrary state $\ket\Xi\in U_{\text{NOHE}}\left( \mathscr{H}^{(\log N)} \right)$, then there exists a sequence of adaptive SQPMs that implements the transformation
\begin{equation}\label{SQPM_inversion}
    C_\text{NOHE}^{(\bm{s}_0)} \ket{\Xi} \xrightarrow[\;\;\;\;\;\text{SQPMs}\;\;\;\;\;]{} U^{-1}_{\text{NOHE}} \ket{\Xi},
\end{equation}
upon the application of a final Pauli correction; moreover, the protocol is always completed in $T_f=2\log N-3$ measurement rounds. 

\subsection{Resource-state based implementation}
\noindent This same protocol for inverting $U_\text{NOHE}$ can as well be executed with the aid of a resource-state of the form
\begin{equation}\label{Phi_2}
    \ket{\Phi^{(2)}} = \frac{1}{2^{N-1}}\sum_{\bm{Y}, \bm{Z}} (-)^{\bm{Y}\cdot\bm{Z}} \ket{\bm{Y}}_{\mathcal{D}'}\otimes C_\text{NOHE}\ket{\bm{Z}}_{\mathcal{P}\cup\mathcal{F}},
\end{equation}
where the indexes run over $\bm{Y}, \bm{Z}\in\left\{0,1\right\}^{N-1}$ and $|\mathcal{D}'|=N-1$. Therein we omit the classical output of $C_\text{NOHE}^{(\bm{s}_0)}$ and simply write $C_\text{NOHE}$ because manifestly any $\bm{s}_0$ is equally well-suited. This state can thus be assembled by first preparing $N-1$ Bell pairs, and then applying $C_\text{NOHE}^{(\bm{s}_0)}$ to one half with any arbitrary $\bm{s}_0$. For the inversion, we employ the scheme in Fig.~6(d) of Methods: we first teleport the input state in the resource state via BMs between the input qubits and $\mathcal{D}'$, collecting the measurement outcomes in a binary vector $\bm{s}_{-1}$; then, we perform SQPMs on $\mathcal{P}$ similarly to what discussed above, but now also accounting for $\bm{s}_{-1}$. Again, this is allowed by the fact that  $C_\text{NOHE}$ is Clifford, implying that $|\Phi^{(2)}\rangle$ is a \emph{stabiliser} state.  

\subsection{Compact (linearised) resource state}\label{Compact (linearised) resource state}
\begin{figure}[b!]
    \centering
    \includegraphics[width=0.9\linewidth]{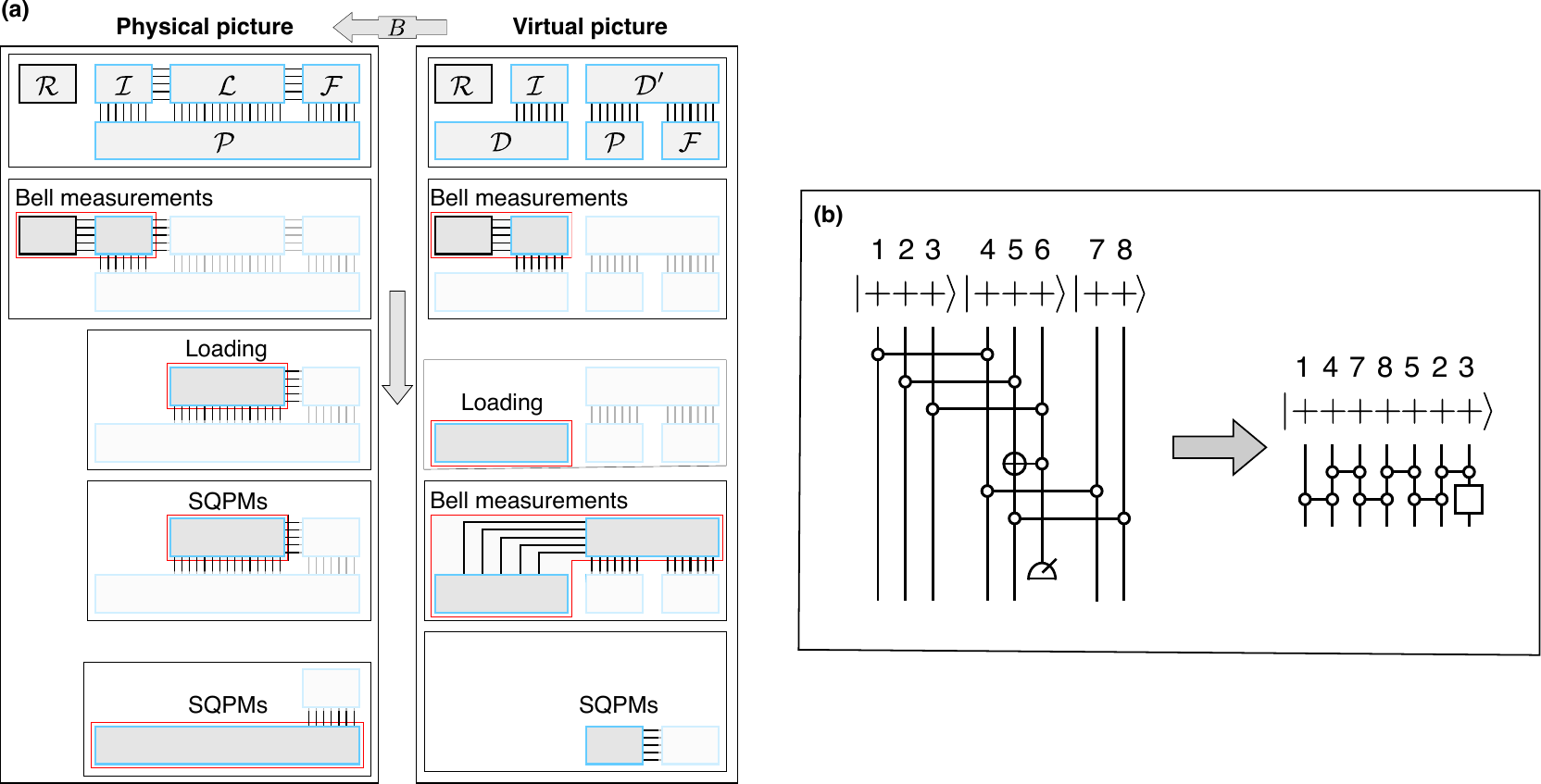}
    \caption{(a) Sketch of the query protocol. To the left, we schematise the induced evolution in the physical Hilbert space $\mathscr{H}_\mathcal{R}\otimes\mathscr{H}_{\mathcal{A}}$, where $\mathcal{A}=\mathcal{I}\cup\mathcal{L}\cup\mathcal{P}\cup\mathcal{F}$; to the right, we show the corresponding evolution in virtual space. Here boxes represent sets of qubits (either physical or virtual), and bonds between them denote the presence of entanglement. The two pictures are connected by the action of the contractive map $B:\mathscr{H}_\mathcal{D}\otimes\mathscr{H}_{\mathcal{D}'}\rightarrow\mathscr{H}_\mathcal{L}$, detailed in Fig.~7 of the Methods section, which essentially implements a partial Bell measurement between $\mathcal{D}$ and $\mathcal{D}'$. The right panel also traces the steps of the `prototypal Clifford query protocol'. (b) Optimisation of the resource-state $|\Phi^{(2)}\rangle$. To the left, the minimal building block according to Figs.~6(c,d) in Methods, which corresponds to $|\Phi^{(2)}\rangle$ for $N=2$. To the right, the result of the Clifford manipulation discussed in the supplemental Section~\ref{Compact (linearised) resource state}, which reduces it to a linear \emph{cluster state} (up to a Hadamard gate on qubit $3$). 
    } 
    \label{fig_M4}
\end{figure}

\noindent The definition of $|\Phi^{(2)}\rangle$ used so far can be drastically optimized via standard Clifford methods. We first discuss the simple case of $N=2$, where we show that $\ket{\Phi^{(2)}}$ can be reduced to a \emph{linear} graph-state. For this, consider the qubit labeling in Fig.~\ref{fig_M4}(b). Before the $X$ measurement on qubit $6$, the state is characterised by the following stabilisers: $X_1Z_4$, $X_4Z_1Z_7$, $X_5Z_2Z_8$, $X_7Z_4Z_8$, $X_8Z_5Z_7$, $X_2Z_5Z_6$, $X_3Z_6$ and $X_6Z_3X_5Z_8$. Measuring $X_6$ substitutes the last three stabilisers with $X_2X_3Z_5$, $\pm X_6$, $\pm Z_3X_5Z_8$ respectively, with the sign set by the measurement outcome. Thus, a $X_6=-1$ outcome is accounted for by applying $X_3$ (and $Z_6$). Importantly, we note the emergence of a \emph{linear} graph state according to the order $1-4-7-8-5-2-3$, up to a Hadamard $H_3$ on qubit $3$. More precisely, if we apply $H_3$, then the stabiliser group on the unmeasured qubits is generated by $X_1Z_4$, $Z_1X_4Z_7$, $Z_4X_7Z_8$, $Z_7X_8Z_5$, $Z_8X_5Z_2$, $Z_5X_2Z_3$, $Z_2X_3$. Therefore, $|\Phi^{(2)}\rangle$ can be prepared with a measurement-free approach with the 1D circuit in Fig.~\ref{fig_M4}(b). Note that this corresponds to Fig.~2(c) in the main text.\\
\indent By iterating this, for any $N$ $|\Phi^{(2)}\rangle$ is prepared via simple Clifford operations which merge linear \emph{cluster states} - in fact preparing a RS for the execution of a circuit in standard measurement-based quantum computation (MBQC). However, crucially for the processing the SQPMs bases remain adaptive. This is in sharp contrast with standard MBQC with cluster states, where Clifford circuits are executed in $\mathcal{O}(1)$ time by measuring all Pauli operators in a single round~\cite{PhysRevLett.86.5188, PhysRevA.68.022312}. Differently, here, while still the RS is equivalent to a cluster state, and while also the measurement bases are Pauli, still they are adaptive - highlighting the non-Cliffordness of the operation we are inverting. This traces back to the fact that here the adaptive basis is chosen between two orthogonal bases, $X$ and $Z$.

\newpage

\section{QRAM resource-state}
\noindent Formally, $\ket\Phi$ is specified in Fig.~7 of Methods: we prepare the RSs $\ket{\Phi^{(1)}}$ and $\ket{\Phi^{(2)}}$ defined in Eqs.~\eqref{Phi_1} and~\eqref{Phi_2} respectively, and perform (partial) BMs. Specifically, referring to the notation in Eq.~\eqref{Phi_1}, we only measure the first $N-1$ qubits of $\mathcal{D}$. After this, we finally apply a Pauli correction to $\mathcal{P}\cup\mathcal{F}$, conditioned on the measurement outcomes. Despite the BMs rendering the circuit stochastic, the Pauli correction can be chosen to make it deterministic (see the section below); for now, we formally define $\ket\Phi$ as the output of this circuit when all measurement outcomes are $+1$, and no correction is applied. Thus,  
\begin{equation}\label{Phi_P}
    \ket{\Phi} = B\Bigg[\ket{\Phi^{(1)}}\otimes\ket{\Phi^{(2)}}\Bigg],
\end{equation}
where we defined the contraction $B:\mathscr{H}^{(3N-2)}\rightarrow\mathscr{H}^{(N)}$ as the linear map
\begin{equation}\label{contraction_P}
   B = \sum_{\bm{x}}\sum_{\bm{j},\bm{k}} (-)^{\bm{j}\cdot\bm{k}} D\left(\bm{x},\bm{j}\right) \ket{\bm{x}}_\mathcal{L} \bra{\bm{j}}_\mathcal{D}\bra{\bm{k}}_{\mathcal{D}'}.
\end{equation}
Therein, $\bm{x}\in\left\{0,1\right\}^N$ indexes the computational basis on $\mathcal{L}$, while $\bm{j},\bm{k}\in\left\{0,1\right\}^{2N-1}$ label the computational basis on $\mathcal{D}$ and $\mathcal{D}'$ respectively; moreover, the term $D(\bm{x},\bm{j})\equiv\prod_{\alpha=0}^{N-1}\delta_{x_\alpha, j_{N-1+\alpha}}$ forces a correspondence between $\bm{x}$ and the last $N$ components of $\bm{j}$. The action of $B$ is also schematized in the top inset of Fig.~7 of Methods.

\subsection{Virtual and physical pictures}
\noindent When we execute the query, $\ket\Phi$ is already prepared; it has support on a set of qubits $\mathcal{A}$ of dimension $|\mathcal{A}|=5N-3$, and belongs to the Hilbert space $\mathscr{H}_\mathcal{A}\cong \mathbb{C}^{2^{|\mathcal{A}|}}$. The construction in Fig.~7 of Methods suggests to partition $\mathcal{A}$ in disjoint subsets as $\mathcal{A}=\mathcal{I}\cup\mathcal{L}\cup\mathcal{P}\cup\mathcal{F}$, of sizes $|\mathcal{I}|=\log N$, $|\mathcal{L}|= N$, $|\mathcal{P}|=2\left(2N-\log N-2\right)$ and $|\mathcal{F}|=\log N + 1$; the ancillary Hilbert space is thus partitioned as
\begin{equation}
    \mathscr{H}_\mathcal{A} = \mathscr{H}_\mathcal{I} \otimes \mathscr{H}_\mathcal{L} \otimes \mathscr{H}_\mathcal{P} \otimes \mathscr{H}_\mathcal{F}.     
\end{equation} 
We will refer to this as the `physical' Hilbert space, as it describes the quantum state of the qubits $\mathcal{A}$, which are physically present on hardware during the query. Formally, $\ket\Phi$ can be written in the computational basis as $\ket{\Phi}=\sum_{i_0,\dots,i_Q}\Phi_{i_0,\dots,i_Q}\ket{i_0,\dots,i_Q}$, where $Q\equiv| \mathcal{A}|-1$ and we set the tensor entries as
\begin{equation}
    \Phi_{i_0,\dots,i_Q} = \sum_{\bm{j},\bm{k}} (-)^{\bm{j}\cdot\bm{k}} D(\bm{x},\bm{j}) \Phi^{(1)}_{i_0,\dots,i_M;\bm{j}}\; \Phi^{(2)}_{\bm{k}; i_{M+N+1},\dots,i_Q};
\end{equation}
therein, we also set $M=\log N -1$ and we identify $\bm{x}\equiv\left(i_{M+1},\dots,i_{M+N}\right)$. We find it useful to associate the contracted mute indexes $(\bm{j},\bm{k})$ with \emph{virtual} degrees of freedom; these belong to the virtual Hilbert space $\mathscr{H}_\mathcal{D}\otimes\mathscr{H}_{\mathcal{D}'}$, which is not associated to physical qubits. The interplay between the physical picture and the  virtual one is then mediated by the map $B:\mathscr{H}_\mathcal{D}\otimes\mathscr{H}_{\mathcal{D}'}\rightarrow \mathscr{H}_\mathcal{L}$; many aspects of the query will be most easily understood in the virtual picture. 

\subsection{Deterministic resource-state preparation}
\noindent This formal definition of $\ket\Phi$ is apparently non-deterministic due to the BMs. Importantly, we can always efficiently cope with this randomness by adding a Pauli correction at the end, making the output deterministic. This is allowed by the fact that one of the two states we are `merging' via BMs, namely $|\Phi^{(2)}\rangle$, is a stabiliser state; thus, by implementing standard propagation of Pauli operators to the right-side of the circuit in Fig.~7 of Methods, we eventually end up with a simple Pauli byproduct on $\mathcal{F}$.  \\

\newpage

\section{Query protocol: details}
\noindent Our proposed query, detailed in Fig.~3 of the main text, proceeds in three steps, which can be put in correspondence with those highlighted in Fig.~\ref{fig_rewriting}(a). However, in our implementation these are executed `virtually' with the aid of $\ket\Phi$ by employing the techniques introduced above. Specifically, it is convenient to use the abstract map $B$ in Eq.~\eqref{contraction_P} to connect the physical operations to a virtual-space dynamics, whose steps are more explicit. For what follows, it is useful to refer to Fig.~\ref{fig_M4}(a) for keeping track of this correspondence.\\
\indent The query begins as follows. The address-state $\ket{\psi}=\sum_{\bm{x}}\psi_{\bm{x}}\ket{\bm{x}}$ is initially supported on the $\log N$ qubits composing the QPU register $\mathcal{R}$; the resource-state $\ket\Phi$ is already assembled and stored on qubits $\mathcal{A}$.\\
\indent For step (i), we perform BMs between corresponding qubits in $\mathcal{R}$ and $\mathcal{I}$. The resulting state has the form $\ket{\Psi_{\bm{m}}}\ket{\Phi_{\bm{m}}(\psi)}=\mathbb{P}^{(\bm{m})}\ket{\psi}\ket{\Phi}$, where $\mathbb{P}^{(\bm{m})}$ is a projector describing the action of the BMs, $\ket{\Psi_{\bm{m}}}$ is the projected state of Bell pairs on $\mathcal{R}\cup\mathcal{I}$, and $\ket{\Phi_{\bm{m}}(\psi)}$ is the conditional state on $\mathcal{L}\cup\mathcal{P}\cup\mathcal{F}$. By propagating $\mathbb{P}^{(\bm{m})}$ through $B$, we can rewrite the nontrivial part of the resulting state as
\begin{equation}
    \ket{\Phi_{\bm{m}}(\psi)} = B\; \Bigg[\ket{\Phi^{(1)}_{\bm{m}}}\otimes \ket{\Phi^{(2)}}\Bigg],
\end{equation}
where $\ket{\Phi^{(1)}_{\bm{m}}(\psi)}\in\mathscr{H}_\mathcal{D}$ is supported on the virtual qubits $\mathcal{D}$. From GT, this has the form
\begin{equation}
    \ket{\Phi^{(1)}_{\bm{m}}(\psi)} = VP_{\bm{m}} \ket{\psi}\ket{0};
\end{equation}
thus, to all effects we have implemented in virtual space the first step of the rewriting in Eq.~\eqref{U_QRAM_blocks}. \\
\indent For step (ii), we now proceed by loading the data. This is achieved via the \emph{loading} operation $W_D(\bm{m})$ on $\mathcal{L}$:
\begin{equation}\label{W_m}
    W_D(\bm{m}) = \bigotimes_{l = 0}^{N-1}  Z^{D_{\bm{b}\oplus\mu^{-1}(l)}}_{l} ,
\end{equation}
where, as before, we indicated by $l\in\mathcal{L}$ the qubits in $\mathcal{L}\subset\mathcal{A}$. By applying this, in the real picture we obtain the state $\ket{\Phi_{\bm{m}}\left(\psi|D\right)} = W_D(\bm{m})\ket{\Phi_{\bm{m}}\left(\psi\right)}$; by propagating $W_D(\bm{m})$ through $B$, we can write it as
\begin{equation}\label{Phi_m(psi_M)}
    \ket{\Phi_{\bm{m}}\left(\psi|D\right)} = B\Bigg[ \ket{\Phi^{(1)}_{\bm{m}}\left(\psi|D\right)}\otimes \ket{\Phi^{(2)}}\Bigg];
\end{equation}
herein, also $|\Phi^{(1)}_{\bm{m}}(\psi|D)\rangle\in\mathscr{H}_\mathcal{D}$ is supported on the virtual qubits $\mathcal{D}$, and reads
\begin{equation}\label{Phi_m_1_psi_M}
    \ket{\Phi^{(1)}_{\bm{m}}(\psi|D)}= W_D(\bm{m})VP_{\bm{m}} \ket{\psi}\ket{0}.
\end{equation}
Therefore, we realized step (ii) of Eq.~\eqref{U_QRAM_blocks} in virtual space.\\
\indent For step (iii), we complete our protocol by inverting the virtual NOHE. We first measure all the qubits in $\mathcal{L}$ in the $X$ basis, collecting the measurement outcomes $\bm{M}$; crucially, the resulting conditional state reads
\begin{equation}
    \mathbb{P}_\mathcal{L}^{(\bm{M})} \ket{\Phi_{\bm{m}}\left(\psi|D\right)} = C_\text{NOHE}^{(\bm{M'})} \ket{\Phi^{(1)}_{\bm{m}}\left(\psi|D\right)},
\end{equation}
where $\mathbb{P}^{(\bm{M})}_\mathcal{L}$ is a projector acting on $\mathcal{L}$ which describes these measurements. Here, $\bm{M'}$ is calculated efficiently by simply propagating Pauli operators though the Clifford circuit. This equation must be read as follows: the state $\ket{\Phi_{\bm{m}}\left(\psi|D\right)}$ has support on qubits $\mathcal{L}\cup\mathcal{P}\cup\mathcal{F}$ as defined in Eq.~\eqref{Phi_m(psi_M)}; the state $|\Phi^{(1)}_{\bm{m}}\left(\psi|D\right)\rangle$ belongs to the virtual space $\mathscr{H}_\mathcal{D}$ as specified by Eq.~\eqref{Phi_m_1_psi_M}; the operation $C_\text{NOHE}$ acting on it executes a map from virtual to physical space of the form $\mathscr{H}_\mathcal{D}\rightarrow\mathscr{H}_{\mathcal{P}\cup\mathcal{F}}$. \\

\indent Further, from Eq.~\eqref{Phi_m_1_psi_M} we note that we always have $|\Phi^{(1)}\left(\psi|D\right)\rangle\in U_\text{NOHE}\left(\mathscr{H}^{(\log N+1)}\right)$. Thus, the same concept of Eq.~\eqref{SQPM_inversion} holds here as well: we can design a sequence of SQPMs on $\mathcal{P}$ such that, upon the application of a final Pauli operation, we get
\begin{equation}
    \mathbb{P}_\mathcal{L}^{(\bm{M})}\ket{\Phi_{\bm{m}}\left(\psi|D\right)} \xrightarrow[\;\;\;\;\;\text{SQPMs}\;\;\;\;\;]{} U_\text{NOHE}^{-1}\ket{\Phi_{\bm{m}}^{(1)}\left(\psi|D\right)}.
\end{equation}
The right-hand side state in the equation above is supported on the remaining unmeasured physical qubits $\mathcal{F}$. In particular, from Eq.~\eqref{Phi_m_1_psi_M}, it follows that we can retrieve the correct output of the QRAM query, i.e. $\ket{\psi_D}$, by simply applying the Pauli operator $P_{\bm{m}}^{-1}$ to the first $\log N$ qubits of $\mathcal{F}$, and a Hadamard $H$, which is Clifford, to the last qubit. 

\newpage

\section{Built-in noise resilience}\label{noise_resilience}

\noindent Here, we discuss the noise-resilience of our proposed QRAM implementation. This includes the theoretical motivation as to why the NOHE features the observed built-in noise resilience, the numerical certification of this favorable behaviour, and the specific technical analysis which explains it. This supplemental section is organised as follows.
\begin{itemize}
    \item[\textbf{A.}] We discuss the main figures of merit we are interested in, including the query fidelity and the meaning of `polylogarithmic infidelity scaling' and `noise-resilience'.
    \item[\textbf{B.}] We introduce and motivate the error models that we consider.
    \item[\textbf{C.}] We briefly summarise the background knowledge concerning the noise-resilience of QRAM. 
    \item[\textbf{D.}] We extensively discuss the analytically methods used to analytically understand the built-in noise-resilience of our and other QRAM approaches; here we also provide heuristic analysis for specific models. 
    \item[\textbf{E.}] We describe and theoretically prove the numerical reductions made in order to efficiently simulate QRAM at large system sizes. 
    \item[\textbf{F.}] We present and discuss several numerical results which are not shown in the main text of our work. 
\end{itemize}

\subsection{Framework and definitions}
\noindent As stated in Methods, we are eventually interested in the following figure of merit. Let $D\in\left\{0,1\right\}^{N}$ denote any arbitrary configuration of the memory, and let $\ket{\psi}\in\mathscr{H}^{(\log N)}$ be any arbitrary address state in the Hilbert space of $\log N$ qubits. We denote by $\ket{\psi_D}\in\mathscr{H}^{(1+\log N)}$ the ideal output of QRAM; that is, if $\ket{\psi} = \sum_{\bm{x}} \psi_{\bm{x}}\ket{\bm{x}}$, then $\ket{\psi_D} = \sum_{\bm{x}}\psi_{\bm{x}}\ket{\bm{x}}\ket{D_{\bm{x}}}$. Moreover, given a specific error model, let us denote by $\rho_D(\psi)$ the density matrix describing the faulty output, where we trace away any employed ancillary qubit and only keep the $\log N +1$ qubits supporting the address and the bus. Then, the \emph{query fidelity} is defined as
\begin{equation}\label{query_fidelity_definition}
    F_\text{QRAM} = \frac{1}{2^N}\sum_D\int_\mathscr{H}\mathrm{d}\psi \; \langle \psi_D | \rho_D(\psi) | \psi_D \rangle ,
\end{equation}
where the summation and the integral average uniformaly across any possible memory configuration and any possible input state. In practice, this is the average fidelity of the QRAM output. A formal definition of the integral can be found in Section~\ref{math methods}.\\
\indent We assume that the error model can be captured by a single parameter, the error-rate $\varepsilon$, which can be thought of as the probability of having an error on a qubit per time step. In this work, we are interested in both fully fault-tolerant and not error-corrected settings; thus, when unspecified, $\varepsilon$ can be interpreted equivalently as the error-corrected logical error-rate in a logical computation, or as the physical error-rate in a physical computation, respectively. When explicitly working in the QEC framework, $\varepsilon_\text{ph}$ shall specify the underlying physical error, while the logical error will be denoted by $\varepsilon$. We recall that in standard QEC settings, the two are related the one to another as $\varepsilon \lesssim \mathcal{O}(\varepsilon_\text{ph}^{d/2})$, where $d$ is the code distance; in the paradigmatic surface code, the physical resources to realise a logical qubit of distance $d$ scale as $\mathcal{O}(d^2)$. \\
\indent In general, the query fidelity will be a function of $\varepsilon$ and $N$; we find it convenient to expand it in powers of $\varepsilon$ as\begin{equation}
    F_\text{QRAM} = 1 - \sum_{k\geq 1} A_k(N)\varepsilon^k,
\end{equation}
absorbing the dependance on $N$ in the coefficients $A_k(N)$. When we refer to the scaling of $F_\text{QRAM}$ with the memory size, we are interested in the behaviour of the $A_k(N)$. A priori, we do not make any strong theoretical assumption on the form of these coefficients. Intuitively, it is reasonable to expect them to share the same qualitative behaviour (e.g., polynomial or logarithmic); more precisely, a sound expectation is that roughly $A_k(N) \sim A_1^k(N)$. \\
\indent Clearly, for relevant applications we are only interested in the case in which $F_\text{QRAM}$ is close to unit. This can be expected to occur when $\varepsilon \ll 1 /A_1(N)$, in which case the query fidelity is well approximated by the leading order 
\begin{equation}
    F_\text{QRAM} \simeq 1 - A_1(N)\varepsilon + \mathcal{O}(\varepsilon^2).
\end{equation}
Considering that this is the regime of practical interest, in the following we will frequently focus specifically on this leading order term. As a formal definition, we are therefore interested in the quantity
\begin{equation}\label{A_1_limit}
    A_1(N) = \lim_{\varepsilon \rightarrow 0^+} \frac{1-F_\text{QRAM}}{\varepsilon},
\end{equation}
which (as we discuss below) can be rigorously calculated by numerically extracting the limit for small $\varepsilon$.\\
\indent Another understanding of our interest in the first-order term is that it sets the scale at which the fidelity vanishes; specifically, given an error-rate $\varepsilon$, one can expect $F_\text{QRAM}$ to be heavily suppressed at memory sizes of the scale $N\sim A_1^{-1}(1/\varepsilon)$, where $A_1^{-1}$ informally represents an inversion of $A_1(N)$. Conversely, one needs $\varepsilon \ll 1/A_1(N)$ for the fidelity to be close to unit. In a QEC setting, this can be reformulated as follows: given an underlying physical error-rate $\varepsilon_\text{ph}$ and a target query infidelity $\varepsilon_\text{QRAM} = 1 - F_\text{QRAM}$, the code distance $d$ for achieving such performance  must scale as
\begin{equation}\label{distance_generic}
    d \simeq \frac{\log\big( A_1(N)/\varepsilon_\text{QRAM}\big)}{\log(1/\varepsilon_\text{ph})},
\end{equation}
which ensures that the logical error-rate is $\varepsilon\ll 1/A_1(N)$.

\subsection{Error models}\label{SM_section_error_models}
\noindent Here, we discuss the various error models that we consider in this work. We note that in general, the noise-resilience of QRAM is expected to hold for any reasonable model, where by `reasonable' we subtend local in space and time. However, the specific scaling (e.g., the exponent $\alpha$ in Eq.~\eqref{Hann_inequality}) can depend on the specific model. Moreover, when it comes to quantitative estimation for realistic (present and future) processors, it is useful to specialise to error models that capture the actual physics as closely as possible. As a final note, we remark that some of the following error models can be understood both as describing physical faults, as well as logical faults; others, are more specific to one of the two settings. We will thus specify section by section which point of view is most suitable for the treated model.    
\subsubsection{Continous depolarisation errors}
\noindent The first error model assumes that all qubits continuously undergo a depolarisation channel. This type of error affects \emph{all} the qubits on the platform, regardless on whether they undergo any operation, or they are just idly waiting to be utilised. Specifically, at any time step, each single qubit has a probability $\varepsilon$ to completely depolarise - i.e., to be substituted with the completely mixed state $\mathbb{1}/2$. This is equivalent to the fact that at any time step, each qubit undergoes (independently) a random Pauli error with probability $3\varepsilon/4$, drawn uniformly from the set $\left\{X, Y, Z\right\}$. The update rule of the CD model thus reads, for the single qubit,
\begin{equation}
    \mathcal{E}(\rho) = \left(1-\frac{3\varepsilon}{4}\right)\rho + \frac{\epsilon}{4}\bigg(X\rho X + Y\rho Y + Z\rho Z \bigg).
\end{equation}
In practice, this map must be applied continuously to all the qubits independently, e.g. by inserting a layer of depolarisation to the entire processor after each layer of parallel gates. \\
\indent At the physical level, this model describes well physical systems where the decoherence of idle qubits is comparable to the two-qubit error rates; we however note that this is typically not the case in current quantum processors. In a QEC setting, on the other hand, due to the continuous syndrome extraction, the notion of `idle' logical qubit becomes oddly defined, as all qubits can be continuously involved in gates for the syndrome extraction, even when not involved in any logical gates. In such scenarios, the CD model can capture the logical errors on logical qubits, and $\varepsilon$ can be understood as the logical error rate. \\
\indent Finally, we note that the CD error model is generally stronger than those introduced below: it continuously affects all the qubits in all the Pauli bases, and moreover at the physical level it never causes qubits to leak out of the computational basis, preventing e.g. erasure conversion. Thus, we find that this setting is of significant interest as it is expected to identity a worst-case scenario, and arguably sets the strongest bounds on the achievable performances. For this reason, we chose to discuss this specific error model in the main text.

\subsubsection{Operational Pauli errors}
\noindent We note that typically current QC platforms are much more prone to errors during gates, then to decoherence of idle qubits. That is, the gate infidelity is much larger than the typical decoherence rate. This is for instance the case of neutral atoms. Specifically, after any two-qubit gate we insert a depolarisation channel affecting the qubits involved in the gate. \\
\indent We note that, assuming that the most faulty operations are indeed the two-qubit gates, within this setting errors are actually correlated. More precisely, when two qubits undergo an entangling gate, the probability that both are affected by an error is first-order in the error rate, $\mathcal{O}(\varepsilon)$, and not second-order. To model this, we first assign an overall probability $\mathcal{O}(\varepsilon)$ that the entire gate is faulty; when this is the case, we then sample with equal weight single- and two-qubit errors, thereby accounting for strongly correlated errors as well. \\
\indent This error model is well suited for describing physical implementations without QEC. In the QEC setting, it can describe settings where `idle' qubits are assumed to be to all the effec
ts errorless, and are therefore not subjected to syndrome extraction and error-correction - that is, they do not continuously undergo syndrome extraction. 

\subsubsection{Leakage and loss errors}
\noindent In several important platforms, the dominating errors during a computation are better described by leakages out of the computational subspace $\left\{\ket{0}, \ket{1}\right\}$. In general, this occurs when the physical system implementing the qubit undergoes an error, such that it is now dark to any subsequent operation. One paradigmatic example is \emph{qubit loss}: e.g. in neutral-atoms, trapped-ions or photonic processors the atom, ion or photon can be physically lost. In this case, any operation addressing it will not have any effect. Moreover, also in platforms such as superconducting circuits, where the units cannot be physically lost, nevertheless a qubit can be erroneously left e.g. in a higher excited level of the transmon. Both these errors eventually result in the system behaving as if the faulty qubit was absent. For the case of neutral atoms, below we also discuss how such losses impact subsequent gates involving the lost atoms.  \\
\indent The type of error above can be described as follows. We introduce a leakage state $\ket{l}$, which in practice can represent any configuration out of the computational subspace, including loss. If $\varepsilon$ is the leakage (or loss) probability, in the simplest setting the error map then reads 
\begin{equation}\label{losses_Kraus}
    \mathcal{E}(\rho) = (1-\varepsilon)\rho + \varepsilon \ket{l}\bra{l};
\end{equation} 
note that in this model it is assumed that the leakage is symmetric for the two computational basis states $\ket{0}$ and $\ket{1}$. 
\\
\indent Finally, it is important to discuss how qubits which have leaked affect the computation. Specifically focussing on neutral atoms, the leakage state $\ket{l}$ can be interpreted as the atom being either physically lost, or in a state which is the result e.g. of a decay from the Rydberg state. In the latter case, one can assume that any driving which is applied subsequently to the atom will be off-resonant, such that it does not affect the atom. Thus, whatever operation one attempts to implement, the atom will always remain in $\ket{l}$. Moreover, considering how two-qubit gates are implemented, it is straightforward to see that the gate will proceed as if the leaked atom was initially in the $\ket{0}$ state. More specifically, the native two-qubit gate for neutral atom is the conventional $CZ=\mathbb{1}-2\ket{11}\bra{11}$ gate, which is implemented by driving the transition between the computational $\ket{1}$ state and a highly-excited Rydberg state $\ket{r}$; thus, the state $\ket{l}$ not being affected by the driving results in the gate having the same result as if the leaked atom was in the $\ket{0}$ state. That is, if one of the two involved atoms has leaked, it implements a $Z$ gate on the other one; if both have leaked, it implements an identity. In summary, in the numerical simulations the state $\ket{l}$ must be treated as follows: if the qubit is supposed to undergo a single-qubit rotation, the atom remains in $\ket{l}$; if the qubit is part of a $CZ$ gate, the atom is treated as effectively being in the $\ket{0}$ state. 

\subsubsection{Heating}
\noindent Atom heating can be a major limitation in the execution of deep circuits with neutral atoms. In practice, heating causes atom loss, but with a progressively increasing loss probability as the computation goes on (due to the atoms reaching higher and higher temperatures). Indeed (assuming paradigmatic experiments based on Alkali atoms), atoms tend to heat up while the tweezers are turned off during the gates: this is majorly due to ballistic isotropic expansion of the atom. As a result, the temperature of the atom increases during this `drop-time', lowering the recapture probability. Hence, after a critical number of two-qubit gates, the recapture probability is small enough that losses become a major source of error, in fact limiting the achievable circuit depths. \\
\indent In our simulations, we model atom loss from heating through a classical description of the atom's motional degrees of freedom - which we find to provide sound estimates for the experimental regimes of interest. More specifically, we assume that when the tweezer is turned off, the atom's motion is well described by a 3D Maxwell-Boltzmann distribution in energy, $p(E)\propto E^2 e^{-E/k_BT}$, where $T$ is the initial temperature. We consider a drop time $\tau_g\simeq 0.3\mu\mathrm{s}$, corresponding to typical gate times. During this time, the ballistic isotropic expansion of the atom causes an increase in the position variance, $\Delta\sigma_x^2 \propto k_BT\tau_g^2/m $, where $m$ is the mass of the atom. As a result, when the tweezer is turned on again after the gate, the mechanical energy is increased accordingly. The atom is only trapped back if the resulting mechanical energy is smaller than the trap depth $U_0$; if this is the case, the larger energy effectively yields an increased temperature $T'$. Iterating this therefore results in an increasing loss probability as the circuit progresses. To model such losses, we employ the same map as in Eq.~\eqref{losses_Kraus}, with the difference that the parameter $\varepsilon$ is now dynamically changed (individually for each atom) according to the concepts above, to account for the number of drops each atom has already experienced at each step.

\subsubsection{Phase-biased errors}
\noindent Finally, we explore the case of faults biased towards $Z$ errors. For this, we consider error maps of the form 
\begin{equation}
    \mathcal{E}(\rho) = (1 - \frac{3\varepsilon}{4})\rho + \frac{\varepsilon}{4} \bigg( r_xX\rho + X r_yY\rho Y + r_zZ\rho Z \bigg),
\end{equation}
where the weights $r_x$, $r_y$ and $r_z$ represent the fractions of errors in the Pauli directions $X$, $Y$ and $Z$ respectively, and therefore satisfy $r_x+r_y+r_z=1$. Within this generic setting, we are specially interested in modeling biases towards the $Z$ axis. We therefore set $p_x=p_y$ and vary the bias parameter $\beta_z$, defined as
\begin{equation}
    \beta_z = \frac{r_z}{r_x + r_y},
\end{equation}
such that $\beta_z = 0.5$ recovers the isotropic depolarisation discussed above, while $\beta_z\rightarrow +\infty$ corresponds to pure dephasing. 

\subsection{Background concepts: noise-resilience of QRAM}
\noindent Quantum-coherent lookup techniques to generic datasets necessarily require a space-time circuit volume which is at best $\mathcal{O}(N\log N)$ at the abstract level. Here, by `abstract' we simply mean that in a QEC setting, this is the \emph{logical} space-time volume; as we will discuss extensively, the physical space-time volume required to implement such logical circuit depends significantly on the decomposition. This includes the QRAM implementations discussed in our work, but also so-called `quantum read-only memory' (QROM) approaches~\cite{babbush2018encoding}.\\
\indent Given such circuit volume, it is natural to expect that the query fidelity generally decays (at least) polynomially in $N$. To be more precise, in general we expect that $A_1(N)=\mathcal{O}(\text{poly} N)$, signaling an overall decay of the fidelity at scales $N\sim\text{poly}(1/\varepsilon)$. This is generally the case of standard circuit decompositions of QRAM, as well as the mentioned QROM. However, a crucial, well assessed result~\cite{hann2021resilience} is that specific QRAM constructions exist, such that
\begin{equation}\label{Hann_inequality}
    A_1(N) \lesssim CT_\text{tot} \log^{\alpha -1} N, 
\end{equation}
where $C$ is a constant, $\alpha\geq 1$ is some exponent and $T_\text{tot}$ is the circuit depth. Typically, for lookup methods featuring such property, one also has $T_\text{tot} = \mathcal{O}(\log N)$, so that eventually $A_1(N)=\mathcal{O}(\log^\alpha)N$. We will refer to this favorable polylogarithmic scaling as charachterising `noise-resilient' lookup protocols, as opposed to protocols with $A_1(N)=\text{poly} N$~\cite{babbush2018encoding}. Importantly, this exponential separation also translates directly to QEC, e.g. through Eq.~\eqref{distance_generic}.  \\
\indent We remark that one crucial feature of noise-resilient protocols is that they employ $\mathcal{O}(N)$ ancillary qubits. While this is not sufficient for noise resilience, it however seems to be \emph{necessary}. Indeed, since the circuit has $\mathcal{O}(N\log N)$ space-time volume, it will suffer of $\mathcal{O}(\varepsilon N\log N)$ faults on average; if one only employed $\mathcal{O}(\log N)$ qubits, the probability that one among the output qubits has an error can be roughly estimated as $\mathcal{O}(\varepsilon N)$, which directly precludes any possibility of polylogarithmic error scaling. In contrast, if one employs $\mathcal{O}(N)$ ancillas, \emph{in principle}, there is a chance that the major part of the entropy is eventually left on the ancillas, which opens to the possibility of polylogarithmic error scaling. In practice, this is realised by circuit decmopositions of $U_\text{QRAM}$ which highly constrain the propagation of faults. The paradigmatic example of this is the bucket-brigade protocol~\cite{giovannetti2008quantum}; our circuit decomposition for the NOHE achieves the same (just in a way that can be most easily combined with neutral atom architectures).

\subsection{Theoretical origin of the noise-resilience of QRAM: analytical approach}
\noindent The noise-resilience of our QRAM architecture is shown through extensive numerical simulations. On the theoretical side, the origin of the noise resilience of QRAM protocols following the BB construction~\cite{giovannetti2008quantum} is well understood theoretically~\cite{hann2021resilience}. Specifically, such noise-resilience is directly linked to information flow (and, specifically, error-propagation) through the circuit. As we have shown above, our circuit-level decomposition of the NOHE is equivalent to the BB, and it therefore follows that our architecture benefits of the very same concepts at the circuit level. \\
\indent In this section, we review and provide theoretical insight in the noise-resilience of QRAM. For this, we first show how the main bounds to the query fidelity are analytically derived. Then, focusing on specific relevant error models, we heuristically discuss how such bounds explain our observed favorable polylogarithmic infidelity scaling. At the technical level, our analysis conceptually follows the one presented in Ref.~\cite{hann2021resilience}. 

\subsubsection{Main setting}\label{SM_noise_analysis1}
\noindent To set the ground, we start by recalling the main figures of merit, and elaborating on some basic aspects. Given a dataset $D=\left\{D_{\bm{x}}\right\}\in\left\{0,1\right\}^N$ and an address state $\ket{\psi}=\sum_{\bm{x}}\psi_{\bm{x}}\ket{\bm{x}}$, an ideal answer to the QRAM query is provided by the state
\begin{equation}
    \ket{\psi_D}=\sum_{\bm{x}}\psi_{\bm{x}}\ket{\bm{x}}\otimes\ket{D_{\bm{x}}}.
\end{equation}
For each dataset $D$ stored in the memory, let us now consider a generic Kraus map, specified by a set of Kraus operators $\left\{K_{D, k}\right\}$, such that $\sum_kK_{D, k}^\dag K_{D, k}=\mathbb{1}$; we write a faulty query to the memory as
\begin{equation}
    \ket{\psi} \xrightarrow[]{\;\;\;\text{faulty query to} \;\mathcal{M}\;\;\;}\rho_\text{out}(\psi|D) = \sum_k K_{D, k}\ket{\psi}\bra{\psi}K^\dag_{D, k};
\end{equation}
thus, we refer to the index $k$ as labeling `error configuration'. In general, $K_{D,k}$ also acts on the ancillas, so that the term $K_{D,k}\ket{\psi}\bra{\psi}K^\dag_{D,k}$ also has support on $\mathcal{O}(N)$ ancillas. We are thus interested in the marginal state, when these ancillas are traced out; we therefore define
\begin{equation}
    \rho(\psi|D)=\text{Tr}_\text{rest}\Big[\; \rho_\text{out}(\psi|D) \;\Big],
\end{equation}
where $\text{Tr}_\text{rest}$ performs the partial trace over any ancillary qubit.\\
\indent When the dataset $D$ is specified and the address state is $\ket\psi$, the fidelity of the query can be defined as the quantity
\begin{align}
    F_\text{QRAM}(\psi|\mathcal{M}) & = \langle \psi_\mathcal{M} | \rho(\psi|\mathcal{M}) |\psi_\mathcal{M} \rangle =  \nonumber \\
    & =\sum_{\alpha, k} |\langle \psi_\mathcal{M}, \alpha | K_{\mathcal{M},k} |\psi \rangle |^2,
\end{align}
where $\left\{\ket{\alpha}\right\}$ is any orthonormal basis on the ancillary Hilbert space. To capture the generic reliability of our QRAM architecture, we are interested in evaluating the following query fidelity:
\begin{equation}
    F_\text{QRAM} = \frac{1}{2^N}\sum_\mathcal{M}\int_{\mathscr{H}}\!\! \mathrm{d}\psi\; F_\text{QRAM}(\psi|\mathcal{M});
\end{equation}
therein, we average over all possible memory configurations (the factor $2^{-N}$ is the related normalization), and also over all possible states in the Hilbert space. We provide formal details on the average $\int_\mathscr{H}\mathrm{d}\psi$ in Section~\ref{Math1}.\\
\indent  For simplicity, in the following it is convenient to restrict to `unitary' error models; that is, with $K_{D,k}^\dag K_{D,k}=p_{D,k} \mathbb{1}$ for $p_{D,k}\geq 0$ and $\sum_kp_{D,k}$=1. This includes e.g. the error models introduced in Section~\ref{SM_section_error_models}, such as conventional depolarising error models, as well as dephasing channels. Typically, scaling results which hold in this restricted case also extend to more general settings, including e.g. decay and damping errors~\cite{hann2021resilience}; but unitary errors are generally simpler to treat analytically. Moreover, for the neutral atom implementation that we discuss more explicitly, these are the error models of most relevance (see the discussion below). In this assumption, it is also useful to perform a discrete stochastic unraveling of the Kraus map, by defining the conditional normalized states $\ket{\psi_{D,k}} =  K_{D,k}\ket{\psi}/\sqrt{p_k}$; then, the map reads
\begin{equation}
    \rho_\text{out} (\psi|D) = \sum_k p_k \ket{\psi_{D,k}}\bra{\psi_{D,k}},
\end{equation}
and the query fidelity takes the convenient form
\begin{align}
    F_\text{QRAM} & = \frac{1}{2^N}\sum_D\int_\mathscr{H}\!\!\mathrm{d}\psi \; \sum_k p_k F_k(\psi|D) = \nonumber \\
    & = \frac{1}{2^N}\sum_D \sum_k p_{D,k} F_k(D) ,
\end{align}
with the conditional fidelities
\begin{equation}
    F_k(\psi|D) = \sum_\alpha |\langle \psi_D,\alpha| \psi_{D,k} \rangle |^2 \;\;\;\;\; \text{and} \;\;\;\;\; F_k(D) = \int_{\mathscr{
    H}}\!\! \mathrm{d}\psi F_k(\psi|D) ;  
\end{equation}
we recall that here $\left\{\ket{\alpha}\right\}$ is an arbitrary basis on the ancillary Hilbert space. 

\subsubsection{Bounding the query fidelity: methods}\label{SM_noise_analysis2}
\noindent We now enter the noise-resilience analysis of QRAM calls. We are first interested in bounding the conditional fidelity 
\begin{equation}
    F_k(\psi|D) = \sum_\alpha |\langle \psi_D,\alpha| \psi_{D,k} \rangle |^2;
\end{equation}
for simplicity, in this section we will drop any index related to the memory $D$, and consider both the address $\psi$ and the error configuration $k$ to be fixed. \\
\indent Analogously to Ref.~\cite{hann2021resilience}, the key technical point in our error-resilience analysis is the following: for each error configuration $k$, we define a set of \emph{good addresses} $\mathcal{G}_k\subset\left\{0,1\right\}^{\log N}$; intuitively, this comprises all those (classical) addresses $\bm{x}\in\left\{0,1\right\}^{\log N}$, which are queried correctly and \emph{coherently} even under error configuration $k$. More precisely, $\mathcal{G}_k$ is defined in such a way that if the address only contained components from $\mathcal{G}$, i.e. if it was of the form $\ket\psi=\sum_{\bm{x}\in\mathcal{G}_k}\psi_{\bm{x}}\ket{\bm{x}}$, then $K_k$ would act on it equally (or very similarly) to the ideal query to the QRAM. The formal definition of $\mathcal{G}_k$ is given below.\\
\\
\noindent \textbf{Set of good addresses: definition.} First, let $\ket{A}$ be any generic state in the ancillary Hilbert space $\mathscr{H}_\mathcal{A}$; that is, 
\indent \begin{equation}
    \ket{A} = \sum_\alpha A_\alpha\ket{\alpha} \in\mathscr{H}_\mathcal{A}.
\end{equation}
Now, for each $\ket{A}$ we define the following set of classical indexes:
\begin{equation}
    \mathcal{G}_{k,\ket{A}} = \left\{  \bm{x}\in\left\{0,1\right\}^{\log N} \;\;\;\text{such that}\;\;\; K_k\ket{\bm{x}}=\sqrt{p_k}\ket{\bm{x}}\ket{D_{\bm{x}}}\ket{A}\right\}.
\end{equation}
Finally, we pick the state $\ket{A^*_k}\in\mathscr{H}_\mathcal{A}$ which maximises the size of $\mathcal{G}_{k,\ket{A}}$, i.e. such that
\begin{equation}
    \Big|\mathcal{G}_{k,\ket{A^*_k}} \Big| = \text{max} \bigg\{ |\mathcal{G}_{k,\ket{A}} |\;\;\;\text{for}\;\;\; \ket{A}\in\mathscr{H}_\mathcal{A}  \bigg\}.
\end{equation}
We then define $\mathcal{G}_k\equiv\mathcal{G}_{k,\ket{A^*_k}}$. We note that in general there will be several disjoint subsets of addresses which can be queried coherently in superposition; thus, \emph{a priori}, there is no reason as to why $\ket{A_k^*}$, and thus $\mathcal{G}_k$, should be unique. For the following, in case multiple choices are possible, it is sufficient to choose one, with no consequence on our results.\\
\\
\noindent \textbf{Generic bound.} With this, we straightforwardly get the following form for the conditional state:
\begin{equation}
    \ket{\psi_k} = \frac{1}{\sqrt{p_k}} \bigg[ \sqrt{p_k} \Big( \sum_{\bm{x}\in\mathcal{G}_k}\psi_{\bm{x}}\ket{\bm{x}}\ket{D_{\bm{x}}} \Big)\ket{A_k} + \sum_{\bm{x}\notin\mathcal{G}_k}\psi_{\bm{x}}K_k\ket{\bm{x}} \bigg],
\end{equation}
which allows us to separate the portion which is queried coherently from the faulty one. We can now choose a basis $\left\{\ket{\alpha}\right\}$ such that $\ket{A_k}\in\left\{\ket{\alpha}\right\}$; it follows that
\begin{align}
    F_k(\psi|D) & = \sum_\alpha |\langle\psi_{D}, \alpha|\psi_{k} \rangle |^2 \geq \nonumber \\
    & \geq |\langle\psi_{D}, A_k|\psi_{k} \rangle |^2 = \nonumber\\
    & =\frac{1}{p_k}\Bigg|\sqrt{p_k} \sum_{\bm{x}\in\mathcal{G}_k}|\psi_{\bm{x}}|^2 + \langle \psi_D, A_k | \sum_{\bm{x}\notin\mathcal{G}_k}\psi_{\bm{x}}K_k\ket{\bm{x}} \Bigg|^2,
\end{align}
where we simply used the positivity of the terms in the sum. \\
\indent We are now clearly interested in bounding the portion of the wavefunction which is not queried correctly; in Section~\ref{Math2}, we use simple linear algebra to show that the following bound always holds:
\begin{equation}\label{eq66}
    \bigg|\langle \psi_D, A_k | \sum_{\bm{x}\notin\mathcal{G}_k}\psi_{\bm{x}}K_k\ket{\bm{x}} \bigg| \leq  \sqrt{p_k}\Big[ 1 - \sum_{\bm{x}\in\mathcal{G}_k}|\psi_{\bm{x}}|^2 \Big].
\end{equation}
Using this, we are now ready to bound the conditional query fidelity: combining the bound above with standard inequalities, in Section~\ref{Math3} we show that
\begin{equation}\label{eq67}
   F_k(\psi|D) \geq \Bigg[ 2\sum_{\bm{x}\in\mathcal{G}_k}|\psi_{\bm{x}}|^2 - 1 \Bigg]^2 \Theta\Bigg( \sum_{\bm{x}\in\mathcal{G}_k}|\psi_{\bm{x}}|^2 - \frac{1}{2} \Bigg).
\end{equation}
Importantly, note that this bound is only nontrivial when at least one half of the address wavefunction is queried coherently, as highlighted by the Heaviside step function. \\
\indent We are now interested in performing the averages over the error configurations $k$, the memory configurations $D$, and the state $\psi$. We start from the latter, i.e., from the conditional fidelity
\begin{equation}
    F_k(D) = \int_\mathscr{H} \mathrm{d}\psi\; F_k(\psi|D).
\end{equation}
In principle, with the methods developed in the mathematical inserts, we can perform exactly the integration over the right-hand side of the bound given so far. However, more conveniently, we can use the well known variance inequality,
\begin{equation}
    \int_\mathscr{H} \mathrm{d}\psi\; f^2(\psi) \geq \bigg( \int_\mathscr{H} \mathrm{d}\psi\; f(\psi) \bigg)^2
\end{equation}
to further bound $F_k(D)$. Specifically, also using the fact that the Heaviside function is idempotent, i.e. $\Theta^2(x)=\Theta(x)$, we get the inequality 
\begin{equation}
    F_k(D) \geq 4\Bigg[  \int_\mathscr{H} \mathrm{d}\psi\; \bigg( \sum_{\bm{x}\in\mathcal{G}_k}|\psi_{\bm{x}}|^2 - \frac{1}{2} \bigg) \;\Theta \bigg( \sum_{\bm{x}\in\mathcal{G}_k}|\psi_{\bm{x}}|^2 - \frac{1}{2} \bigg)  \Bigg]^2.
\end{equation}
Evaluating this is rather technical. With the methods in Sections~\ref{Math4},~\ref{Math5} and~\ref{Math6} we calculate that
\begin{equation}
    \int_\mathscr{H} \mathrm{d}\psi\; \bigg( \sum_{\bm{x}\in\mathcal{G}_k}|\psi_{\bm{x}}|^2 - \frac{1}{2} \bigg) \;\Theta \bigg( \sum_{\bm{x}\in\mathcal{G}_k}|\psi_{\bm{x}}|^2 - \frac{1}{2} \bigg) \geq \frac{|\mathcal{G}_k|}{N} -\frac{1}{2}; 
\end{equation}
we note that this bound can be improved by adding a Heaviside function to the right, as one can easily see that the integral must always be positive; however, for our purposes this will be sufficient. With this, using the variance inequality we can find the bound given in the Methods:
\begin{align}\label{bound_general}
    F_\text{QRAM} & \geq 4 \sum_k p_k \Bigg[\frac{|\mathcal{G}_k|}{N} -\frac{1}{2}\Bigg]^2 \geq \nonumber \\
    & \geq 4 \Bigg[\frac{1}{N} \sum_k p_k |\mathcal{G}_k| -\frac{1}{2}\Bigg]^2.
\end{align}
This expression will be the basis for the theoretical heuristic analysis below.

\subsubsection{Origin of noise-resilience: calculation}\label{estimates_noise_resilience}
\noindent The core fact underlying the noise-resilience of QRAM is that in general, for realistic noise-models, the fraction of `good' addresses only decreases polylogarithmically in the memory size for small error rates,
\begin{equation}\label{heuristic_expectaion}
    \frac{1}{N} \sum_k p_k |\mathcal{G}_k| \geq 1 - \varepsilon\text{polylog} N + \mathcal{O}(\varepsilon ^2),
\end{equation}
for small values of the error rate $\varepsilon$. We show this explicitly in the paragraphs below, also specialising to explicit error models. From this, also using the Bernoulli inequality, one directly gets
\begin{equation}
    \frac{1-F_\text{QRAM}}{\varepsilon} \leq \text{polylog}N + \mathcal{O}(\varepsilon),
\end{equation}
such that, invoking Eq.~\eqref{A_1_limit}, the leading term in $N$ specifying the query fidelity scaling in the high-fidelity regime is 
\begin{equation}
    A_1(N) \leq \text{polylog} N
\end{equation}
for some polylogarithmic function. In general, the latter can depend on the error model.\\
\indent From the discussion above, it is therefore clear that understanding the origin of the noise-resilience of QRAM can be reduced to proving Eq.~\eqref{heuristic_expectaion}. To gain heuristic insight in it, we can proceed with the following argument. For simplicity, here we have in mind the continuous depolarisation error model described above, such that any component of our QRAM architecture is always subjected to faults at every time step, even if idle. \\
\indent To start, let $\eta$ be the probability that one qubit involved in the NOHE suffers an error during the whole procedure; that is, $\eta$ will scale as
\begin{equation}\label{eta_definition}
    \eta \simeq 1 - (1-\varepsilon)^{T_\text{tot}},
\end{equation}
where $\varepsilon$ is the error-rate in the models of Section~\ref{SM_section_error_models}, and the exponent considers the fact that the entire  QRAM protocol involves $T_\text{tot}$ logical steps, with  $T_\text{tot} = \mathcal{O}(\log N)$. Next, recalling that we are interested in the low-error regime, we can separate the error-events $k$ based on the number of errors occurred. More precisely, note that we are eventually interested in calculating the limit~\eqref{A_1_limit}: this correpsonds to truncating to the scenarios where at most one error has occurred in the whole architecture, while we neglect any two-fault event. This motivates considering an expansion of the form
\begin{align}\label{G_expansion}
    \mathbb{E}\Big[ |\mathcal{G}_k| \Big] & = \sum_k p_k |\mathcal{G}_k| = \nonumber \\
    & = (1-\eta)^{\nu} N  + \nu\eta(1-\eta)^{\nu -1}  \mathbb{E}\Big[|\mathcal{G}_{k|1}|\Big] + \dots ,
\end{align}
where $\nu = \mathcal{O}(N)$ is the total number of potential error locations, and $ \mathbb{E}\Big[|\mathcal{G}_{k|1}|\Big]$ is the expected value of $|\mathcal{G}_{k}|$, conditioned on the fact that only one error occurred. For concreteness, note that $\nu$ can depend nontrivially on the specific error model (see the discussion below).\\
\indent To estimate $\mathbb{E}\Big[|\mathcal{G}_{k|1}|\Big]$, we proceed as follows. First, recall that this corresponds to the expected number of errorless queried memory locations in the assumption that only one error has occurred. For simplicity, here we focus on the continuous depolarisation model: thus, each qubit undergoes an error with the very same probability. Thus, under the condition that one error \emph{has} occurred, since $2N-1$ qubits are involved in the circuit, each has undergone an error with probability $1/(2N-1)$. Now, recalling the division in sectors, we note that an error affecting one qubit in sector $K$ will affect exactly $N/2^K$ memory locations - thus, leaving $N-N/2^K$ errorless.\\
\indent Let us now assume that the only one error has occurred in sector $K$. During the circuit, this will spread across the sector; it is easy to show that on average, accounting for this spreading, the number of qubits in sector $K$ which are affected by the error is $2^K(K+1)$. To understand this, one can proceed as follows. An error in the first two qubits will necessarily affect all the $2^K$ qubits in the sector. An error in the following $2$, will only affect half of it, thus $2^{K-1}$. The next four qubits in the sector will only affect one fourth, thus $2^{K-2}$, and so one. In practice, the average number of qubits in sector $K$ which are affected by the error originating in sector $K$ is estimated as
\begin{equation}
    2\times 2^K + \sum_{j=1}^{K-1}2^j\times 2^{K-j} = 2^K(K+1)
\end{equation}
as claimed. In addition, we note that errors can also propagate to previous sectors $J<K$; however, such errors need not be taken in account in our calculation, as it is easy to see that they will necessarily propagate forward to affect the same memory locations that we are already counting. \\
\indent Combining the facts above, it follows that the expected number of errorless locations can be estimated as
\begin{align}\label{heuristics1}
    \mathbb{E}\Big[|\mathcal{G}_{k|1}|\Big] & \simeq \frac{1}{2N-1}\sum_{K=0}^{\log N} 2^K \Bigg[N - (K+1)\frac{N}{2^K}\Bigg] \nonumber \\
    & = N - \frac{N}{2N-1} \frac{(1+\log N)(2+\log N)}{2} \simeq N - \frac{1}{4}\log^2 N.
\end{align}
Plugging this in Eqs.~\eqref{G_expansion} and~\eqref{bound_general}, to leading order in $\eta$ we therefore derive
\begin{equation}\label{heuristics2}
    F_\text{QRAM} \geq \Bigg[ 1 - \frac{\nu \eta}{2N-1} (1+\log N)(2+\log N) \Bigg]^2 + \mathcal{O}(\eta^2). 
\end{equation}
Invoking the Bernoulli inequality and Eq.~\eqref{A_1_limit}, we therefore get the expression
\begin{equation}
    A_1(N) \leq \frac{2\nu}{2N-1} (1+\log N)(2+\log N) \lim_{\varepsilon\rightarrow 0^+}\frac{\eta}{\varepsilon},
\end{equation}
where we recall that $\eta$ is the only part in the right-hand side above that depends on $\varepsilon$. From Eq.~\eqref{eta_definition} we get $\lim_{\varepsilon\rightarrow 0} \eta/\varepsilon = T_\text{tot}$, therefore yielding 
\begin{equation}
    A_1(N) \leq \frac{2\nu T_\text{tot}}{2N-1} (1+\log N)(2+\log N). 
\end{equation}
To conclude our proof, we now recall that $T_\text{tot} = \mathcal{O}(\log N)$, and that in general one has $\nu \sim \mathcal{O}(N)$. From this, it is obvious that the dominator $2N-1$ will suffice to compensate the scaling with $N$ of $\nu$, only leaving polylogarihmic terms in the expression above. \\
\\
\noindent \textbf{Bounding the leading exponent.} To be more concrete, we can focus on the continuous depolarisation model, and provide a bound to the leading exponent. More precisely, assuming a behaviour of the form $A_1(N) \sim \log^\alpha N$ for large $N$, where $\alpha$ is the dominant exponent, we are now interested in bounding $\alpha$. For this, note that in the continous depolarisation model, if $Q=\mathcal{O}(N)$ qubits are involved in total, then the number of potential error locations is $\nu = Q T_\text{tot} = \mathcal{O}(N\log N)$. Here, we recalled that in this model all qubits can undergo an error at \emph{any} time. Thus, in such setting one gets
\begin{equation}
    \frac{\nu}{2N-1} = \mathcal{O}(\log N).\;\;\;\;\;\;\;\;\; \text{(continous depolarisation)}
\end{equation}
Combining with the scaling $T_\text{tot}=\mathcal{O}(\log N)$, one therefore ends up with an estimate of the form $\alpha \lesssim 4$. Otherwise, considering models where faults dominantly arise from gates, e.g. the model of operational Pauli errors, one would rather consider $\nu=\mathcal{O}(N)$. From this, one would get 
\begin{equation}
    \frac{\nu}{2N-1} = \mathcal{O}(1),\;\;\;\;\;\;\;\;\; \text{(operational errors)}
\end{equation}
which suggests $\alpha \lesssim 3$ for this model. However, we note that in our numerics we observe significantly smaller dominating exponents, suggesting that the bounds above, while instructive, are overestimating the impact of errors in the circuit. 

\subsection{Numerical simulation of QRAM: methods}
\noindent Here, we discuss the numerical methods used to simulate our QRAM approach at meaningful scales. In general, the numerical challenge is that QRAM requires the application of a quantum circuit on $\mathcal{O}(N)$ qubits, which are formally described by a Hilbert space of dimension $2^{\mathcal{O}(N)}$. Apparently, this makes numerical simulations rapidly intractable, as the required resources seem to be exponential in the size $N$ of the memory. \\
\indent Importantly, standard \emph
{circuit-level} decompositions of QRAM, while complex, can actually be simulated efficiently on a classical computer - i.e., with $\text{poly} N$ resources - with methods discussed e.g. in Refs.~\cite{hann2021resilience, nest2010simulatingquantumcomputersprobabilistic}. These numerical methods also apply to our circuit decomposition of QRAM as combination of $V$, $W_D$ and $V^{-1}$, with the circuit diagrams shown e.g. in Fig.~\ref{non_parallel_NOHE} or in Fig.~\ref{parallel_NOHE}. However, such methods do not easily combine with our \emph{measurement-based} implementation.  \\
\indent Our strategy for the numerical simulation of QRAM is therefore as follows: (i) First, we show that our measurement-based implementation has \emph{exactly the same} noise propagation properties as in our circuit-level implementation; (ii) Once this is well assessed, we numerically simulate the circuit-level implementation, to extract the fidelity scaling also in our fully measurement-based scheme.  \\
\indent In the sections below, we present these concepts according to the following scheme: 
\begin{enumerate}
    \item We briefly review how circuit-level QRAM can be efficiently simulated; 
    \item We discuss the challenges of adapting the same methods to our measurement-based implementation; 
    \item We show that numerically calculating the query fidelity in our measurement-based implementation can be reduced to the simulation of the related circuit construction.   
\end{enumerate}
Together, these points explain the series of numerical reductions that we implemented in order to simulate large-scale instances of our QRAM architecture. 

\subsubsection{Efficient simulation of circuit-based QRAM}
\noindent Circuit-based QRAM can be simulated efficiently on a classical computer by combining the following facts: 
\begin{itemize}
    \item While $\mathcal{O}(N)$ qubits are involved in total, the \emph{input} has nontrivial support on $\mathcal{O}(\log N)$ qubits only.
    \item The circuit can be written in terms of Pauli, controlled-not and and Toffoli gates only.
\end{itemize}
The second point implies that computational basis states are always mapped to computational basis states - that is, the number of nonzero amplitudes in the computational basis decomposition remains constant. Together with the first point, this implies that we only need to keep track of $\mathcal{O}(N)$ amplitudes, even though the full Hilibert space has dimension $2^{\mathcal{O}(N)}$. More precisely, at step $t$ the quantum state is of the form
\begin{equation}
    \ket{\Psi(t)} = \sum_{\bm{x}\in\mathcal{B}(t)}\psi_{\bm{x}(t)}\ket{\bm{x}(t)},
\end{equation}
where $\mathcal{B}(t)\subset\left\{0,1\right\}^{\mathcal{O}(N)}$ is a time-dependent subset of the computational basis of $\mathcal{O}(N)$ qubits, but crucially has an exponentially smaller size $|\mathcal{B}(t)| = 2N$. \\
\indent Importantly, even within the error models that we consider, both these points remain valid for a single trajectory of our Monte Carlo simulation, allowing for efficient numerics in the noisy setting as well. This is because we always consider Pauli errors and employ a quantum trajectory approach to implement the related Kraus maps, which simply require the stochastic application of Pauli operators. While the case of qubit losses may seem to be more involved, as it is not decomposed in Pauli Kraus operators, it can also be implemented efficiently, by simply equipping all the qubits in our simulation with an additional `flag' bit, which specifies whether the qubit was lost or not.  \\
\\
\noindent \textbf{Challenges for measurement-based QRAM.} To \emph{directly} simulate the \emph{full} measurement-based version protocol, for $V^{-1}$ one needs to implement Pauli measurements and feedforward operations on the resource-state $\ket{\Phi^{(2)}}$. While \emph{per se} this can be efficiently simulated (as $\ket{\Phi^{(2)}}$ is a stabiliser state), it poses a crucial challenge when combined with the simulation of $V$ for QRAM detailed above. Indeed, exactly because it is a stabiliser state, $\ket{\Phi^{(2)}}$ has support on the full computational basis of all the $\mathcal{O}(N)$ involved ancillary qubits. This can be efficiently coped with through the stabiliser formalism within a fully Clifford setting, but it is not naively compatible with the framework above necessary for $V$. To be precise: one could straightforwardly exploit the stabiliser formalism to directly simulate our Clifford measurement-based implementation of $V^{-1}$, only \emph{if} the input state was a stabiliser state itself; but unfortunately, this is not the case (indeed, it is a highly non-stabiliser state), preventing us from employing this strategy. While such simulation is not necessarily numerically \emph{impossible}, it would further developments on the numerical side; since our methods allow to rigorously calculate the fidelity otherwise, we leave such developments to future work, potentially targeting larger-scale simulations e.g. within complex quantum algorithms. 

\subsubsection{Equivalence of error propagation between circuit-based and measurement-based QRAM}
\noindent Here, we show that our measurement-based implementation of $V^{-1}$ has the very same information flow and error propagation as the related circuit-based version. To set the ground, we recall that, in standard measurement-based quantum computation (MBQC), the propagation of errors through the gadgets realising a given circuit primitive is equivalent to the propagation through the circuit primitive itself - a direct consequence of the specific gadget construction~\cite{PhysRevA.68.022312}. Here, however, we employ a non-standard MBQC to invert $V$ on its image in the codomain $V(\mathscr{H})$: this is a \emph{destructive} protocol and, while it can be related to a circuit primitive acting on $V(\mathscr{H})$, it might not be clear a priori to what extent all the properties of the circuit primitive translate to the MBQC setting.\\
\indent More precisely, the only potential concern is that the hypothesis that our measurement-based protocol fully implements $V^{-1}$ is actually imprecise: as discussed in our manuscript, the fact that we can execute this task in a Clifford way is due to the fact that we are only inverting $V$ on its image, and to all the effects we are exploiting measurements to implement an irreversible (and thus, non unitary) operation. One therefore needs to be careful when appealing to the equivalence between circuit-based and measurement-based quantum computation, as this is typically formulated and formally proved for unitary circuits.\\
\indent We now argue that the very same error-propagation of the reversible circuit QRAM also applies in our measurement-based, irreversible case. Conceptually, this can be understood as follows. In the circuit implementation, one introduces ancillary qubits, which should ideally be left disentangled from the output. In a faulty scenario, such ancillas actually remain entangled, such that tracing them out leads to decoherence on the output. In practice, tracing out is equivalent to allowing the environment to `measure' the ancillas; we recall that, by definition, the outcome is independent on such measurement basis. In our measurement-based protocol, we \emph{directly} measure the ancillas. As we show below, this is completely equivalent to enforcing a certain desired measurement basis to the environment, and assuming to have access to the information on the measurement outcome. While we make use of such information in subsequent steps, this automatically generates the very same fault on the remaining qubits. \\
\indent In the next paragraphs, we support our insight by directly demonstrating that the propagation of a certain error has the \emph{exact same effect} in the non-Clifford circuit implementation of our NOHE primitive, as well and in our measurement-based Clifford implementation. We consider two distinct cases as examples of this: in the first two sections below, we analyse the effect of an error which does not result in the faulty interpretation of a measurement outcome; in the third section, we consider the opposite case, when an error also leads to a faulty measurement. With this, on the one hand we show that our measurement-based implementation, for what concerns the error propagation, is equivalent to the circuit-level implementation of the BB; on the other hand, we support our insight that, as a consequence, the same error-resilience is expected. \\
\\
\begin{figure}[t!]
    \centering
    \includegraphics[width=0.9\linewidth]{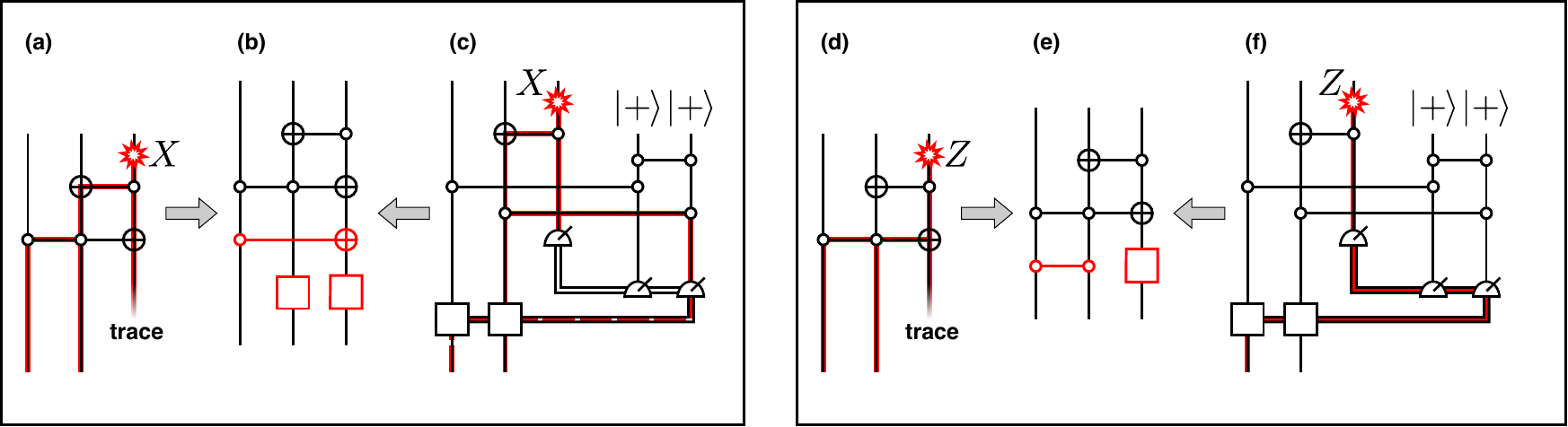}
    \caption{ \textbf{Equivalence between circuit and measurement-based primitives.} (a,b,c) Here we confront the propagation of an $X$ error on the third qubit through the circuit (non-Clifford) primitive in (a) and the (Clifford) measurement-based primitive in (c). (a) Here, the error propagates as in Eq.~\eqref{propagation1}; ideally, the third qubit is should then be left in $\ket{0}$, and is therefore traced out. (c) Here, we actively measure the third qubit in the $X$ basis. Note that this error does not affect the outcome of the first measurement, but nevertheless it propagates due to the previous operations. Moreover, there can be a faulty feedforward from the measurement of the fifth qubit to the first qubit, if the outcome of the first measurement is $-1$ - as stated by Eq.~\eqref{propagation2}. (b) Both scenarios (a) and (b) eventually result in the same error map given in Eq.~\eqref{errormap_case1}. (d,e,f) Here, we consider the propagation of a $Z$ error on the third qubit. (d) The $Z$ error does not propagate through the first controlled-not, but it propagates through the subsequent Toffoli gate as in Eq.\eqref{propagation3}. (f) Here, the $Z$ error does not propagate directly to the left qubits. However, it results in an erroneous interpretation of the measurement outcome, and therefore propagates to the ancillas to the right by introducing an error in the measurement basis; this propagates to the output qubits via the resulting faulty feedforward. (e) Again, one can see that (d) and (f) eventually result in the same error model, corresponding to a $CZ$ gate between the two output qubits. 
    } 
    \label{fig_errors}
\end{figure}

\noindent \textbf{Error-propagtaion in the non-Clifford circuit-based implementation.} We consider the circuit primitive in Fig.~\ref{fig_errors}(a), which in our circuit we use to invert the Friedkin gate (under the assumption that the third qubit must end up in the $\ket{0}$ state). Thus, ideally the output would read
\begin{equation}
    \ket{\Psi_\text{ideal}} = \Big[ \ket{\psi_0}\ket{0}_1 + \ket{\psi_1}\ket{1}_1  \Big]\ket{0}_3;
\end{equation}
here, $\ket{\psi_{0}}$ and $\ket{\psi_{0}}$ are the marginal states (also including qubit $2$), conditioned on the first qubit being in $\ket{0}$ or $\ket{1}$, respectively. These marginal states also include any other qubit in the platform, and from normalization they must respect $||\psi_0||^2+||\psi_1||^2=1$. \\
\indent To illustrate the key point, let us suppose an $X$ error occurs on qubit $3$. Then, the propagation through the circuit primitive reads
\begin{equation}\label{propagation1}
    X_3 \longrightarrow X_2X_3 CX_{1,3},
\end{equation}
where $CX_{1,3} = \mathbb{1}_{1,3} - 2\ket{1_1, -_3}\bra{1_1, -_3}$ is the controlled-not gate, with qubit $1$ acting as control and qubit $3$ being the target. The faulty output of the circuit thus reads
\begin{align}\label{errormap_case1}
    \ket{\Psi_\text{error}} & =  \ket{\psi'_0}\ket{0}_1\ket{1}_3 + \ket{\psi'_1}\ket{1}_1  \ket{0}_3 = \nonumber \\
    & = \frac{\ket{\psi'_1}\ket{1}_1 + \ket{\psi'_0}\ket{0}_1}{\sqrt{2}} \ket{+}_3 + \frac{\ket{\psi'_1}\ket{1}_1 - \ket{\psi'_0}\ket{0}_1}{\sqrt{2}} \ket{-}_3, 
\end{align}
where we set $\ket{\psi'_{0,1}}=X_2\ket{\psi_{0,1}}$, and we rewrote the expression as in the last line for reasons that will become clear soon. \\
\indent We are now interested in tracing out qubit $3$, which should ideally be disentangled. Recalling that the trace can be performed on any arbitrary orthonormal basis of qubit $3$, it is convenient to do it in the $X$ basis. This results in the following density matrix: 
\begin{align}
    \rho & = \text{Tr}_3 \Big[\ket{\Psi_\text{error}}\bra{\Psi_\text{error}}\Big] = \nonumber \\
    & =  \frac{1}{2} X_2 \ket{\Psi_\text{ideal}}\bra{\Psi_\text{ideal}} X_2 +  \frac{1}{2} Z_1X_2 \ket{\Psi_\text{ideal}}\bra{\Psi_\text{ideal}} Z_1X_2.
\end{align}
This can be interpreted as follows: with probability $1/2$ the error is $X_2$, and with probability $1/2$ is is $Z_1X_2$. To all effects, this can be understood by thinking at the environment measuring qubit $3$ in the $X$ basis, getting the outcome $s=\pm 1$ with uniform probabilities, and therefore resulting in a conditional error in the output state.\\
\\
\noindent \textbf{Error-propagation in the Clifford measurement-based implementation.} We now analyse the propagation of the very same error (an $X$ error on qubit $3$), for our measurement-based, Clifford gadget. For this, recall that the protocol now introduces an ancillary Bell pair on qubits $4$ and $5$, and proceeds in three steps: (i) We apply $CX_{3,2}$, $CZ_{1,4}$ and $CZ_{2,5}$ (in this order, note that the first does not commute with the third), where $CZ_{i,j}=\mathbb{1}-2\ket{1_i,1_j}\bra{1_i,1_j}$ is the controlled-$Z$ gate; (ii) We measure qubit $3$ in the X basis, registering the value $s=0$ if the outcome is $+1$, and $s=1$ if it is $-1$; (iii) If $s=0$ we measure qubits $4$ and $5$ in the $Z$ basis, and if $s=1$ we measure them in the $X$ basis, collecting the outcomes $\alpha$ (qubit $1$) and $\beta$ (qubit $2$) with the same convention above; (iv) We apply the byproduct correction
\begin{equation}\label{SM_noise_byproduct}
    Z_1^{(1-s)\alpha + s\beta}\otimes Z_2^{(1-s)\beta + s\alpha}
\end{equation}
on qubits $1$ and $2$. In the absence of errors, this produces the same result as the circuit primitive above (under the crucial assumption that qubit $3$ must end in the $\ket{0}$ state). \\
\begin{figure*}[t!]
    \begin{minipage}{1.0\textwidth}
        \includegraphics[width=\textwidth]{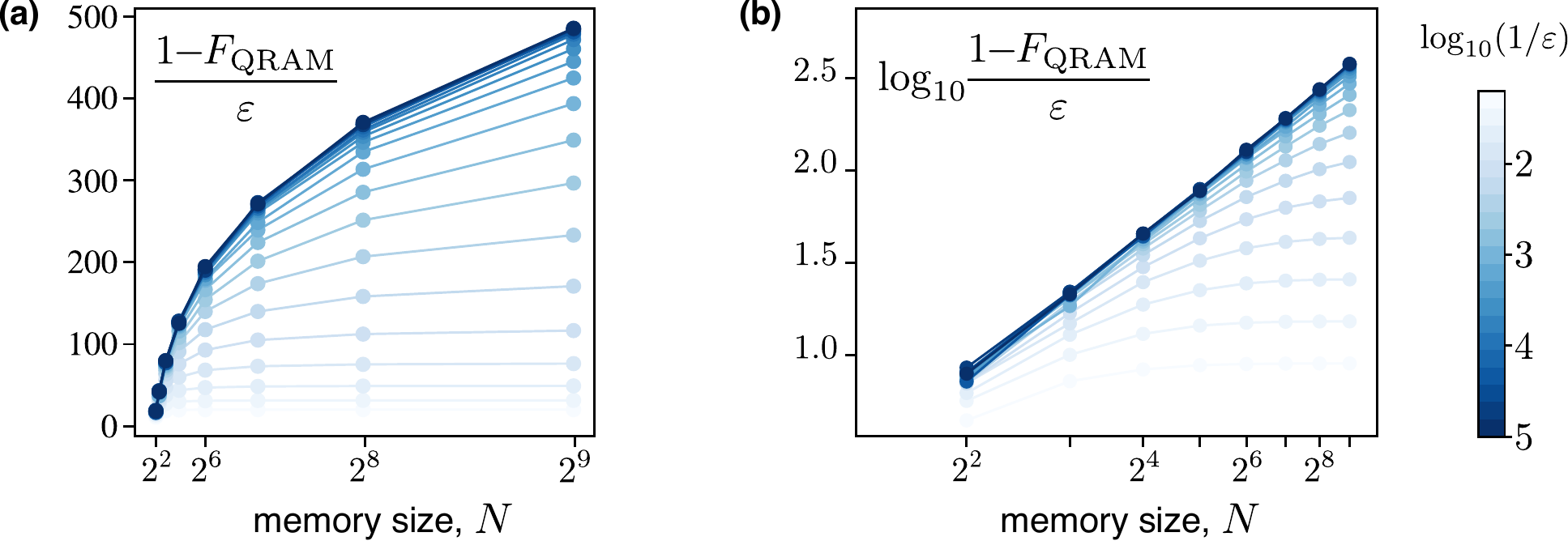}
    \end{minipage}%
\caption{ \textbf{Continuous depolarisation model.} (a) Query infidelity $\varepsilon_\text{QRAM} \equiv 1-F_\text{QRAM}$ in our QRAM scheme rescaled by the error rate $\varepsilon$, for various values of such rate. Collapse to a common line for small error rates certifies that we are numerically reaching a reliable limit $\varepsilon\rightarrow 0^+$ for the regimes considered. (b) The same plot is shown in a log-loglog plot: that is, on the vertical axis we display linearly $\log_{10}((1-F_\text{QRAM})/\varepsilon)$, while in the horizontal axis we display linearly $\log_{10}\log N$. Thus, collapse to a straight line probes the polylogarithmic dependence $(1-F_\text{QRAM})/\varepsilon \simeq C\log^\alpha N$ for small $\varepsilon$. }
\label{fig_CD}
\end{figure*}
\indent Let us now consider the same error as before; that is, an $X$ error on qubit $3$ prior to the execution of the protocol. We first look at the propagation of the error through step (i), i.e. before any measurement is performed:
\begin{equation}
    X_3 \xrightarrow[]{\;\;\;\text{step (i)}\;\;\;} X_2X_3Z_5. 
\end{equation}
Next, we proceed with step (ii) and measure qubit $3$ in the $X$ basis, registering the outcome $s$. Note that the error $X_3$ commutes with the measurement basis, and thus it does not affect the measurement outcome. We can now consider the two possible outcomes separately. 
\begin{itemize}
    \item If we get the outcome $s=0$ (that is, we measure $+1$), step (iii) predicts to measure qubits $4$ and $5$ in the $Z$ basis. In this case, note that the measurement outcomes of this round of measurements is not influenced by the error, as the measurement basis commutes with it. Thus, the feedforward proceeds exactly, and we are eventually only left with the error $X_2$. 
    \item If, instead, we get the outcome $s=1$, for step (iii) we must measure qubits $4$ and $5$ in the $X$ basis. Crucially, the error now influences the measurement outcome on qubit $5$: since the error anticommutes with this measurement basis, $(X_5)(X_2X_3Z_5)=-(X_5)(X_2X_3Z_5)$, we will interpret the outcome $\beta$ erroneously, and therefore the error propagates in the byproduct correction in Eq.~\eqref{SM_noise_byproduct}. It is simple to see that this results in an additional $Z$ error on qubit $1$; thus, in this case the error propagated by the measurement-based protocol is $Z_1X_2$.
\end{itemize}
In summary, it is easy to see that, depending on the registered outcome $s$, the error propagation reads
\begin{equation}\label{propagation2}
    X_3 \longrightarrow Z_1^sX_2. 
\end{equation}
To conclude our argument, we note that the outcomes $s$ are uniformly distributed, i.e. we get $s=0$ or $s=1$ with $1/2$ probability each; this is a typical property of such measurement-based protocols, and in this case it can also be verified by straightforward calculation. Thus, in summary again we can get an error $X_2$ or $Z_1X_2$ with uniform probability, which is \emph{exaclty} what happened in the case above.  \\
\noindent \textbf{Comparison when the fault impacts the measurement outcome.} Finally, we consider a more complicated, yet crucial comparison. That is, we now consider an initial $Z$ error on qubit $3$. This has the very important feature that, in our Clifford measurement-based implementation, this error in principle \emph{does} impact the outcome of the $X$ measurement on qubit $3$: in fact, it flips it. As such, it can be regarded as a different type of error with respect to the one above, and we are interested in analysing it separately. While it apparently seems more demanding, here it is actually simpler to prove that the propagation of the error through the measurement-based gadget is again exactly the same as in the circuit primitive.\\
\begin{figure*}[t!]
    \begin{minipage}{1.0\textwidth}
        \includegraphics[width=\textwidth]{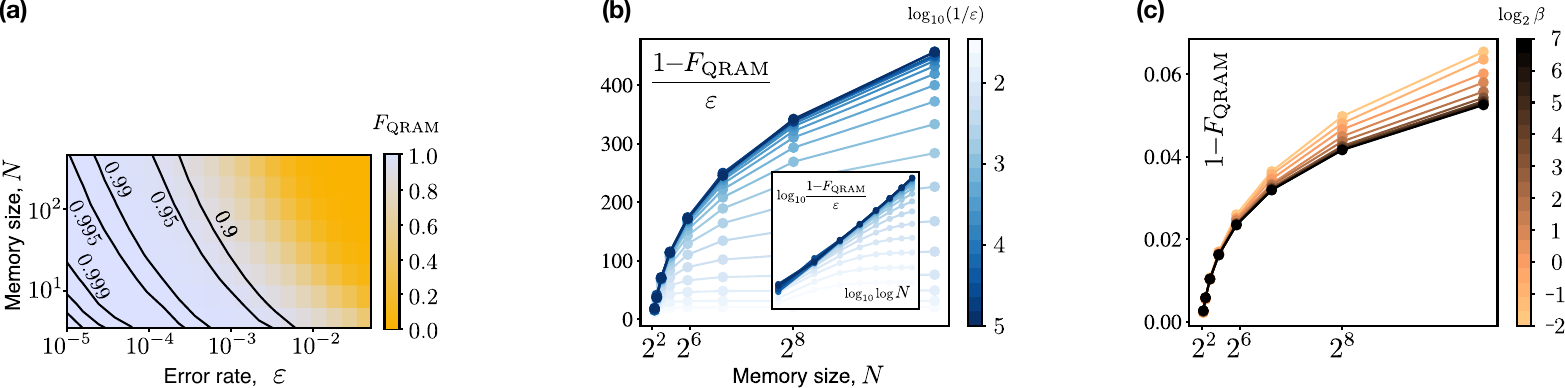}
        
    \end{minipage}%
\caption{ \textbf{Operational Pauli errors and biased noise.} (a) Query fidelity $F_\text{QRAM} $ in our QRAM scheme under two-qubit gate errors, accounting for correlations in the faults. (b) Query infidelity $\varepsilon_\text{QRAM} \equiv 1-F_\text{QRAM}$ rescaled by the error rate $\varepsilon$, for various values of such rate (again for two-qubit gate errors, with correlated faults); the inset log-loglog plot certifies the logarithmic behaviour in the limit of small $\varepsilon$. (c) Query fidelity for error models with bias $\beta=p_z/(p_x+p_x)$ towards phase errors, assuming a total error rate $\varepsilon = 10^{-4}$.     }
\label{fig_bias_correlated}
\end{figure*}

\indent To see this, let us first consider the propagation through the circuit primitive. Here, it is easy to see that
\begin{equation}\label{propagation3}
    Z_3\longrightarrow Z_3CZ_{1,2}. 
\end{equation}
Importantly, we now recall that at the end of the circuit, ideally qubit $3$ ends in the $\ket{0}$ state; this implies that the component $Z_3$ in the error above acts trivially, and simply leaves the output state with the correlated error $CZ_{1,2}$.\\
\indent To analyse the effect of the same error in the Clifford, measurement-based paradigm, it is convenient to continue on the same track as above. More precisely, we note the following important property: the fact that \emph{after} the circuit gadget qubit $3$ is left in the $\ket{0}$ state - that is, the $+1$ eigenvalue of $Z_3$ -, implies that (ideally) the input, before applying the gadget, must have been a $+1$ eigenstate of the operator $Z_3 CZ_{1,2}$ (this does not determine the three-qubit state as the eigenspace is degenerate). This implies that the error $Z_3$, on this subspace, is \emph{equivalent} to $CZ_{1,2}$ - that is, it acts the same on the input,
\begin{equation}
    Z_3\ket{\psi_\text{in}} = CZ_{1,2}\ket{\psi_\text{in}},
\end{equation}
where $\ket{\psi_\text{in}}$ is the input state. Thus, to understand the propagation of $Z_3$ through the measurement-based primitive, we can instead just analyse the propagation of $CZ_{1,2}$. This is now easy to analyse, as it commutes with all the operations. Thus, we can conclude that also in the measurement-based primitive the error $Z_3$, when propagated through, results in the error $CZ_{1,2}$ on the output - which is exactly the same as in the circuit primitive. 

\subsection{Numerical simulation of QRAM: results}
\begin{figure*}[t!]
    \begin{minipage}{1.0\textwidth}
        \includegraphics[width=\textwidth]{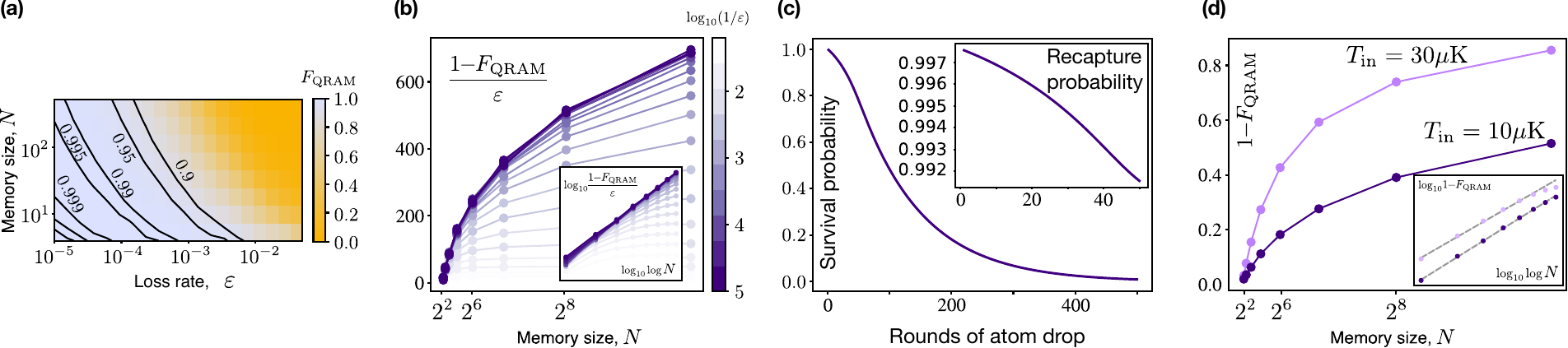}
        
    \end{minipage}%
\caption{ (a) Query fidelity $F_\text{QRAM} $ in our QRAM scheme in presence of atom losses, as a function of loss rate $\varepsilon$ and memory size $N$. (b) Query infidelity $\varepsilon_\text{QRAM} \equiv 1-F_\text{QRAM}$ rescaled by the loss rate rate $\varepsilon$, for various values of such rate; the inset log-loglog plot certifies the logarithmic behaviour in the limit of small $\varepsilon$. (c) Survival probability of an atom after $t$ rounds of two-qubit gates (atom drops). The inset shows the recapture probability when the tweezer is turned on, as a function of the number of previous recaptures; note that this is nonconstant due to heating of the atom during the drop time. Results in this plot are obtained assuming an initial cooling temperature $T_\text{in}=30\mu\mathrm{K}$. (d) Query fidelity, affected by losses due to heating, assuming standard parameters for processors based on Rb atoms. The two curves correspond to initial temperatures of $30\mu\mathrm{K}$ (brighter line) or $10\mu\mathrm{K}$ (darker line), respectively.   }
\label{fig_losses}
\end{figure*}

\noindent Below, we present our numerical results concerning the performance of our QRAM architecture in the error models described in Section~\ref{SM_section_error_models}.  For the error models analysed, we considered error-rates ranging from $\varepsilon \simeq 10^{-5}$ to $\varepsilon \simeq 0.1$. Our offset of convergence for the Monte Carlo simulation is set to $\sigma = 10^{-5}$, where $\sigma^2$ is the sample variance over all the Monte Carlo sampled trajectories, achieving this convergence precision across all the analysed parameters. These calculations have been performed using the LEO HPC infrastructure of the University of Innsbruck.\\
\\
\noindent \textbf{Continuous depolarisation.} Numerical results are shown in Fig.~\ref{fig_CD}, which complements Fig.4(d,e) of the main text. From the fit, we deduce a leading exponent $\alpha \simeq 2.2$. We note that this is significantly lower than the predicted bound $\alpha \leq 4$. \\
\\
\noindent \textbf{Operational Pauli errors.} Numerical results are shown in Fig.~\ref{fig_bias_correlated}(a,b). As expected, our simulations confirm the persistence of the built-in noise resilience of our architecture also in this case. Moreover, we note that also with correlation, we find higher fidelities with respect to the continuous depolarisation model, which therefore remains a sound worst-case benchmark. However, scaling-wise we find a dominant exponent $\alpha\simeq 2.2$ compatible with continuous depolarisation. \\
\\
\noindent \textbf{Leakage and loss errors. } Numerical results are displayed in Figs.~\ref{fig_losses}(a,b). There, we confirm again that the built-in noise-resilience of our protocol still persists when accounting for qubit losses, resulting again in a favorable polylogarithmic scaling of the fidelity. In this case, we find a slightly larger dominant exponent, $\alpha \simeq 2.5$.  \\
\\

\noindent \textbf{Heating.} For concreteness, we consider Rb atoms initially cooled at $T\simeq 10 - 30 \mu \mathrm{K}$ and 3D tweezer traps of depth $U_0/k_B\simeq 1\mathrm{mK}$, with trapping frequencies $\omega_x=\omega_y\simeq 150\mathrm{kHz}$ in the polar directions, and $\omega_z = 30\mathrm{kHz}$ in the axial direction. To model the ballistic expansion during two-qubit gates, we assume a drop time $\tau_\text{drop}\simeq 0.3\mu\mathrm{s}$. Fig.~\ref{fig_losses}(c) shows the survival probability of an atom as a function of the number of executed two-qubit gates; the nonlinear effect of the heating can be observed in the inset of the figure, where we display the recapture probability at step $t$, conditioned on the atom having survived all the previous $t-1$ steps. Assuming two-qubit gates in the $99\%$ fidelity regime, losses already become a dominating source of error after few rounds of two-qubit gates, with survival probability dropping below $95\%$ after around $16$ rounds.\\
\indent The consequences of this effect on QRAM queries are reported in Fig.~\ref{fig_losses}(d). As expected, heating results in a faster fall off of the fidelity; crucially, as shown in the inset, it still decays polylogarithmically~\footnote{Note that the deviation from the linear fit in the log-loglog plot for $N\gtrsim 256$ is simply due to the system approaching the low-fidelity regime, rather than the noise-resilience breaking down.}, highlighting that also this effect can be coped efficiently with QEC. We note that with by lowering the initial cooling to an initial temperature $T_\text{in}=10\mu\mathrm{K}$, which is within current experimental capabilities, one achieves remarkably improved fidelities, still above the $60\%$ range up to $N=256$; for $N=512$, we find $F_\text{QRAM}\simeq 49\%$. Note that in the figure inset, the line for $T_\text{in}=30\mu\mathrm{K}$ starts deviating from the linear fit for large $N$, due to the system being already in the low-fidelity regime. 
\\
\\
\noindent \textbf{Phase-biased errors.} Numerical results are reported in Fig.~\ref{fig_bias_correlated}(c). We find a rather smooth improvement of the fidelity as we approach large values of $\beta$, while for $\beta = 0.25$, corresponding to dominant $X$ and $Y$ errors, the fidelity is slightly lower than in the isotropic case ($\beta = 0.5$).\\
\\

\newpage

\section{QRAM factory}

\noindent The QRAM factory has the task of continuously providing fresh and high-fidelity copies of the RS $\ket{\Phi}$ to the processor, in order to support repeated queries. For this, we design a factory that incorporates four operational levels:  
\begin{itemize}
    \item \textbf{Factory operational level 1: Toffoli factories.} \\ A series of parallel factories continuously assembles Toffoli states, which are then injected in factory layer 3. This operates in continous-mode, in parallel with all the other layers. The space allocated to this factory layer is $S\leq\mathcal{O}(d^2 N)$, where $d$ is the code distance, and the cycle runtime is $\mathcal{O}(\text{polylog}\log N)$. 
    \item \textbf{Factory operational level 2: long-range entanglement transport.} \\
     An efficient rearrangement subroutine continously distributes Bell pairs in a specific configuration, so to build the entanglement connectivity necessary for layer 3. This also operates in continouus mode, in parallel with the other layers. One cycle of this layer comprises $\mathcal{O}(1)$ steps of long-range atom movement in total, and has overall rate $\mathcal{O}(TN^{1/4})$, where $T$, as defined in tge main text, is the time needed for moving one atom across a minimal trap distance. One cycle manipulates $\mathcal{O}(N)$ qubits. 
    \item \textbf{Factory operational level 3: Resource-state assembly.} \\
    A measurement-based circuit of depth $\mathcal{O}(\log N)$ assembles one copy of $\ket{\Phi}$, consuming the Toffoli states produced in layer 1 to inject Toffoli gates, and the Bell pairs produced in layer 2 for implementing effective long-range gates. This has a cycle rate $\mathcal{O}(\tau\log N)$ and, while operating, it consumes $\mathcal{O}(N)$ Toffoli and Bell states. This operates in parallel with layers 1 and 2, and is combined with layer 4 as detailed below. 
    \item \textbf{Factory operational level 4: try-until-succeed post-selection.} \\
    Layer 3 is combined with an error-detection and post-selection procedure in the first steps of the assembly of $\ket{\Phi}$, which are set by an integer $m\ll \log N$, called the \emph{factory parameter}. Whenever an error is detected at any of these first steps, the employed qubits are discarded, and layer 3 starts again. 
\end{itemize}
Below, we present and discuss these layers one by one. 

\subsection{Factory layer 1: T-state and Toffoli factories}

\noindent Here, we analyse the cost related to the injection of non-Clifford gates in the QRAM factory, which is necessary to provide the required amount of non-stabiliserness to the RS $\ket{\Phi}$. Below, we first count the number of non-Clifford resources. Then, we discuss that, even though $\mathcal{O}(N)$ $T-$state factories are employed to fuel the QRAM factory with properly distilled resources, nevertheless this does \emph{not} result in a time-overhead with respect to the online Clifford query. This crucial fact supports our fundamental claim that the factory rate is comparable to the query runtime, such that continuous operation of the QRAM factory is compatible with the logical clock of the QPU. 

\subsubsection{Toffoli and T-gate cost}
\noindent In a fault-tolerant implementation, the cost is evaluated in terms of basic non-Clifford gates. We note that the only non-Clifford subroutine in our QRAM design is the preparation of the part of the RS related to $|\Phi^{(1)}\rangle$, which can be implemented e.g. via $U_\text{NOHE}$ in Eq.~\eqref{U_OHE_decomposition}. Here, one can immediately recognize that the native non-Clifford gate is the Toffoli gate; the Toffoli count for implementing $U_\text{NOHE}^{(N)}$ is given by
\begin{equation}
    \sum_{K=0}^{\log N -2}\sum_{J=K+1}^{\log N -1} 2^K = N - \log N -1,
\end{equation}
where we used the fact that $\sum_{K=0}^{n}K2^K=(n-1)2^{n+1}+2$. For querying a memory of size $N$, we need to implement $U_\text{NOHE}^{(2N)}$, implying a total Toffoli count 
\begin{equation}\label{Toffoli count}
    n_\text{Toffoli} = 2N - \log N -2.
\end{equation}
In principle, each Toffoli gate can be implemented fault-tolerantly with an ad-hoc distillation procedure~\cite{PhysRevA.63.052314, PhysRevA.87.032321, Haah_2018}. \\
\indent We note that standard paradigms often rely on the injection of $T-$gates, and thus employ distillation of $T-$states~\cite{PhysRevA.71.022316, PhysRevA.86.052329}. In this respect, the Toffoli gate can be decomposed in the Clifford + $T$ setting in several ways, depending on the figure of merit that is more convenient to optimize. Importantly, in our setting any non-Clifford gate is implemented offline, in the QRAM factory. This suggests that it might be convenient to sacrifice circuit-depth, to minimize the $T-$cost. In this spirit, we envision decompositions as e.g. in Ref.~\cite{PhysRevA.87.022328}, where a Toffoli gate is decomposed in $4$ gates of the $T-$type, plus Clifford operations. 

\subsubsection{T-state factory parameters}
\noindent Here, we dicuss the parameters for the $T-$state factories. For this, let $1-\varepsilon_0$ be the initial fidelity at which noisy logical $T-$states are prepared; this will arguably be of order $\mathcal{O}(\varepsilon_\text{ph})$, where $\varepsilon_\text{ph}$ is the physical two-qubit gate fidelity. Moreover, let $\varepsilon_T$ be the distilled fidelity of the $T-$states which are provided as outputs by the factory. Recall that, from the error-scaling of the query fidelity, if we want to achieve a target $F_\text{QRAM} = 1 -\varepsilon_\text{QRAM}^*$, then the distilled fidelity needs to be of order $\varepsilon_T = \mathcal{O}(\varepsilon_\text{QRAM}^*/\log^\alpha N)$. \\
\\
\noindent For simplicity, we consider the paradigmatic Bravyi-Kitaev $15\rightarrow 1$ factory scheme~\cite{PhysRevA.71.022316}. Then, recall that distillation is only successful if $\varepsilon_0 <\varepsilon_\text{th} \simeq 0.31$. Then, to distill one $T-$state to precision $\varepsilon_T$ one needs to input to the factory $\lambda = \text{poly} (\log (\varepsilon_\text{th}/\varepsilon_T)/\log (\varepsilon_\text{th}/\varepsilon_0))$, implying 
\begin{equation}
    \lambda_\text{QRAM} = \mathcal{O}(\text{polylog}(\log N))
\end{equation}
noisy $T-$states are required by the factory to distill one high-fidelity $T-$state usable for QRAM. These states need to be passed through $k \sim \log (\log(\varepsilon_\text{th}/\varepsilon_T)/\log (\varepsilon_\text{th}/\varepsilon_0))$ rounds of distillation; thus, the (logical) time required to prepare one high-fidelity $T-$state is
\begin{equation}
    k_\text{QRAM} = \mathcal{O}(\log \log \log N).
\end{equation}

\subsection{Factory layer 2: dynamical rearrangement subroutine}
\noindent We now detail the rearrangement subroutine which allows us to efficiently prepare the BPD needed for assembling $\ket\Phi$. Below, we describe two strategies. The first is the most intuitive, allowing to understand the main concept. The second is an optimised one, which is quadratically faster than the former. 

\subsubsection{Basic rearrangement strategy}
\noindent Our first rearrangement protocol is divided in exactly three rearrangement steps, independently on $N$. Remarkably, each step fully respects state-of-the-art constraints on parallel coherent transport with acousto-optical deflectors (AODs); namely, that columns and rows are transported rigidly. The following protocol refers to Fig.~8 in the Methods.  \\
\indent As an underlying structure, we consider $8\left(2N-\log N - 2\right)$ static traps generated by a spatial-light modulator (SLM), spatially arranged as shown in Fig.~8(b). Specifically, we employ a hierarchy of two groupings: First, we divide the traps in $\log N$ main \emph{sectors}, labeled by $K=0,1,\dots,\log N -1$; each sector $K$ is further divided in $K+1$ \emph{columns}, labeled by $C=0,1,\dots,K$; moreover, column $C$ in sector $K$ is divided in $2^C$ \emph{stations}, labeled by $S=0,1,\dots, 2^C-1$. For each station, we consider $8$ traps, which are arranged in a $3\times 3$ checkerboard (with a defect in the right-bottom spot); we can thus label the traps within a station by two integers $(x,y)$, so that the right-bottom defect spot is $(1,3)$. Together, the five labels $(K,C,S,x,y)$ identify unambiguously one static trap. If we impose integer cartesian coordinates $(X,Y)$ on the 2D plane, these are related with the adopted labeling as $X=x+3C+3K(K+1)/2$ and $Y=y+3S$.  \\
\indent To start, we position $5\left(2N-\log N - 2\right)$ qubits in the traps, such that $5$ traps are occupied per station; precisely, we leave empty the leftmost column of each station, i.e. the traps $(3,y)$. This can be visualised in the first panel of Fig.~8(c). In each station, the qubits are initialised in the following configuration: 
the one in trap $(2,3)$ is simply in the $\ket{0}$ state, while the others are divided in two entangled pairs of the form $\sum_{i,j}(-)^{ij}\ket{i,j}$, which is a two-qubit cluster state; for common language, we will refer to the involved qubits as 'Bell pairs', as they are locally equivalent. The pairs are such that $(1,1)$ and $(1,2)$ are entangled with $(2,1)$ and $(2,2)$ respectively; thus, the 'entanglement bonds' are horizontal as shown in Fig.~8(c). These pairs are prepared simultaneously in $\mathcal{O}(1)$ time by first initializing the involved qubits in the $\ket{+}$ state, and then moving them to the entangling zone to apply a $CZ$ gate. After this, we proceed with three subsequent layers of rearrangements. We recall that the goal of this rearrangement is to spread the Bell pairs among the stations to realize the needed BPD, namely the one specified by the NBG in Fig.~8(a).  \\
\indent The first layer of rearrangements couples stations within the same sector as shown in Fig.~8(d); this is also displayed in the first transition of Fig.~8(c). Specifically, for each station (except the rightmost ones of each sector) we pick with the acousto-optical deflector (AOD) beams the qubit in trap $(1,2)$, and we move it to a trap of a different station of the neighboring column. Precisely, depending on the starting basis, the arrival trap is either $(1,2)$ or $(1,3)$; we can write this rearrangement as
\begin{equation}
    \left(K, C, S, 1, 2\right) \longrightarrow \left(K, C-1, \left\lfloor\frac{S+1}{2}\right\rfloor, 3, 2+\left\lfloor S/2\right\rfloor\right)
\end{equation}
for $C>0$ and $K>0$. Crucially, note that across the whole construction (i.e., even considering all the sectors together), moved qubits starting with the same vertical coordinate end up with the same vertical coordinate, as well as qubits starting with the same horizontal coordinate end up with the same horizontal coordinate; thus, we can perform this rearrangement with a single AOD step.\\
\indent The second layer or rearrangements couples neighboring sectors as shown in Fig.~8(e); this is also highlighted in Fig.~8(c). Specifically, for all the stations in sector $K$ (except those in the leftmost column), we pick the qubit in trap $(1,1)$, and we move it to trap $(3,1)$ of a corresponding station in the neighboring sector $\mathcal{S}_{K-1}$. Thus, the rearrangement reads
\begin{equation}
    \left(K,C,S,1,1\right) \longrightarrow \left( K-1 , C, S, 3, 1\right)
\end{equation}
for $C<K$ and $K>0$. Again, a crucial feature is that moved qubits starting with the same vertical (horizontal) coordinate end up with the same vertical (horizontal) coordinate; thus, again we can execute this with a single AOD step. \\
\indent The last layer of rearrangements couples the leftmost columns of each sector. Specifically, for each station in the leftmost column of sector $K$ we pick the qubit in trap $(1,3)$, and we move it to a station in the leftmost column of sector $K-1$, as shown in Fig.~8(f) - and as also highlighted in Fig.~8(c); depending on the starting station, the arrival trap is either $(3,2)$ or $(3,3)$. Specifically, we perform the rearrangement
\begin{equation}
   \left(K,C,S,1,1\right) \!\longrightarrow\! \left( K-1 , C-1, \left\lfloor\!\frac{S+1}{2}\!\right\rfloor, 3, 2+\left\lfloor S/2\right\rfloor \right)
\end{equation}
for $K>0$. Also in this third case, again motions are uniform in the vertical (horizontal) coordinates, such that this final layer of rearrangements can be executed in a single AOD step as well. \\
\indent Each of the rearrangements above must be executed in several rounds, but it only features one main step with long-range motion. Moreover (with the notation used in the Methods) it can be easily seen that the largest motion can never exceed a distance $3Nl/2$, which allows us to estimate $T_r\leq 3 \sqrt{3N/2} T $.

\subsubsection{Optimised rearrangement strategy}
\noindent The scheme above can be optimised by compacting the sectors as follows. Note that here we are storing all the columns in parallel, such that the vertical dimension of sector $K$ is of order $\mathcal{O}(2^K)$. From this packing originates the fact that the largest motion spans a distance $\mathcal{O}(N)$. However, the area occupied by sector $K$ is of order $\mathcal{O}(2^K)$, and the columns can therefore be packed in a rectangular area of dimensions $2^{K/2}\times 2^{K/2}$; that is, the dimension of sector $K$ is now of order $2^{K/2}$. Importantly, this implies that the largest motion will span a distance of order at most $\mathcal{O}(\sqrt{N})$. However, in comparison with the scheme above, now each step may contain \emph{two} motions across this maximal distance, resulting in a total of $6$ long-range motions in place of the $3$ above. However, with the new packing the maximal traveled distance at each step becomes $3\sqrt{N}l/2$; it follows that now $T_r\leq 3\sqrt{6}TN^{1/4}$, as stated in the main text.  

\subsection{Factory layer 3: resource-state preparation scheme}
\noindent Any quantum circuit can be implemented via local operations on a set of suitably distributed Bell pairs (Bell-pair distribution, BPD), where the spatial distribution of the pairs depends on the connectivity required by the circuit; this is achieved by simply employing standard teleportation to shuttle the qubit information, and interfacing qubits locally. We now first focus on this paradigm for preparing $\ket\Phi$, and then elaborate on the preparation of the BPD in the paragraph below. This can be understood in two parts: first, we prepare $|\Phi^{(1)}\rangle$ and $|\Phi^{(2)}\rangle$; second, we implement $B$ via partial BMs to assemble $\ket\Phi$. Importantly, the BPD for $|\Phi^{(1)}\rangle$ and $|\Phi^{(2)}\rangle$ is the same, allowing for efficient preparation.   \\
\indent The needed BPD can be understood in terms of the `nested bifurcation graph' (NBG) displayed in Fig.~8(a) of Methods. Therein, as shown in the insets, triangles represent `stations' where qubits are stored; links represent the presence of a Bell pair shared between the triangles. For preparing $|\Phi^{(1)}\rangle$ ad $|\Phi^{(2)}\rangle$, we simply iterate the application of the three-qubit gadgets displayed in Fig.~8(d) and Fig.~8(b) respectively; correspondingly, each station contains six qubits (three to support the teleported-in input, and three to teleport-out the output). Once the BPD is is prepared according to the NBG, preparing $\ket\Phi$ is straightforward; remarkably, while $|\Phi^{(1)}\rangle$ needs $\mathcal{O}(\log N)$ steps to be prepared, instead $|\Phi^{(2)}\rangle$, thanks to its stabiliserness, is prepared with a \emph{single} layer of parallel operations. The only nontrivial part that remains to be discussed is the preparation of the BPD, which we outline below.

\subsection{Factory layer 4: try-until-succeed offline scheme}
\noindent Here, we analyse the QRAM factory protocol based on heralding.

\begin{figure*}[t!]
    \begin{minipage}{1.0\textwidth}
        \includegraphics[width=\textwidth]{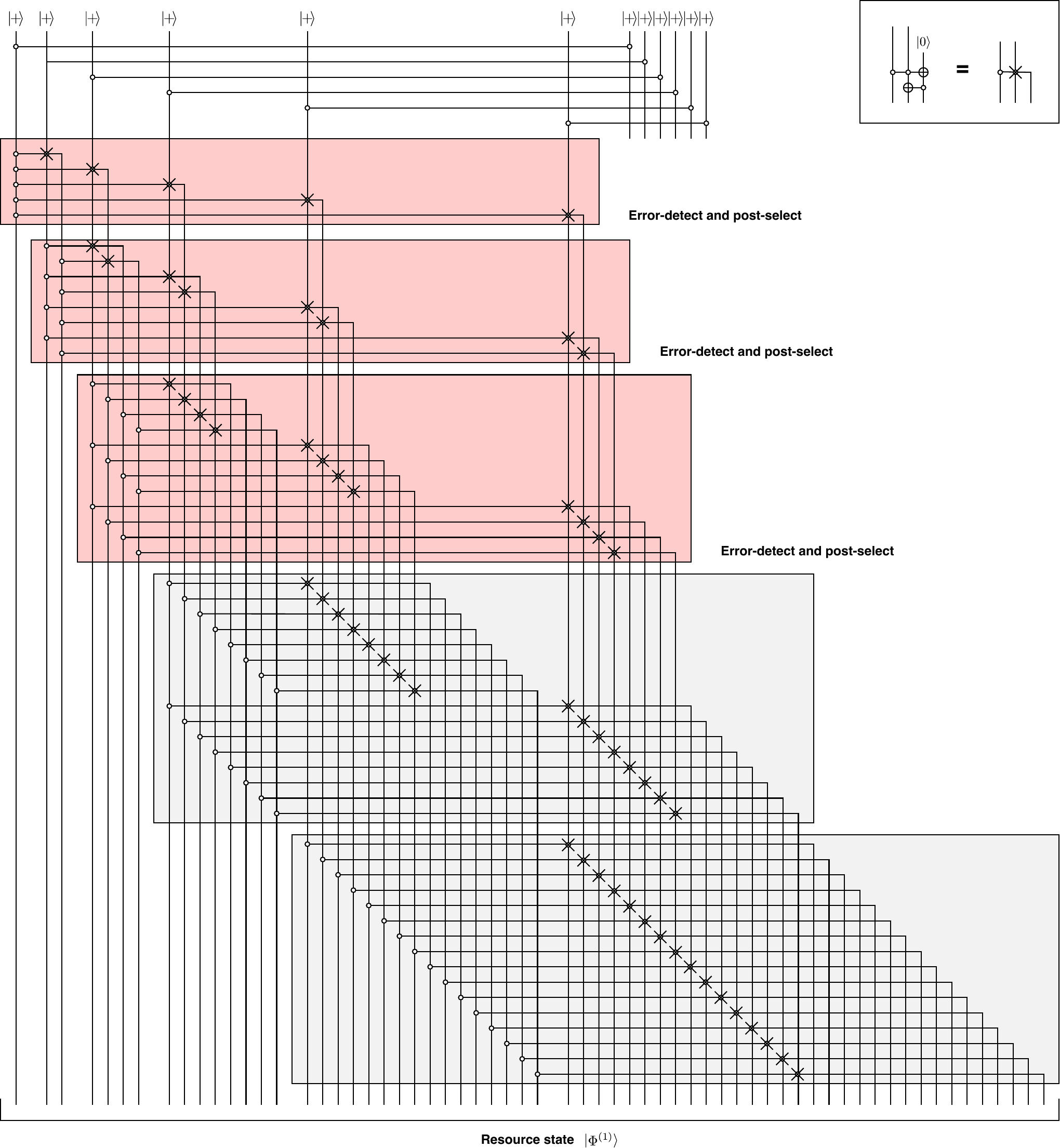}
    \end{minipage}%
\caption{\textbf{Non parallelised QRAM factory circuit}. We implement error-detection and herald on all the gates where the first $m$ address bits act as controls. In the non-parallelised version, this is achieved straightforwardly by simply heralding the first $m$ blocks (shown in red). The figure displays this on the example of $N=$ and $m=3$. }
\label{non_parallel_factory}
\end{figure*}

\subsubsection{Motivation and framework}

\noindent One major advantage of offline state preparations is that one can, in principle, `retry until succeed'. More precisely, if one has a reliable method to \emph{detect} if an error has occurred, then one can decide to just discard all the qubits so-far involved in the failed attempt, and restart again. Crucially, since the preparation is offline, this does not require to restart the full \emph{algorithm}, but just the preparation subroutine. If the failure probability of the subroutine is small enough, then one only needs to try few times: as a result, one can obtain an improved state fidelity, while the total cost (i.e., the resources wasted in the failed attempts) remains small. This concept is at the core of well-known strategies such as `magic statre factories' in fault-tolerant quantum computation, or heralded gates on photoni platforms. \\
\indent There are two main paradigms where such error detection is available, including the physical and the logical, fault-tolerant setting:
\begin{enumerate}
    \item \textbf{Physical level: leakage.} In many physical platforms, the main source of errors is \emph{leakage} out of the computational space: that is, the system which is supposed to represent a qubit leaks to a state which is not in the space spanned by $\ket{0}$ and $\ket{1}$. In this case, the error can in principle be detective, as e.g. projective measurements can distinguish whether the qubit is inside or outside the computational space. Let  $\varepsilon$ be the error probability, and $\varepsilon'$ the joint probability that an error has occurred, but is not revealed through this leakage detection procedure. Then, if $\varepsilon '\ll\varepsilon$, to all the effects discarding an attempt upon detecting an error results in the remaining error probability effectively being suppressed as $\varepsilon\rightarrow \varepsilon '\ll\varepsilon$.
    \item \textbf{Logical level: error-detection.} In a QEC setting, one can leverage the syndrome information to detect the presence of errors. If $d$ is the distance of the code, typically one can detect up to $d-1$ errors, but only correct $\sim d/2$, resulting in a logical error $\varepsilon\sim\varepsilon_\text{ph}^{d/2}$. Upon post-selecting on no detected error instead of attempting to correct, the residual logical error probability becomes $\varepsilon'\sim\varepsilon_\text{ph}^d$, as $d$ errors must occur for the syndrome not to identify none. It follows that the logical error is quadratically enhanced, $\varepsilon'\sim\varepsilon^2\ll\varepsilon$.
\end{enumerate}
In our QRAM architecture, all the complexity is outsourced to the offline preparation of the RS $\ket{\Phi}$. It is therefore natural to ask whether one can take advantage of such preparation being offline, to improve the performance through heralding procedures. \\
\indent However, at first sight, this seems obstructed by the fact that $\ket{\Phi}$ is a large entangled many-body state, with support on $\mathcal{O}(N)$ qubits, whose preparation requires the execution of $\mathcal{O}(N)$ gates. The probability of detecting \emph{no error} throughout the whole preparation of $\ket{\Phi}$ therefore drops down exponentially in $N$. This implies that, to all practical effects, one would never be able to stop the `retry loop', resulting in an exponentially vanishing preparation rate. This naive analysis therefore suggests that, as is typical for many-body entangled states, try-until-succeed protocols should not be possible in our QRAM factory. \\
\indent In contrast with the intuitive analysis above, it turns out that an efficient try-until-succeed protocol \emph{can} be designed for the QRAM resource-state. This apparently surprising result follows directly from the very special structure of the NOHE circuit (or, equivalently, of the bnucket-brigade protocol). The key intuition is that these circuits proceed by incorporating layers comprising an exponentially growing number of gates and ancillary qubits. Thus, if one only heralds the \emph{first} layers, the success rate (probability of not detecting errors) can be kept high, as the heralding is only performed on a very small number of gates. However, from the well-understood noise-resilience mechanism of such protocols~\cite{hann2021resilience}, these layers, even though their weight within the circuit is very small, they actually contribute \emph{extensively} to the overall fidelity. As a consequence, one can boost the fidelity extensively, while keeping a very high success rate. Below, we first present and explain our try-until-succeed protocol, and then we analyse its performance. 

\begin{figure*}[t!]
    \begin{minipage}{1.0\textwidth}
        \includegraphics[width=\textwidth]{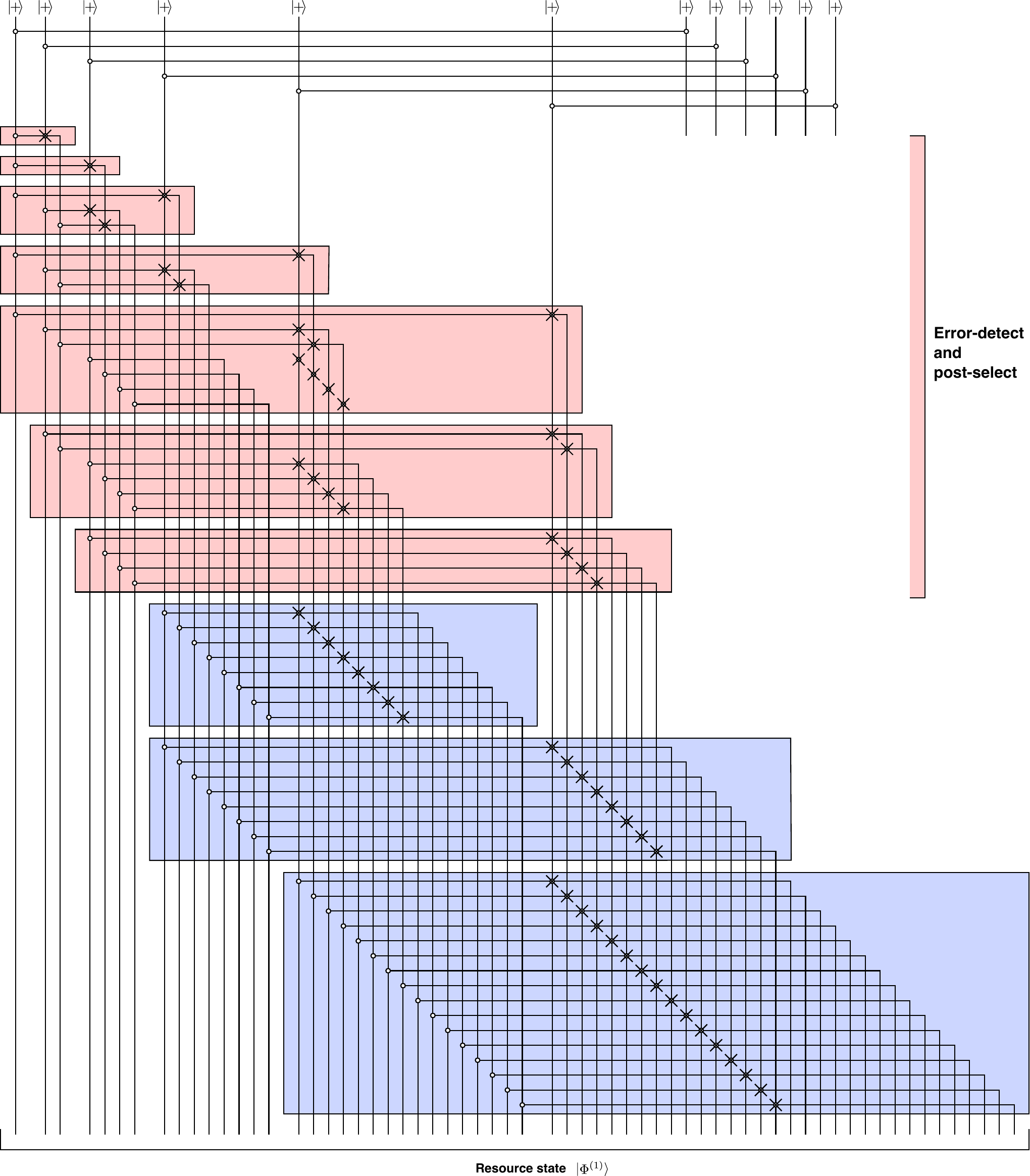}
    \end{minipage}%
\caption{\textbf{Parallelised QRAM factory circuit}. Parallelised version of the heralded QRAM factory of Fig.~\ref{non_parallel_factory}. The number of heralded gates is therefore the same, $\mathcal{O}(M\log N)$. }
\label{parallel_factory}
\end{figure*}

\subsubsection{Protocol}
\noindent The factory parameter $m$ is chosen such that $m\ll\log N$. Moreover, we also choose it such that $M\equiv 2^m \ll 1/\varepsilon$, where $\varepsilon$ is the operational error in the individual steps of the assembly of the RS $\ket{\Phi}$ (thus, the \emph{logical} error in a QEC setting). Then, the idea is as follows: we aim at heralding on the first $m$ bits of the address being queried correctly. As we show below, this will only require to herald $\mathcal{O}(M\log N)$ gates  \\
\\
\noindent \textbf{Non parallelised version.} Consider the circuit decomposition in Fig.~\ref{non_parallel_NOHE}, and recall that it is divided in sectors $K=0,1,\dots$, such that sector $K$ is devoted to the OHE of the $K-$th bit of the address. Also, recall that in the non pralallelised version, each sector $K$ is identified with a circuit block $K$ (in grey in the figure), which contains all those gates where qubits of sector $K$ act as controls. Then, to herald on the first $m$ address bits being queried correctly, we implement error-detection in the first $m$ circuit blocks, heralding on all the gates succeeding. This is shown in Fig.~\ref{non_parallel_factory}.  \\
\indent It is straightforward to calculate that the number of heralded Toffoli gates in this protocol is 
\begin{equation}
    \sum_{K=0}^{m-1}(\log N - K)2^K = \log N (M-1) - (m-2)M + 1 = \mathcal{O}(M\log N);
\end{equation}
as a consequence, the success rate can be estimated as $1-\varepsilon M \log N$, such that by choosing $M\ll 1/\varepsilon$ one can ensure a high factory rate. Moreover, choosing $m\ll \log N$, one can ensure that the fraction of gates that are `wasted' when an attempt is discarded, $\mathcal{O}(M/N)$, remains always small. \\
\\
\noindent \textbf{Parallelised version} For the parallelised version, one needs to be careful: in order to avoid a large resource waste, one shall \emph{first} implement all the heralded gates, and \emph{then} the remaining non heralded ones. While this is straightforward in the non parallelised example above, in the parallelised version it requires to give up a small degree of parallelism. More precisely, Fig.~\ref{parallel_factory} shows explicitly how the circuit in Fig.~\ref{parallel_NOHE} is deformed to implement the heralded factory. The gate counts remain unchanged with respect to the non-parallelised analysis above, and so do the rate estimates. 

\subsubsection{Performance}
\noindent To evaluate how our try-until-succeed factory enhances the query fidelity, we now re-evaluate the estimates in Section~\ref{estimates_noise_resilience}. For this, we note that Eq.~\eqref{heuristics1} needs to be reconsidered, as now the first $m$ address bits are expected gto be error-free (to first order in the error-detection efficiency). Thus, we now get
\begin{align}\label{heuristics1}
    \mathbb{E}\Big[|\mathcal{G}_{k|1}|\Big] & \simeq \frac{1}{2N-M}\sum_{K=m}^{\log N} 2^K \Bigg[N - (K+1)\frac{N}{2^K}\Bigg] \nonumber \\
    & = N - \frac{N}{2N-1} \frac{(1+\log N)(2+\log N)}{2} + \frac{N}{2N-M}\frac{m(1+m)}{2},
\end{align}
where we also considered that the total number of potentially faulty locations is $2N-M$ now. Repeating the calculations of Section~\ref{estimates_noise_resilience}, one can therefore estimate the improvement of the fidelity with respect to Eq.~\eqref{heuristics2} as 
\begin{equation}
    \Delta F_\text{QRAM} \sim \frac{2\nu\eta}{2N-M} m(m+1) - \frac{2\nu\eta(M-1)}{(2N-1)(2N-M)}(1+\log N)(2+\log N). 
\end{equation}
To provide a qualitative estimate, recall that, for small errors, one has $\eta = \mathcal{O}(\varepsilon\log N)$. Considering the continuous depolarisation model, where $\nu=\mathcal{O}(N\log N)$, one therefore estimates
\begin{equation}
  \Delta  F_\text{QRAM} = \mathcal{O}(\varepsilon m^2 \log^2 N).
\end{equation}
Considering the numerically calculated infidelity $1-F_\text{QRAM} \sim \log^\alpha N$, with $\alpha \simeq 2.2$, this therefore represents an extensive improvement. \\
\indent As a final note, however, we recall that our numerical simulations prove unambiguously query fidelities higher than the theoretical bounds given. Similarly, also our numerical results on the heralded QRAM factory (reported in the main text) do not closely follow the bounds given by the theory above. This suggests that our estimates are rather pessimistic, and opens the question on whether a more accurate theoretical understanding of the noise resilience of our NOHE circuit could be provided.

\subsection{Quantum error-correction overhead}

\noindent One crucial result is that the query runtime $T_\text{query}$ is comparable to the factory time $T_\Phi$ needed to assemble one copy of $\ket{\Phi}$; this is specified in Eq.(7) of the main text in a physical setting (i.e., without QEC). Naively, since the query in our protocol is entirely transversal, while the factory requires (i) a large amount of non-Clifford gates and (ii) long-range atom rearrangements, one may expect that this result breaks down when fault-tolerant operation is considered. In contrast, below we show that, in realistic scenarios, the same result also holds in the QEC setting. \\
\indent More precisely, in the following, we will denote by $T'_\text{query}$ and $T'_\Phi$ the error-corrected runtimes of the query and the QRAM factory, respectively. Then, we now argue that
\begin{equation}
    T'_\text{query}\sim T'_\Phi
\end{equation}
also holds. Perhaps surprisingly, we actually find that, for large system sizes, the rescaled difference $(T'_\Phi-T'_\text{query})/T'_\text{query}$ actually grows \emph{slower} in the fault-tolerant setting, than $(T_\Phi-T_\text{query})/T_\text{query}$ in a physical implementation. This counterintuitive result is due to the fact that for large $N$ the factory rate is dominated by the rearrangement time, which only grows with an overhead factor $\mathcal{O}(\sqrt{d})$ with the code distance, while the query runtime grows with a more severe overhead factor $\mathcal{O}(d)$.\\
\\
\noindent \textbf{Fault-tolerant query runtime.} To estimate the time overhead of fault-tolerant operations, we assume that logical operations (e.g., logical gates or logical measurements) are always interleaved by $\mathcal{O}(d)$ rounds of syndrome extraction, where $d=\mathcal{O}(\log\log N)$ is the code distance. Indeed, in standard QEC schemes, this ensures that the syndrome information is sufficient to reliably correct errors. For neutral atoms, neglecting the \emph{physical} gate time, this implies that logical gates and logical measurements will have approximately the same runtime, which we approximate by $\simeq \tau d$, where $\tau$ is the duration of a physical mid-circuit measurement. Considering that Pauli operations are performed virtually, this implies that the the query has a logical runtime
\begin{equation}
    T'_\text{query} \simeq 2\tau d \log N = dT_\text{query}. 
\end{equation}
\\
\noindent \textbf{Fault-tolerant factory runtime.} For the QRAM factory, it is useful to divide the runtime in two parts: one related to the rearrangement and standard Clifford operations, and one related to the $T-$state factories which provide non-Clifford resources to the former. \\
\indent For the evaluation concerning dynamical rearrangement, we focus on the surface code. Here, since one logical qubit is represented by $\mathcal{O}(d^2)$ physical qubits disposed approximately within a square, it follows that all the linear dimensions are directly multiplied by a factor $\mathcal{O}(d)$. As explained above, our evaluation in Eq.(7) of the main text is based on the observation that no atom movement spans a distance larger than $a\sqrt{N}$, where $a$ is the minimal inter-trap distance; this should therefore be substituted with the bound $ad\sqrt{N}$, leading to the modified form
\begin{equation}
    T'_\Phi \lesssim 2\tau d \log N + 3\sqrt{6d}TN^{1/4} + \Delta,
\end{equation}
where $\Delta$ is the additional term due to the $T-$state factories that we consider below. \\
\indent We now consider the overhead related to the distillation of high-fidelity $T-$states, which we quantified by $\Delta$ in the equation above. We recall that for fueling the preparation of one resource-state $\ket\Phi$, we need to prepare $\mathcal{O}(N)$ such $T-$states; thanks to the noise-resilience of the protocol, each needs only to be distilled up to a residual infidelity $\simeq 1/\text{polylog}N$. As also discussed above, for producing each of these states we need $\lambda = \mathcal{O}(\mathrm{polylog}( \log N))$ input noisy $T-$states, which we pass through $k = \mathcal{O}(\log\log\log N)$ rounds of distillation. Since all of this should be intended at the logical level, recalling also the observations above, the time needed to distill one single high-fidelity $T-$state is $\mathcal{O}(dk\tau)$, while the space demand is $\mathcal{O}(d^2\lambda)$. Now, let $S$ be the total physical space allocated to $T-$state factories within the QRAM factory. Then, the \emph{rate} at which $T-$states are delivered can be quantified by 
\begin{equation}
    r_T \simeq \frac{S}{d^3\tau \lambda k},
\end{equation}
which can be understood as the number of high-fidelity $T-$states produced inside the QRAM factory, per unit of (physical) time. Now, let us consider the following (non optimal) strategy for the QRAM factory: we \emph{first} prepare all the necessary $T-$states, and \emph{then} we proceed with the assmebly of $\ket\Phi$. Then, we are free to use the entire space allocated to the QRAM factory, for preparing $T-$states, which implies that $S$ is only upper bounded by $S\leq\mathcal{O}(d^2N)$. Since we need $n_T = \mathcal{O}(N)$ such $T-$states, the time needed for producing them can be estimated as 
\begin{equation}
    \Delta = \frac{n_T}{r_T} \simeq d\tau\lambda k = \mathcal{O}(\tau\mathrm{
    polylog}\log N). 
\end{equation}
\\
\noindent \textbf{Conclusion.} Inserting this in the evaluations above, we find that 
\begin{equation}
    \frac{T_\Phi'-T_\text{query}'}{T_\text{query}'} \simeq \frac{1}{\sqrt{d}}\frac{T_\Phi-T_\text{query}}{T_\text{query}} + \mathcal{O}\bigg(\frac{\mathrm{polylog}(\log N)}{\log N}\bigg) = \frac{T}{\tau}\mathcal{O}\bigg(\frac{N^{1/4}}{\log N \sqrt{\log\log N}}\bigg) + \mathcal{O}\bigg(\frac{\mathrm{polylog}(\log N)}{\log N}\bigg).
\end{equation}
From this, one can see that the discrepancy related to the rearrangement is actually (weakly) suppressed rather than enhanced, for large memory sizes; this is a consequence of the fact that while the runtime of logical operations incurs a multiplicative factor $\mathcal{O}(d)$, the rearrangement time only incurs a factor $\mathcal{O}(\sqrt{d})$. On the other hand, in the QEC setting there is an additional overhead term due to the $T-$state factories, which is however also suppressed for large memories.\\
\begin{table}[t]\label{table}
	\centering
	\caption{Fault-tolerant estimates, with target query fidelities $F_{\rm QRAM}\approx 0.99$ (left) and $F_{\rm QRAM}\approx 0.999$ (right). We consider two estimation approaches: ``opt'' uses the numerically extracted scaling for $A_1(N)$; ``cons'' uses the worst-case analytical bound. Parameters: $\tau=500~\mu\mathrm{s}$, $T=33~\mu\mathrm{s}$, $\Delta\simeq 15d\tau$, and $T'_{\rm rearr}\approx \sqrt d\,(3\sqrt6\,T\,N^{1/4})$. Times are always expressed in $\mathrm{ms}$.}
	\label{tab:ft_timing_99_vs_999}
	\vspace{0.5em}

	\renewcommand{\arraystretch}{1.10}
	\setlength{\tabcolsep}{3.0pt}

	\begin{minipage}[t]{0.49\linewidth}
		\centering
		\textbf{$F_{\rm QRAM}\approx 0.99$}\par\vspace{0.25em}
		\begin{tabularx}{\linewidth}{@{} l l S[table-format=2.0] S[table-format=4.0] S[table-format=3.1] S[table-format=3.1] S[table-format=4.1] S[table-format=1.3] @{}}
			\toprule
			{$N$} & {Estimation} & {$d$} & {$T'_q$} & {$\Delta$} & {$T'_r$} & {$T'_\Phi$} & {\textbf{Ratio} $T'_\Phi/T'_q$} \\
			\midrule
			$2^{13}$ & opt  & 27 &  351 & 202.5 & 12.0  &  565.5 & 1.611 \\
			$2^{13}$ & cons & 39 &  507 & 292.5 & 14.4  &  813.9 & 1.605 \\
			\addlinespace
			$2^{24}$ & opt  & 31 &  744 & 232.5 & 86.4  & 1062.9 & 1.429 \\
			$2^{24}$ & cons & 45 & 1080 & 337.5 & 104.1 & 1521.6 & 1.409 \\
			\bottomrule
		\end{tabularx}
	\end{minipage}\hfill
	\begin{minipage}[t]{0.49\linewidth}
		\centering
		\textbf{$F_{\rm QRAM}\approx 0.999$}\par\vspace{0.25em}
		\begin{tabularx}{\linewidth}{@{} l l S[table-format=2.0] S[table-format=4.0] S[table-format=3.1] S[table-format=3.1] S[table-format=4.1] S[table-format=1.3] @{}}
			\toprule
			{$N$} & {Estimation} & {$d$} & {$T'_q$} & {$\Delta$} & {$T'_r$} & {$T'_\Phi$} & {\textbf{Ratio} $T'_\Phi/T'_q$} \\
			\midrule
			$2^{13}$ & opt  & 35 &  455 & 262.5 & 13.7  &  731.2 & 1.607 \\
			$2^{13}$ & cons & 47 &  611 & 352.5 & 15.8  &  979.3 & 1.603 \\
			\addlinespace
			$2^{24}$ & opt  & 39 &  936 & 292.5 & 97.0  & 1325.5 & 1.416 \\
			$2^{24}$ & cons & 53 & 1272 & 397.5 & 113.0 & 1782.5 & 1.401 \\
			\bottomrule
		\end{tabularx}
	\end{minipage}
\end{table}\\

\subsubsection{Quantitative estimations}
\noindent We now explicitly estimate the QEC overhead. First, as discussed above, for the long-range rearrangement time we will consider the following conservative replacement:
\begin{equation}
	T'_{\rm rearr}\approx \sqrt d\,T_{\rm rearr}
	\lesssim \sqrt d\,\bigl(3\sqrt6\,T\,N^{1/4}\bigr).
	\label{eq:FTrearr}
\end{equation}
To obtain informative estimations, we benchmark two representative memory sizes spanning the regimes: $N=2^{13}$ (kB-scale) and $N=2^{24}$ (MB-scale). While the latter would arguably be out of reach for current single-module computers, it is still informative to appreciate the persistence of $T_\Phi \sim T_\text{query}$ at all relevant scales. For target query fidelities in the $\sim 99\%$ regime (i.e.\ $1-F_{\rm QRAM}\sim 10^{-2}$), we estimate that (with a standard surface code and the physical error parameters used in our feasibility estimates) distances
\[
	d(N=2^{13})\simeq 39,\qquad d(N=2^{24})\simeq 45
\]
would be sufficient. Then, using $\tau=500~\mu{\rm s}$, we obtain
\begin{align}
	T'_{\rm query}(N=2^{13})
	&\approx 2(0.5{\rm ~ms})\cdot 39\cdot 13
	\simeq 507{\rm ~ms},\\
	T'_{\rm query}(N=2^{24})
	&\approx 2(0.5{\rm ~ms})\cdot 45\cdot 24
	\simeq 1080{\rm ~ms}.
\end{align}
To estimate the distillation overhead, we consider the standard Bravyi--Kitaev $15\!\to\!1$ factory scheme and the physical-error regime used in our feasibility estimates; we note that, in such framework, one round of distillation suffices for the $\sim 1/{\rm polylog}(N)$ target accuracy here, so $\lambda=15$, $k=1$, and thus
\begin{align}
	\Delta(N=2^{13})
	&\simeq 15\,d\,\tau
	\simeq 15\cdot 39\cdot 0.5{\rm ~ms}
	=292.5{\rm ~ms},\\
	\Delta(N=2^{24})
	&\simeq 15\,d\,\tau
	\simeq 15\cdot 45\cdot 0.5{\rm ~ms}
	=337.5{\rm ~ms}.
\end{align}
Finally, using $T=33~\mu{\rm s}$ in Eq.~\eqref{eq:FTrearr} gives
\begin{align}
	T'_{\rm rearr}(N=2^{13})
	&\approx \sqrt{39}\,(3\sqrt6)(33~\mu{\rm s})\,2^{13/4}
	\simeq 14.4{\rm ~ms},\\
	T'_{\rm rearr}(N=2^{24})
	&\approx \sqrt{45}\,(3\sqrt6)(33~\mu{\rm s})\,2^{24/4}
	\simeq 104.1{\rm ~ms}.
\end{align}
Putting the pieces together yields
\begin{align}
	T'_{\Phi}(N=2^{13})
	&\approx 507{\rm ~ms}+292.5{\rm ~ms}+14.4{\rm ~ms}
	\simeq 814{\rm ~ms},\\
	T'_{\Phi}(N=2^{24})
	&\approx 1080{\rm ~ms}+337.5{\rm ~ms}+104.1{\rm ~ms}
	\simeq 1522{\rm ~ms},
\end{align}
so that $T'_{\Phi}/T'_{\rm query}\approx 1.6$ for $N=2^{13}$ and $\approx 1.4$ for $N=2^{24}$. Thus, even at the prefactor level and for explicit FT parameters chosen to reach $\sim 99\%$ query fidelity, the offline factory time remains comparable (within an order-one factor) to the online query time, supporting the pipelining argument for sequential QRAM calls. Table~\ref{table} summarises our results, also considering a higher target fidelity of $F_\text{QRAM}\simeq 99.9\%$.

\newpage

\section{Implementation in current neutral-atom experiments}

\noindent Here, we evaluate the feasibility of our proposal within currently fully-operating neutral-atom experiments. For this, we mainly consider Refs.~\cite{Bluvstein_2023, bluvstein2025architectural}. We therefore assume a quantum processor featuring $\sim 500$ neutral-atoms in movable optical tweezers, with two-qubit gate errors $\varepsilon\simeq 5\times 10^{-3}$ and non-destructive mid-circuit measurements. To model errors, we assume the pessimistic setting of continuously depolarising errors (which is the error-model giving the lowest fidelities in our simulations). The numbers reported below are derived form our numerical simulations presented above.  

\subsection{Physical level implementations}

\noindent With our scheme, the processor described above can demonstrate \emph{physical} QRAM calls with query fidelities above $F_\text{QRAM}\gtrsim52\%$ for memory sizes up $N = 32$, by employing $\lesssim 160$ atoms overall. For smaller-scale implementations, e.g. $N=4,8,16$, we estimate query fidelities $F_\text{QRAM}\simeq 91\%, 80\%, 66\%$ respectively; for more ambitious system sizes, e.g. $N=64, 128$, we find $F_\text{QRAM}\simeq 35\%, 26\%$ respectively. Considering the latter cases, we note that still $N=64$ is largely within present capabilities, as it would require operation on $\lesssim 320$ atoms overall Similarly, $N=128$ would require full operation on $\lesssim 640$ atoms - a number that has not been demonstrated yet, but does not seem to be out of reach for current leading experiments.  

\subsection{Logical implementations}

\noindent While fully error-corrected implementations of complex quantum routines such as QRAM have not been demonstrated to date, we note that all the fundamental ingredients for envisioning such implementations are currently present. We therefore find it instructive to also estimate the potential for demonstrations with QEC. For this, we assume a standard surface code implementation, such that the logical error is suppressed as $\varepsilon_L \simeq \alpha \Lambda^{-(d+1)/2}$, with $\alpha \simeq 0.1$~\cite{beverland2022assessing, Gidney_2021} and $\Lambda=\varepsilon_\text{th}/\varepsilon$, where we pick the threshold to be $\varepsilon_\text{th}\simeq 1\%$~\cite{PhysRevA.89.022321}. Recalling that the surface code needs $\sim 2d^2$ atoms to operate one logical qubit (considering both data qubits, as well as ancillas for stabiliser measurements), assuming $\sim500$ atoms as above, QEC experiments with $d=3,5,7$ can already demonstrate e.g. one error-corrected QRAM call for memories of sizes $N\simeq 16, 8, 4$ respectively. Note that this assumes that $T-$state factories are implemented following the scheme depicted in our new version of the Supplemental Material (also see the reply to Referee 1). A physical error $\varepsilon\simeq 5\times 10^{-3}$ as above is below threshold, therefore allowing to probe exponential suppression of the error query infidelity as $d$ is increased from $d=3$ to $d=5$ and $d=7$. 

\newpage

\section{Comparison with Quantum Read-Only Memory}

\noindent In Ref.~\cite{babbush2018encoding}, Babbush and colleagues introduce a construction for implementing coherent table-lookups, which is known as `quantum read-only memory' (QROM). While the target task is conceptually the same of QRAM (just slightly adapted for specific quantum chemistry implementations), QROM is a sharply different construction, whose key goal is to reduce the system size requirement. In practice, QROM is implemented through a `unary iteration' procedure, described in detail in Ref.~\cite{babbush2018encoding}; as a result, it operates with $\mathcal{O}(\log N)$ qubits, but has a circuit depth scaling as $\mathcal{O}(N\log N)$. At the circuit level, the number of Toffoli gates is $\mathcal{O}(N)$ as for QRAM. At hte circuit level, QROM can therefore be understood as a method where space and time are traded in the opposite way as in QRAM, while maintaining approximately the same logical circuit volume. \\
\indent In this section, we analyse QROM in detail, and compare it to our QRAM proposal. We show that, despite featuring the same \emph{circuit-level} space-time volume and $T-$count, an exponential separation exists between the two approaches. This is due to the  fundamental error suppression mechanism built into our proposal, which results in an exponentially better error-scaling than Ref.~\cite{babbush2018encoding}. As a consequence, our protocol can operate with exponentially lower code distances and QEC resources. In terms of $T-$state factories, this implies that our proposed QRAM can tolerate exponentially less precise $T-$states, thereby significantly alleviating the $T$-state factory effort. For instance, exponentially fewer `noisy' $T-$states need to be provided to a factory, and exponentially less distillation rounds are needed, in order to distill one single logical $T-$state usable for a query.\\
\indent In our comparison, we are mostly interested in the following figure of merit. Let $\mathcal{V}$ be the total space-time circuit volume in a fault-tolerant implementation. Then, since for both QRAM and QROM the circuit volume is $\mathcal{O}(N\log N)$, we define the space-time volume \emph{overhead} as
\begin{equation}
    \xi = \frac{\mathcal{V}}{N\log N}. 
\end{equation}
In practice, this quantity represents the \emph{physical} space-time volume of a \emph{logical} volume unit, during a fault-tolerant implementation. 

\subsection{Impact of errors on QROM}

\begin{figure*}[t!]
    \begin{minipage}{1.0\textwidth}
        \includegraphics[width=\textwidth]{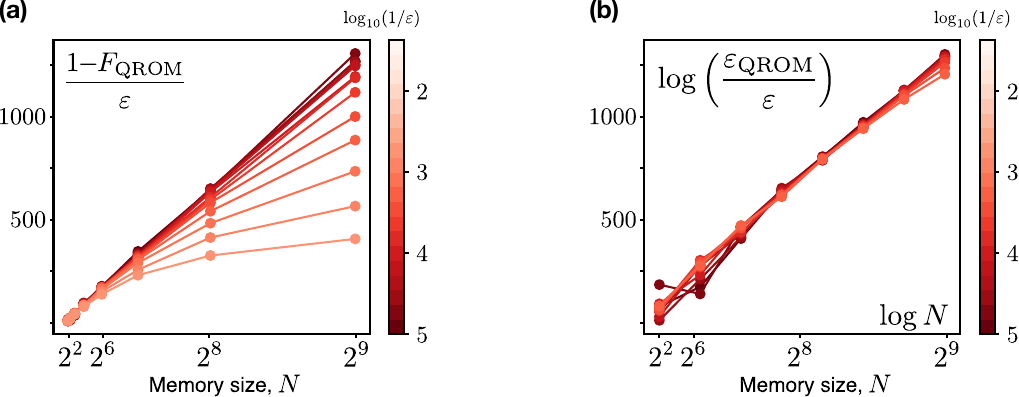}
    \end{minipage}%
\caption{\textbf{QROM fidelity scaling}. (a) The QROM infidelity $1-F_\text{QROM}$, rescaled by the two-qubit error rate $\varepsilon$. For small values of $\varepsilon$, collapse to a straight line probe an infidelity scaling of the form $F_\text{QROM}= 1 - A\varepsilon N$, where $A\equiv \lim_\varepsilon (1-F_\text{QROM}/\varepsilon)$ is a constant. (b) For comparison with our QRAM results, we report that the scaling $\propto N$ is confirmed by a log-log plot. Note that this is different from our QRAM plots, where collapse to a straight line was only achieved in log-loglog plots.   }
\label{fig_QROM_fidelity}
\end{figure*}

\noindent To start our comparison, we simulate the QROM circuit under Pauli gate errors, in fact repeating the same analysis made for QRAM in the sections above. We define the QROM query fidelity analogously as we did for QRAM in Eq.~\eqref{query_fidelity_definition}, i.e., as the average output fidelity, averaged over all possible memory instances, and over all possible input addresses. Again, we can expand it in powers of the error rate $\varepsilon$ as
\begin{equation}
    F_\text{QROM} = 1 - \sum_{k\geq 1} A_k^{\text{QROM}}(N) \varepsilon^k,
\end{equation}
thereby capturing the scaling in $N$ through the coefficients $A^\text{QROM}_k(N)$. Analogoulsy to our QRAM analysis, we simulate QROM and calculate $F_\text{QROM}$ for values of $\varepsilon$ ranging form $10^{-2}$ to $10^{-5}$. Results for QROM are reported in Fig.~\ref{fig_QROM_fidelity}. More precisely, in Fig.~\ref{fig_QROM_fidelity}(a) we display the query error $\varepsilon_\text{QROM} = 1 -F_\text{QROM}$, divided by the logical operational error $\varepsilon$. That is, the plot shows $\varepsilon_\text{QROM}/\varepsilon$ as a function of the memory size $N$; different curves correspond to decreasing values of $\varepsilon$. For small $\varepsilon$, the curves collapse to the leading coefficient in the expansion above, dfined as
\begin{equation}
    A^\text{QROM}_k = \lim_{\varepsilon \rightarrow 0^+} \frac{1-F_\text{QROM}}{\varepsilon}.
\end{equation}
Such collapse clearly hints to a polynomial behaviour with $N$ for small errors; that is, in the regime of large fidelity, one has 
\begin{equation}
    A_1^\text{QROM}(N) = \text{poly} N
\end{equation}
Our numerical results clearly suggest that the polynomial is a linear function, such that
\begin{equation}\label{QROM_fidelity}
    F_\text{QROM} = 1 - A\varepsilon  N + \mathcal{O}(\varepsilon^2)
\end{equation}
for some constant $A$. Such scaling is also further certified by the logarithmic plot in Fig.~\ref{fig_QROM_fidelity}(b). We note that these results agree with the theoretical predicitons of Ref.~\cite{hann2021resilience}, which conjectured such fidelity suppression for QROM. Moreover, we note that this is exactly what one would expect from a circuit of depth $\mathcal{O}(N)$.\\
\indent In practice, this marks the following sharp difference between our proposed QRAM implementation QROM. In a fault-tolerant setting, with our QRAM architecture, for implementing high-fidelity queries to a memory of size $N$, one must push QEC to suppress the logical error bellow a scale $\varepsilon \ll 1/\text{polylog N}$; this is discussed extensively in the QRAM factory section above. In contrast, for QROM one must reach the exponentially more precise regime $\varepsilon\ll 1/\text{poly} N$. As we will comment below, this has significant consequences in terms of QEC resources, where such exponential separation leads to dramatic differences in the fault-tolerance cost, space-time volume and non-Clifford factory overhead.

\subsection{Space-time volume overhead for fault-tolerant QROM}

\noindent We now analyse the space-time volume overhead required by QROM for implementing error-corrected high-fidelity queries. 

\subsubsection{Fault-tolerant operation overhead}
\noindent From the fidelity scaling in Eq.~\eqref{QROM_fidelity}, it follows that to implement the QROM subroutine with a given target precision $\varepsilon_\text{QROM}$, QEC must be pushed to suppress the logical error below the scale of $\varepsilon\ll \varepsilon_\text{QROM}/N$. In turn, considering e.g. a surface code implementation, this implies that the needed code distance must scale with the memory size as 
\begin{equation}
    d_\text{QROM} = \mathcal{O}( \log N).
\end{equation}
Considering that $\log N\equiv n$ is the size of the address register in the QPU, this implies that QROM needs $\mathcal{O}(n^2)$ physical qubits just to represent one single logical qubit, such that $n$ logical qubits are represented by $\mathcal{O}(n^3)$ physical resources. Moreover, we note that since logical operations are typically interleaved by $\mathcal{O}(d)$ rounds of syndrome extraction, this also implies that a single logical QROM step has effective depth scaling polynomially with the register size: every single logical operation requires depth $\mathcal{O}(n)$. In conclusion, the actual space-time volume required by the QROM circuit therefore reads
\begin{equation}
    \mathcal{V}^{\text{circuit}}_\text{QROM} = \mathcal{O}( N\log^4 N ),
\end{equation}
where we multiplied the physical circuit volume by a factor $\mathcal{O}(d^3)$ to account for the overheads discussed above~\footnote{Here, we assume a surface code implementation, such that the space-time volume overhead can be estimated as $\mathcal{O}(d^3)$.}. It is also useful to interpret this number as follows: the space-time volume per databit required by QROM scales polynomially in the register size, $\mathcal{V}'_\text{QROM}/N = \mathcal{O}(\text{poly } n)$. Moreover, this only accounts for the fault-tolerant implementation of the query, without considering the continuous operation of $T-$state factories.

\subsubsection{Factory overhead}
\noindent Similar concepts apply for the overhead associated with $T-$state factories. We assume standard distillation techniques~\cite{PhysRevA.71.022316}, such that the space-time overhead necessary to distill $T-$states to precision $\varepsilon$ scales as $\text{polylog}(1/\varepsilon)$. Importantly, to implement QROM also the distilled $T-$states must be precise to the scale $\varepsilon_\text{QROM}/N$; that is, the number of noisy $T-$states consumed for realising one single $T$ state employable in QROM scales as $\text{polylog}(N)$. More precisely, while the number of physical circuit-level $T-$gates required by QROM is $ \mathcal{O}(N)$, the total number of noisy $T$ states \emph{consumed} in order to do so (accounting for the magic state factories) actually scales like
\begin{equation}
    n_{T, \text{QROM}} = \mathcal{O}( N\text{polylog}N); 
\end{equation}
in other words, for querying a QROM with an address of $n$ qubits, one must consume $\mathcal{O}(\text{poly} n)$ noisy $T$ states for \emph{each} of the $\mathcal{O}(N)$ required $T$ gates. In addition, for distilling $T-$states up to such precision, one needs to implement
\begin{equation}
    k_\text{QROM} = \mathcal{O}(\log \log N )
\end{equation}
iterations of the distillation procedure. 

\subsubsection{Total overhead}
\noindent Combining the concepts above, the total space-time volume overhead for QROM is 
\begin{equation}
    \mathcal{V}_\text{QROM} = \mathcal{O}( N\log N \text{polylog} N ).
\end{equation}
The overhead factor therefore scales as 
\begin{equation}
    \xi_\text{QROM} = \mathcal{O}(\text{polylog} N). 
\end{equation}

\subsection{Comparison}
\noindent In sharp contrast with the evaluation above for QROM, in the case of QRAM the space-time volume for fault-tolerant operation reads $\mathcal{V}_\text{QRAM} = \mathcal{O}(N\log N \text{polylog} \log N)$, as a result of the code distance being $d_\text{QRAM} = \text{polylog} \log N$. Thus, the overhead factor reads
\begin{equation}
    \xi_\text{QRAM} = \mathcal{O}(\text{polylog}\log N). 
\end{equation}
Comparing the two approaches therefore yields
\begin{equation}
    \xi_\text{QROM} \sim \exp \xi_\text{QRAM};
\end{equation}
that is, the resource overhead factor for QROM is exponentially larger than that of QRAM. For a rough comparison, note that in practice one has approximately
\begin{equation}
    \mathcal{V}_\text{QROM} \sim \log^3 N \mathcal{V}_\text{QRAM}.
\end{equation}
That is, with the same space-time volume required for a QROM query, one could implement roughly $\sim \log^3 N$ QRAM queries, where we recall that $\log N$ is the size of the main QPU register.

\newpage

\section{The phase quantum RAM}\label{phase_QRAM}
\noindent Here, we discuss the implementation of a different type of query, which is defined as follows: 
\begin{equation}
    U_\text{pQRAM} \Bigg[ \sum_{\bm{x}}\psi_{\bm{x}}\ket{\bm{x}}\Bigg] = \sum_{\bm{x}}(-)^{D_{\bm{x}}}\psi_{\bm{x}}\ket{\bm{x}}.
\end{equation}
In practice, here there is no bus qubit, and the memory databit $D_{\bm{x}}$ is imprinted in the phase of the quantum state. This type of oracle is the one employed in many pioneering quantum algorithms~\cite{nielsen2010quantum, deutsch1992rapid, bernstein1993quantum, grover1996fast}, including Grover's algorithm for quantum search~\cite{grover1996fast}. \\
\indent One straightforward way to implement $U_\text{pQRAM}$ within our setting, is to recognise that to all effects, if the input state $\ket{\psi}$ is defined on $\log N$ qubits as usual, then we simply have
\begin{equation}\label{phase oracle equality}
    U_\text{QRAM} \Big[\ket{\psi, -}\Big] = \Big[U_\text{pQRAM}\ket{\psi}\Big]\ket{-}.
\end{equation}
That is, we can proceed with the very same protocol as before, but by initialising the bus qubit in the state $\ket{-}$ instead of in $\ket{0}$. Above, we proceed as done so far: since the bus qubit is still initialised in a known state, this only requires to modify the definition of the RS $\ket{\Phi}$ - more precisely, in the part concerning $|\Phi^{(1)}\rangle$. \\
\indent Eq.~\eqref{phase oracle equality} can be understood as follows. Recall that by initialising the bus qubit in $\ket{0}$, the pointer qubit in the $N$ OHE qubits is then in the $\ket{+}$ state; that is, all the OHE qubits are in $\ket{0}$, except the pointer, which is in $\ket{+}$. Also, recall that this arises from the Hadamard gate in Eq.~\eqref{V_HU}, which rotates the bus as $H\ket{0}=\ket{+}$. By starting with the pointer in $\ket{-}$, this first Hadamard gate rotates it as $H\ket{-}=\ket{1}$, implying that the pointer in the OHE qubits will now be a $\ket{1}$ state. More precisely, all the OHE qubits will be in $\ket{0}$, except the pointer at location $\mu(\bm{x})$, which is in $\ket{1}$. If we now implement $W_D$, this will still act nontrivially on all the OHE qubits, except for the pointer; but now, on the pointer it will simply provide a phase, conditional on the stored databit: $Z^{D_{\bm{x}}}\ket{1}=(-)^{D_{\bm{x}}}\ket{1}$. This phase will then be inherited by the entire state once the NOHE is inverted. Moreover, since the pointer has not changed its state, when inverting the NOHE the bus will just be left disentangled from the output state, and specifically it will end up in $\ket{-}$. This therefore successfully implements the phase QRAM, and exploits all of our previous results.

\subsection{Halving the T cost for the phase QRAM}

\noindent Importantly, for $U_\text{pQRAM}$ we can also achieve a significant reduction in the $T$ cost. This is due to the following reason. Let us consider all those operations in $U_\text{NOHE}$ which involve the OHE qubits; that is, the Friedkin gates where the last $N$ qubits (i.e., sector $\log N$ in the intuitive description) are targets. Note that this comprises $N-1$ Toffoli gates. Also, recall that the third qubit in these Friedkin gates is always initialysed in $\ket{0}$, so that for the input we can effectly only consider the first two qubits. Crucially, as opposed to the previous case, now these two input qubits cannot belong to any computational state of $\mathscr{H}^{(2)}\sim\mathbb{C}^4$. Indeed, from the NOHE structure, it is straightforward to realise that if the first qubit is in the $\ket{1}$ state, then, at this step of the computation, the second qubit must be supporting the bus, i.e., be in the state $\ket{1}$ as well. Thus, the input actually belongs to a smaller subspace of $\mathscr{H}^{(2)}$, which is spanned by the basis $\left\{\ket{00}, \ket{01}, \ket{11}\right\}$ - and thus has dimension $3$. The entire input of the Friedkin gate therefore belongs to the subspace of $\mathscr{H}^{(3)}$ spanned by $\left\{\ket{000}, \ket{010}, \ket{110}\right\}$. We now note that on this subspace, the Friedkin gate can actually be realised with a Clifford procedure, by instead implementing $(C_1X_2)(C_1X_3)$. It can be straightforwardly verified that with this construction, we indeed map
\begin{align}
    \ket{000} & \longrightarrow  \ket{000} \nonumber \\
    \ket{010} & \longrightarrow  \ket{010} \nonumber \\
    \ket{110} & \longrightarrow  \ket{101},
\end{align}
which is the Friedkin gate on the subspace of interest. With this construction, we therefore eliminate the need for the Toffoli gate, resulting in a final count for implementing the phase QRAM
\begin{equation}\label{Toffoli count_pQRAM}
    n_\text{Toffoli} = N - \log N -1,
\end{equation}
corresponding to roughly half the number of Toffoli gates required for the standard QRAM, as given in Eq.~\eqref{Toffoli count}. We note that this reduction here is allowed \emph{not only} by the fact that the input belongs to a three-dimensional subspace, but also - and crucially - by the fact that the three vectors spanning this space also belong to the computatinal basis, which diagonalises the Toffoli gate. Indeed, also in the case of conventional QRAM the input for the Friedkin gates considered here belongs to a three-dimensional space, spanned by $\left\{\ket{000}, \ket{010}, \ket{1+0}\right\}$; however, these vectors do not diagonalise the Toffoli gate, thereby preventing the same construction. 

\newpage

\section{Quantum RAM for quantum memories}

\noindent We now discuss generalisations of QRAM to quantum memories; that is, we consider datasets $D=\left\{\ket{\phi_{\bm{x}}}\right\}$, where now at location $\bm{x}$ a quantum state $\ket{\phi_{\bm{x}}}$ is stored. We will refer to this operation as Quantum Random Access Quantum Memory (QRAQM). Direct extensions of QRAM as defined so far to this domain are not possible, as they incur fundamental limits imposed by the no-cloning theorem~\cite{wootters1982single}. In a nutshell, for \emph{arbitrary} quantum memories we cannot have an operation of the form
\begin{equation}
    \sum_{\bm{x}}\psi_{\bm{x}}\ket{\bm{x}} \otimes \ket{0}\otimes\ket{D} \xrightarrow[]{\text{coherent quantum operation}} \sum_{\bm{x}}\psi_{\bm{x}}\ket{\bm{x}} \otimes \ket{\phi_{\bm{x}}}\otimes\ket{D}, \;\;\;\;\;\;\;\;\;\; \text{(\emph{not} possible)},
\end{equation}
where $\ket{D}$ is the state of the memory, featuring all the stored states $\ket{\phi_{\bm{x}}}$. To understand why this is not possible, it is sufficient to note that it would directly imply that by simply choosing $\ket{\psi}=\ket{\bm{x}}$, we would clone the unknown state $\ket{\phi_{\bm{x}}}$. More generally, we cannot execute a quantum analogue to the classical RAM for quantum memories, without introducing entanglement between the QPU and the dataset. \\
\indent To be more precise, we note that the consideration above, of course, applies to the general case where one can have $0<|\langle\phi_{\bm{x}}|\phi_{\bm{y}}\rangle|<1$; if one had always either parallel or orthogonal states, than this would be clearly equivalent to standard QRAM for classical memories. \\
\indent To formalise the notion of QRAQM, we instead consider the following operation:
\begin{equation}
    U_\text{QRAQM}\Bigg[\sum_{\bm{x}}\psi_{\bm{x}}\ket{\bm{x}} \otimes \ket{0}\otimes\ket{D} \Bigg] = \sum_{\bm{x}}\psi_{\bm{x}}\ket{\bm{x}} \otimes \ket{\phi_{\bm{x}}},
\end{equation}
where now the dataset $D$ is \emph{destroyed} after the query. Indeed, we highlight $\ket{D}$ in the input on the left-hand side to highlight that the dataset is erased during the operation. Note that $\ket{D}=\bigotimes_{\bm{x}}\ket{\phi_{\bm{x}}}$ now cannot be supported on classical bits: it is stored on $N$ memory qubits, which we index by $\bm{x}$. We still implement QRAQM with the tools discussed for QRAM, but we modify the loading $W_D$. That is, we proceed in three steps: (i) We apply $V$; (ii) We implement a modified loading procedure $W$ - note that this does \emph{not} depend on $D$ anymore; (iii) We apply $V^{-1}$. Below, we discuss the modified loading protocol. \\
\indent The loading $W$ for QRAQM proceeds as follows. First, the classically-controlled $Z$ gates become $CZ$ gates, which is still Clifford. More precisely, we apply a $CZ$ gate between each memory qubit $\bm{x}$, and the OHE qubit identifying label $\bm{x}$. Then, we simply measure the databits in the $X$ basis, and store the measurement outcomes $s_{\bm{x}}$ in a classical database $S=\left\{s_{\bm{x}}\right\}$; here, as usual $s_{\bm{x}}=0$ if we measured $+1$, and $s_{\bm{x}}=1$ if we measured $-1$. This teleports the quantum state on the pointer via standard quantum teleportation~\cite{PhysRevLett.70.1895}, up to a byproduct of the form $HZ^{s_{\bm{x}}}$. More specifically, if the initial register input is the computational basis state $\ket{\bm{x}}$, then the OHE qubits will be all left in $\ket{0}$ (because the $CZ$ gate did not affect them, and they are therefore uncorrelated with the memory qubit $\bm{x}$), while the pointer qubit at location $\bm{x}$ will be left in the state $HZ^{s_{\bm{x}}}\ket{\phi_{\bm{x}}}$. \\
\indent When after the loading $W$ we invert the NOHE, the final output will be left in the state
\begin{equation}
     \sum_{\bm{x}}\psi_{\bm{x}}\ket{\bm{x}} \otimes \ket{0}\otimes\ket{D} \longrightarrow\sum_{\bm{x}}(-)^{s_{\bm{x}}}\psi_{\bm{x}}\ket{\bm{x}} \otimes \ket{\phi_{\bm{x}}},
\end{equation}
\begin{equation}
     \sum_{\bm{x}}\psi_{\bm{x}}\ket{\bm{x}} \otimes \ket{0}\otimes\ket{D} \longrightarrow\sum_{\bm{x}}\psi_{\bm{x}}\ket{\bm{x}} \otimes Z^{s_{\bm{x}}} \ket{\phi_{\bm{x}}},
\end{equation}
which is almost the target one: the bus qubit now stores the desired quantum states, and the quantum memory is erased by the measurements. To complete the protocol, we now need to eliminate the unwanted phases $s_{\bm{x}}$. For this, it is sufficient to use the phase QRAM introduced in the section above, with the classical dataset $S$ stored previously. Indeed, note that, defining $\ket{\phi_{\bm{x}}}=\phi_{\bm{x},0}\ket{0} + \phi_{\bm{x},1}\ket{1}$, the state above can be written as 
\begin{equation}\label{partial_output_QRAQM}
    \sum_{\bm{x}}\psi_{\bm{x}}\ket{\bm{x}} \otimes \ket{\phi_{\bm{x}}} = \sum_{\bm{x},y}(-)^{s_{\bm{x}}y} \phi_{\bm{x},y} \psi_{\bm{x}} \ket{\bm{x}}\ket{y},
\end{equation}
where $y\in\left\{0,1\right\}$; similarly, the desired output is in the convenient form
\begin{equation}\label{finall_output_QRAQM}
    \sum_{\bm{x}}\psi_{\bm{x}}\ket{\bm{x}} \otimes \ket{\phi_{\bm{x}}} = \sum_{\bm{x},y}\phi_{\bm{x},y} \psi_{\bm{x}} \ket{\bm{x}}\ket{y}.
\end{equation}
Crucially, the two states in Eqs.~\eqref{partial_output_QRAQM} and~\eqref{partial_output_QRAQM} are related by a call to the phase QRAM discussed in Section~\ref{phase_QRAM}, with a classical memory $D'$ of size $2N$ with entries $D_{\bm{x}, y} = s_{\bm{x}}y$. Thus, we can complete the query to QRAQM with a final call to the phase QRAM.\\
\indent In practice, QRAQM is therefore equivalent to the subsequent call to a standard QRAM of size $N$ (with a slight Clifford modification of the loading step), followed by a call to a phase QRAM of size $2N$, whose entries are set by measurement outcomes in the previous QRAM call. In terms of non-Clifford cost, note that the standard QRAM call requires $2N-\log N -2$ Toffoli gates; the phase QRAM consumes roughly half the Toffoli gates per memory cell [see Eq.~\eqref{Toffoli count_pQRAM}], but in this case the memory size is doubled, yielding an equal contribution $2N - \log N - 2 $. Thus, the overall Toffoli count for QRAQM with this scheme is $4N - 2\log N - 4$, corresponding to twice the cost of QRAM.

\newpage

\section*{Additional details}

\section{Nested one-hot encoding: mechanism and intuition}
\noindent Below, we elaborate on the NOHE encoding in a more intuitive way, explaining the key guiding principles and connecting the circuit to a simple protocol. Finally, we explain how our circuit can be understood in terms of a `nested one-hot encoding bifurcation graph' - a graphical depiction which specifies the needed connectivity. 

\begin{figure*}[t!]
    \begin{minipage}{1.0\textwidth}
        \includegraphics[width=\textwidth]{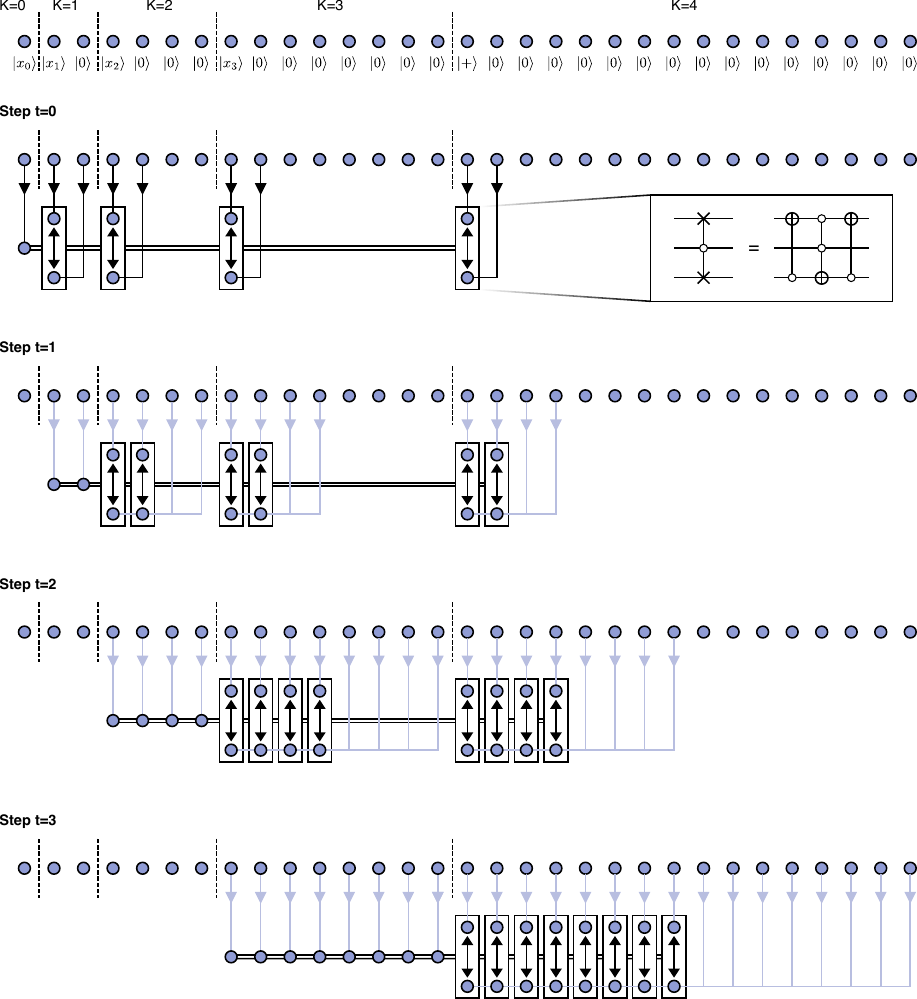}
    \end{minipage}%
\caption{\textbf{Intuition behind the NOHE}. Graphical representation of the intuitive protocol detailed in the text. At each step $t$, one sector $K$ triggers Friedkin gates (specified in the inset) in sectors $K'>K$, such that each qubit of $K$ only controls one Friedkin gate per $K'$.}
\label{fig_intuitive_OHE}
\end{figure*}

\subsection{Main idea of nested one-hot encoding}\label{main_idea_NOHE}
\noindent Here, we review the concepts above with a more intuitive approach, complementing the formal discussion with a qualitative construction. To this end, let us consider an arbitrary bitstring $\bm{x}=(x_0,\dots,x_{\log N -1})$ consisting of $\log N$ bits. As mentioned in methods, within the standard OHE framework this is represented via a string of $N$ bits, such that all are set to $0$, except a `hot' pointer bit, at position $\mu(\bm{x})=\sum_K2^Kx_K$, which is set to $1$. Specifically, 
\begin{equation}
    \bm{ohe}(\bm{x}) = \left( 0,0,\dots,1,\dots,0,0 \right),
\end{equation}
where the $1$ is at position $\mu(\bm{x})$. Note that $\mu(\bm{x})$ is nothing more than the integer whose binary representation is given by $\bm{x}$; it thus establishes a one-to-one correspondence, and can be inverted on its image. For our quantum OHE, we are interested in encoding the computational basis states $\ket{\bm{x}}$ via 
\begin{equation}
    \ket{\text{OHE}(\bm{x})} = \ket{0,0,\dots,+,\dots,0,0},
\end{equation}
where now the $\ket{+}$ state is at position $\mu(\bm{x})$, in fact playing the role of a quantum pointer. Importantly, note that in general $\langle \text{OHE}(\bm{x})|\text{OHE}(\bm{y})\rangle = (1+\delta_{\bm{x},\bm{y}})/2$; that is, two states $\ket{\text{OHE}(\bm{x})}$ and $\ket{\text{OHE}(\bm{y})}$, corresponding to the OHEs of orthogonal states $\ket{\bm{x}}$ and $\ket{\bm{y}}$, are \emph{not} orthogonal. This highlights that we cannot design a quantum operation that deterministically \emph{just} maps $\ket{\psi}\rightarrow\ket{\text{OHE}(\psi)}$ for any input state $\ket{\psi}$, as it would violate the distinguishability of orthogonal quantum states. In practice, the addi2tional term $\ket{\text{NOHE}(\bm{x})}$ appearing in Eq.~\eqref{V_NOHE_SM} above [or Eq.~(3) of the main text] can be understood as a \emph{necessary} byproduct of the OHE operation, which guarantees reversibility.\\
\indent Most importantly, the exact structure of $\ket{\text{NOHE}(\bm{x})}$ originates from a carefully designed protocol, which has crucial features in terms of noise-resilience. In essence, it constrains the propagation of errors in such a way that the resulting infidelity scales as $\mathcal{O}(\text{polylog} N)$, even though the space-time resources employed are $\mathcal{O}(N\log N)$.  \\
\indent We postpone the discussion on the noise-resilience of our protocol to Section~\ref{noise_resilience}. Here, we aim at explaining where the structure of $\ket{\text{NOHE}(\bm{x})}$ claimed in Methods originates from in practice. For this, we now intuitively sketch a protocol for implementing the map $\ket{\psi}\rightarrow\ket{\text{OHE}(\psi)}$; our circuit decomposition in Methods, which we discuss rigorously below, is nothing more than the mathematical formalization of this protocol.\\
\\
\noindent \textbf{Intuitive protocol.} We now outline an intuitive protocol for preparing $\ket{\text{OHE}(\bm{x})}$, which will directly lead to the emergence of $\ket{\text{NOHE}(\bm{x})}$ as well. Our circuit decomposition is inspired by this. A graphical depiction is given in Fig.~\ref{fig_intuitive_OHE}. For simplicity, consider a classical address $\ket{\bm{x}}=\ket{x_0,x_1,\dots,x_{\log N - 1}}$; all that follows generalizes by linearity to an arbitrary superposition state $\ket{\psi}=\sum_{\bm{x}}\psi_{\bm{x}}\ket{\bm{x}}$. We start by considering $\log N+1$ sets of qubits, labeled by $K=0,\dots, \log N$, such that sector $K$ contains $2^K$ qubits. Now, qubit $K$ of the address (i.e., the one supporting $\ket{x_K}$) is identified with the first qubit of sector $K$; the first qubit of the last sector ($K=\log N$) is set to $\ket{+}$; all other qubits are initially set to zero. Thus, the initial configuration is of the form
\begin{equation}
    \Big[\ket{x_0}\Big]\Big[\ket{x_1}\ket{0}\Big]\Big[\ket{x_2}\ket{0}^{\otimes 3}\Big]\Big[\ket{x_3}\ket{0}^{\otimes 7}\Big]\Big[\ket{x_4}\ket{0}^{\otimes 15}\Big]\dots \Big[\ket{x_{\log N -1}}\ket{0}^{\otimes N/2-1}\Big]\Big[\ket{+}\ket{0}^{\otimes N-1}\Big],
\end{equation}
where we used the parenthesis to highlight the different sectors. We now proceed as follows.
\begin{enumerate}
    \item \textbf{Step t = 0.} First, we take the first qubit (the only one in sector $K_c=0$), and we use it as control for a Friedkin gate targeting the first two qubits of each sector $K\geq 1$. That is, if $x_0=0$ nothing changes, while if $x_0=1$ the first two qubits of each sector are swapped. Precisely, assuming $x_0=1$, we get the state
\begin{equation}
    \Big[\ket{x_0=1}\Big]\Big[\ket{0}\ket{x_1}\Big]\Big[\ket{0}\ket{x_2}\ket{0}^{\otimes 2}\Big]\Big[\ket{0}\ket{x_3}\ket{0}^{\otimes 6}\Big]\dots \Big[\ket{0}\ket{x_{\log N -1}}\ket{0}^{\otimes N/2-2}\Big]\Big[\ket{0}\ket{+}\ket{0}^{\otimes N-2}\Big].
\end{equation}
\item \textbf{Step t = 1.} Second, we now use the first qubit of sector $K_c=1$, and use it as control for Friedkin gates between the first and the third qubit of each sector $K\geq 2$; moreover, the second qubit of $K_c=1$ acts as control of Friedkin gates targeting the second and the fourth qubit of each $K\geq 2$. Again, if $x_2=0$, nothing changes, while if $x_1=1$ we are in fact shifting by $2$ sites the only nontrivial qubit in each sector. More precisely, if we previously had $x_0=0$ and now $x_1=1$, we would get the state
\begin{equation}
    \Big[\ket{x_0=0}\Big]\Big[\ket{x_1=1}\ket{0}\Big]\Big[\ket{0}^{\otimes 2}\ket{x_2}\ket{0}\Big]\Big[\ket{0}^{\otimes 2}\ket{x_3}\ket{0}^{\otimes 5}\Big]\dots \Big[\ket{0}^{\otimes 2}\ket{x_{\log N -1}}\ket{0}^{\otimes N/2-3}\Big]\Big[\ket{0}^{\otimes 2}\ket{+}\ket{0}^{\otimes N-3}\Big],
\end{equation}
while if we had $x_0=1$ and now $x_1=1$, instead we would get
\begin{equation}
    \Big[\ket{x_0=1}\Big]\Big[\ket{0}\ket{x_1=1}\Big]\Big[\ket{0}^{\otimes 3}\ket{x_2}\Big]\Big[\ket{0}^{\otimes 3}\ket{x_3}\ket{0}^{\otimes 4}\Big]\dots \Big[\ket{0}^{\otimes 3}\ket{x_{\log N -1}}\ket{0}^{\otimes N/2-4}\Big]\Big[\ket{0}^{\otimes 3}\ket{+}\ket{0}^{\otimes N-4}\Big].
\end{equation}
This can be understood as follows: for $K\geq 2$, the only nontrivial (hot) qubit is shifted to position $\mu(x_0,x_1)=x_0+2x_1$; moreover, note that in sectors $K=0,1$ we have to all effects prepared $\bm{ohe}^{(0)}(\bm{x})$ and $\bm{ohe}^{(1)}(\bm{x})$ respectively.
\item \textbf{Generic step t.} This intuitive protocol for preparing $\ket{\text{OHE}(\bm{x})}$ continues by iterating this procedure: at each step (say, step $t$), we pick all the qubits in sector $K_c=t$, and use them as controls for Friedkin gates in the sectors $K\geq K_c+1$. More precisely, qubit $k$ in sector $K_c$ controls the Friedkin gates targeting qubits $k$ and $k+2^{K_c}$ in each sector $K\geq K_c+1$. After this step, all sectors $K=0,1,\dots,K_c+1$ will be supporting $\bm{ohe}^{(K)}(\bm{x})$; all the others, i.e. for $K\geq K_c+2$ on, will display the only nontrivial qubit, hosting the state $\ket{x_K}$ (or $\ket{+}$ for $K=\log N$), at position $\sum_{J=0}^{K_c}2^Jx_J$. For instance, after step $t=4$, we get a state of the form
\begin{equation}
    \ket{x_0}\ket{\bm{ohe}^{(1)}(\bm{x})}\dots\ket{\bm{ohe}^{(5)}(\bm{x})} \Big[\ket{0}^{\otimes\mu-1}\ket{x_6}\ket{0}^{\otimes 2^6-\mu}\Big]\Big[\ket{0}^{\otimes\mu-1}\ket{x_7}\ket{0}^{\otimes 2^7-\mu}\Big]\dots\Big[\ket{0}^{\otimes\mu-1}\ket{+}\ket{0}^{\otimes N-\mu}\Big],
\end{equation}
where we set $\mu = \sum_{J=0}^42^Jx_J$ for simplicity. 
\end{enumerate}
\indent It follows that after step $t=\log N -1$, the state is exactly the following: all sectors $K=0,1,\dots,\log N -1$ display the state $\ket{\bm{ohe}^{(K)}(\bm{x})}$; the last state the desired $\ket{\text{OHE}(\bm{x})}$. That is, we end up with 
\begin{align}
    \ket{\text{NOHE}(\bm{x},+)} & = \ket{\bm{ohe}^{(0)}(\bm{x})}\ket{\bm{ohe}^{(1)}(\bm{x})}\dots \ket{\bm{ohe}^{(\log N - 1)}(\bm{x})} \ket{\text{OHE}(\bm{x})} = \nonumber \\
    & = \ket{\text{NOHE}(\bm{x})}\ket{\text{OHE}(\bm{x})},
\end{align}
which is what we claimed. 

\begin{figure*}[t!]
    \begin{minipage}{1.0\textwidth}
        \includegraphics[width=\textwidth]{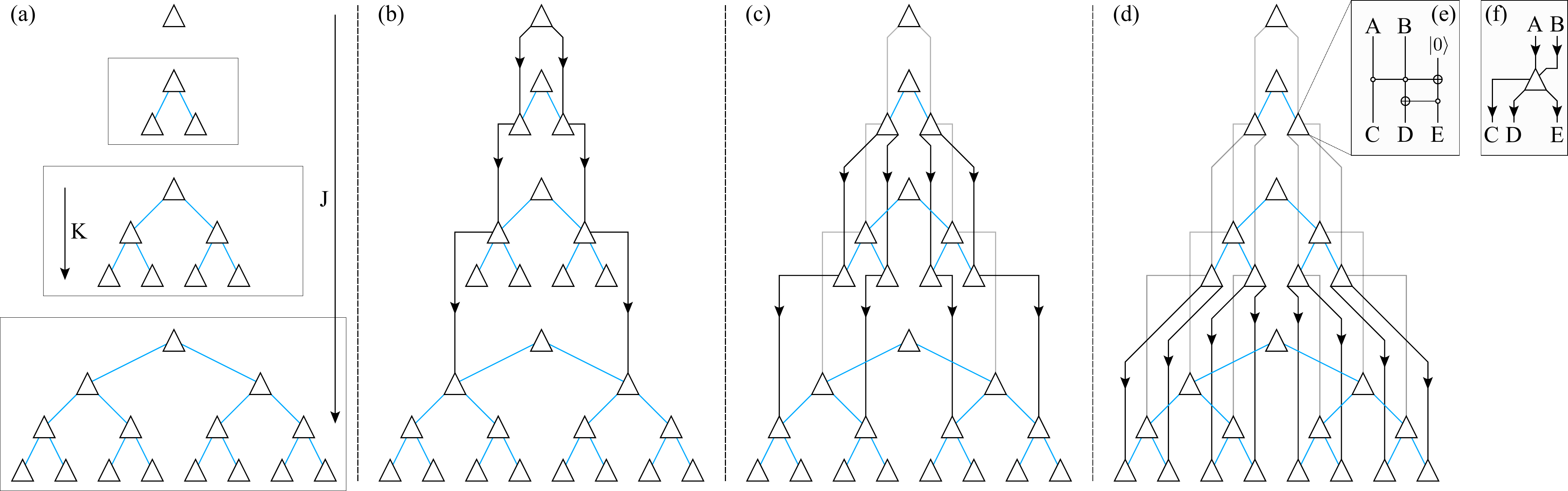}
    \end{minipage}%
\caption{\textbf{Nested bifurcation graph construction.} (a) We start with $\log N$ bifurcation graphs of increasing size; triangles represent vertexes, while blue lines are the edges of the sub-graphs. Here we the specific example is with $N=16$. (b-d) We connect subsequent sub-graphs by adding `cascaded' patterns. Here different shades of black for the edges only help reading the figure. (e) To each vertex in the resulting graph we associate the gadget displayed, which takes as input a two-qubit state and outputs a three-qubit state; specifically, we introduce an additional qubit initialised in the $\ket{0}$ state, and execute a Toffoli gate followed by a controlled-not. (f) Details of how the edges connected to each edge in the graph refer to input and output qubits. }
\label{fig_nested_graph}
\end{figure*}

\subsection{Graphical intuition: the nested bifurcation graph}
\noindent Here, we introduce an understanding of OHE operations in terms of graphs and (classical) logical circuits. This has the threefold purpose of (i) shining light on the circuit decomposition in Eq.~\eqref{U_OHE_decomposition}, (ii) understanding the relation of our NOHE with known QRAM proposals and (iii) preparing for the dynaical rearrangement subroutine that we will employ in the QRAM factory. The following construction proceeds in two steps: First, we define a `nested bifurcation graph' (NBG); Second, we relate this to a logical circuit that characterises the OHE. \\
\indent The NBG is constructed as shown in Fig.~\ref{fig_nested_graph}. We start by drawing an array of binary tree graphs of increasing depth $J=1,2,...,\log N$ as in Fig.~\ref{fig_nested_graph}(a); we label the sub-graphs by $J$, so that sub-graph $J$ features $2^J-1$ vertexes. The edges of these sub-graphs are depicted in blue, and they are \emph{directed}: we interpret each edge as `incoming' from the top and `outgoing' towards the bottom. As a convention, we count the layers of a sub-graph $J$ as $K=0,1,\dots,J-1$ starting from the top, so that layer $K$ contains $2^K$ vertexes. Next, we connect the sub-graphs as shown in Figs.~\ref{fig_nested_graph}(b-d); just for clarity, we depict these new edges in different shades of black, and highlight their orientation. In a nutshell, for every sub-graph $J=1,2,\dots,\log N -1$ we consider the vertexes in the last layer (i.e., $K=J-1$), and for each of these vertexes we start two `cascaded' patterns; these patterns essentially `pass through' layers $K=J$ of each subsequent sub-graph. This eventually results in a graph where bifurcation graphs of increasing size are nested the one with the other.\\
\indent We now relate the NBG with the NOHE as follows. To each vertex in the NBG we associate the logic gadget displayed in Fig.~\ref{fig_nested_graph}(e): this formally takes two inputs and returns three outputs, and in fact realises a reversible version of the logical AND operation with quantum gates. In Fig.~\ref{fig_nested_graph}(f) we show explicitly how the entries of each vertex in a graph are related to the entries of the gadgets.

\newpage

\section{Circuit decmoposition of NOHE}\label{Circuit decmoposition of NOHE}

\noindent Here, we discuss circuit decompositions of the nested one-hot encoding (NOHE) operation. In essence, the following is a formal analysis of the section above.  \\
\\
\noindent \textbf{Formalities.} We introduce an operation implementing the NOHE, i.e. acting on any state $\ket\psi$ on $\log N$ qubits as $\ket\psi\rightarrow\ket{\text{NOHE}(\psi)}$. Since this maps a Hilbert space of dimension $N$ to one of dimension $2^{N-1}$, so far this can be understood as an isometry. To write it as a unitary $U_\text{NOHE}$, we can introduce $N-1-\log N$ ancillary qubits, and first specify the action when these are all initialized in the $\ket{0}$ state: 
\begin{equation}\label{NOHE_action}
    U_\text{NOHE}^{(N)} \ket{\psi}\otimes\ket{0}^{\otimes N-1-\log N} = \ket{\text{NOHE}(\psi)};
\end{equation}
the complete unitary rotation remains undefined, and will be fixed e.g. by specifying the circuit decomposition. As in Methods, here the superscript $N$ indicates the dimension of the input Hilbert space of the isometry, and we will omit it when it is clear from the context. Moreover, we will frequently just write $U_\text{NOHE}\ket{\psi}=\ket{\text{NOHE}(\psi)}$ when the dimensions are clear from the context. Specifically, when introducing a bus qubit, we might simply write
\begin{equation}
    U_{\text{NOHE}} ^{(2N)}\ket{\psi}\ket{+} = \ket{\text{NOHE}(\psi,+)},
\end{equation}
subtending that the input contains $2N - 2 - \log N$ ancillas initialized in the $\ket{0}$ state.\\
\\
\noindent \textbf{Unitary decomposition.} We now formalize the concept above; that is, we discuss the decomposition of $U_{\text{NOHE}}$ in basic gates,
\begin{equation}
    U_\text{NOHE}^{(N)} = \Bigg[\prod_{K=0}^{\log N -2} \prod_{J=K+1}^{\log N-1} \overline{CS}(K|J)\Bigg]\Bigg[ \prod_{K=2}^{\log N-1} S_{K,2^K-1} \Bigg],
\end{equation}
that we introduced in Methods. Recall that here $S_{b,c}\equiv \left(\mathbb{1}_{b,c} + X_bX_c + Y_bY_c + Z_bZ_c \right)/2$ implements a swap of the qubits $b\leftrightarrow c$, i.e. $S_{b,c}\ket{\psi_b}\otimes\ket{\phi_c} = \ket{\phi_b}\otimes\ket{\psi_c}$ for any factorized state, and for $K<J$ we defined 
\begin{equation}\label{CS_KJ}
    \overline{CS}(K|J) = \prod_{\alpha = 2^K-1}^{2(2^K-1)} CS(\alpha | \alpha + 2^J-2^K, \; \alpha + 2^J),
\end{equation}
where 
\begin{equation}\label{cswap}
    CS(a|b,c) = \ket{0}_a\!\bra{0} \otimes\mathbb{1}_{b,c} + \ket{1}_a\!\bra{1} \otimes S_{b,c},
\end{equation}
is the Friedkin gate (controlled-swap). Moreover, recall that we use the ordering convention that $\prod_{k=k_\text{min}}^{k_\text{max}}U_k= U_{k_\text{max}}U_{k_\text{max}-1}\dots U_{k_\text{min}}$. \\
\\
\noindent \textbf{Unitary decomposition: formal proof.} To show that the proposed decomposition indeed acts as desired, i.e. as
\begin{equation}\label{claim 1}
    \Bigg[\prod_{K=0}^{\log N -2} \prod_{J=K+1}^{\log N-1} \overline{CS}(K|J)\Bigg]\Bigg[ \prod_{K=2}^{\log N-1} S_{K,2^K-1} \Bigg] \ket{\psi}\ket{0}^{N-1-\log N} = \ket{\text{NOHE}(\psi)} \;\;\;\;\;\;\; \text{(Claim 1)} 
\end{equation}
for any state $\ket\psi$ on $\log N$ qubits, by linearity it is sufficient to prove it when $\ket\psi$ is an arbitrary computational basis state $\ket{\bm{x}}$. We recall here that $\ket{\text{NOHE}(\psi)}$ is defined as follows: 
\begin{equation}\label{NOHE_def}
    \text{NOHE}(\bm{x}) = \Big(\bm{ohe}^{(0)}(\bm{x}), ..., \bm{ohe}^{(\log N-1)}(\bm{x})\Big),
\end{equation}
where $\bm{ohe}^{(K)}_\alpha(\bm{x})=x_K\delta_{\alpha, \mu(x_0,\dots,x_{K-1})}$ and $\mu(\bm{z})=\sum_Jx_J2^J$ (see Methods).\\
\indent For proving the circuit decomposition above, we proceed as follows. We start by rewriting the decomposition above as
\begin{equation}\label{rewriting 1}
    \Bigg[\prod_{K=0}^{\log N -2} \prod_{J=K+1}^{\log N-1} \overline{CS}(K|J)\Bigg]\Bigg[ \prod_{K=2}^{\log N-1} S_{K,2^K-1} \Bigg]  = \Bigg[ \prod_{J=1}^{\log N - 1} U^{(J\rightarrow J+1)} \Bigg]\Bigg[ \prod_{K=2}^{\log N-1} S_{K,2^K-1} \Bigg],
\end{equation}
where we isolated the part $U^{(J\rightarrow J+1)} = \prod_{K=0}^{J-1}\overline{CS}(K|J)$ for $J\geq 1$. This can be seen straightforwardly by putting together the following two observations. First, that 
\begin{equation}
    \Big[\;\overline{CS}(K|J),\overline{CS}(K'|J')\;\Big]=0 \;\;\;\; \text{whenever} \;\;\;\; J\neq J',
\end{equation}
and independently on $K$ and $K'$. Second, that on the left-hand side of Eq.~\eqref{rewriting 1} the operations $\overline{CS}(K|J)$ are ordered for increasing $J$ (specifically, with smaller $J$ to the right, i.e. acting first). With these considerations, it is simple to see that the gates in the product can be recollected as above.\\ 
\indent We are now interested in proving the following preliminary claim:
\begin{equation}\label{claim 2}
    U^{(J\rightarrow J+1)} \ket{\text{NOHE}(x_0,x_1,\dots, x_{J-1})} \ket{x_J}\ket{0}^{\otimes 2^J-1} =\ket{\text{NOHE}(x_0,x_1,\dots, x_{J})} \;\;\;\;\; \text{(Claim 2)}.
\end{equation}
For this, we directly prove the stronger claim that for $J'\leq J$ we always have
\begin{equation}\label{stronger claim}
     \prod_{K=0}^{J'-1}\overline{CS}(K|J) \ket{\text{NOHE}(x_0,\dots,x_{J-1})} \ket{x_J}\ket{0}^{\otimes 2^J-1} = \ket{\text{NOHE}(x_0,\dots,x_{J-1})} \ket{x_J\bm{ohe}(x_0,x_1,\dots, x_{J'-1})} \ket{0}^{\otimes 2^J-2^{J'}}, 
\end{equation}
which reduces to our Claim 2 in Eq.~\eqref{claim 2} when $J'=J$. To prove Eq.~\eqref{stronger claim}, we proceed by induction on $1 \leq J'\leq J$. First, we can straightforwardly check that it holds for $J'=1$, as
\begin{align}
    \overline{CS}(0|J) \ket{\text{NOHE}(x_0,\dots,x_{J-1})} \ket{x_J}\ket{0}^{\otimes 2^J-1} = \ket{\text{NOHE}(x_0,\dots,x_{J-1})} \Bigg[ (1-x_0)\ket{0,0} + x_0\ket{0,x_J}\Bigg]\ket{0}^{\otimes 2^J-2},
\end{align}
where we used the definition~\eqref{CS_KJ}, and it is easy to recognize that 
\begin{equation}
    (1-x_0)\ket{0,0} + x_0\ket{0,x_J} = (1-x_J)\ket{0,0} + x_J\Bigg[(1-x_0)\ket{0,0}+x_0\ket{0,1}\Bigg] = x_J\ket{\bm{ohe}(x_0)}.
\end{equation}
Next, we move to the inductive step $J'\rightarrow J'+1$. Noting that $\prod_{K=0}^{J'}\overline{CS}(K|J)=\overline{CS}(J'|J)\prod_{K=0}^{J'-1}\overline{CS}(K|J))$, we use the inductive hypothesis to get
\begin{align}\label{decomposition_align1}
    & \overline{CS}(J'|J)\prod_{K=0}^{J'-1}\overline{CS}(K|J))  \ket{\text{NOHE}(x_0,\dots,x_{J-1})} \ket{x_J}\ket{0}^{\otimes 2^J-1} = \nonumber \\
     & \overset{(1)}{=} \; \overline{CS}(J'|J) \ket{\text{NOHE}(x_0,\dots,x_{J-1})} \ket{x_J\bm{ohe}(x_0,x_1,\dots, x_{J'-1})} \ket{0}^{\otimes 2^J-2^{J'}} = \nonumber \\
     & \overset{(2)}{=} \overline{CS}(J'|J) \ket{\text{NOHE}(x_0,\dots,x_{J-1})} \Bigg[(1-x_J)\ket{0}^{2^{\otimes J'}} + x_J\ket{\bm{ohe}(x_0,x_1,\dots, x_{J'-1})} \Bigg] \ket{0}^{\otimes 2^J-2^{J'}} = \nonumber \\
     & \overset{(3)}{=} \ket{\text{NOHE}(x_0,\dots,x_{J-1})}\Bigg[ (1-x_J)\ket{0}^{\otimes 2^{J'+1}} + x_J\ket{\bm{ohe}(x_0,x_1,\dots, x_{J'})}  \Bigg]\ket{0}^{\otimes 2^J-2^{J'+1}}= \nonumber  \\
     & \overset{(4)}{=} \ket{\text{NOHE}(x_0,\dots,x_{J-1})} \ket{x_J\bm{ohe}(x_0,x_1,\dots, x_{J'})}\ket{0}^{\otimes 2^J-2^{J'+1}}.
\end{align}
In the chain of equalities above, we proceeded as follows. For (1) we explicitly used the inductive hypothesis. For (2) we just conveniently rewrite the expression in the parenthesis. For (3), we note that for any state $\ket{\Psi_c}$ on $2^J-1$ qubits, we always have $\overline{CS}(J'|J)\ket{\Psi_c}\ket{0}^{\otimes 2^{J'+1}}=\ket{\Psi_c}\ket{0}^{\otimes 2^{J'+1}}$, and that moreover
\begin{align}
    &\overline{CS}(J'|J) \ket{\text{NOHE}(x_0,\dots,x_{J-1})} \ket{\bm{ohe}(x_0,x_1,\dots, x_{J'-1})}\ket{0}^{\otimes 2^{J'}} = \nonumber \\ 
    & \overset{(3a)}{=} \ket{\text{NOHE}(x_0,\dots,x_{J-1})} \Bigg[ (1-x_{J'})  \ket{\bm{ohe}(x_0,x_1,\dots, x_{J'-1})}\ket{0}^{\otimes 2^{J'}} + x_{J'}  \ket{0}^{\otimes 2^{J'}}\ket{\bm{ohe}(x_0,x_1,\dots, x_{J'-1})}\Bigg] = \nonumber \\
    & \overset{(3b)}{=} \ket{\text{NOHE}(x_0,\dots,x_{J-1})} \ket{\bm{ohe}(x_0,x_1,\dots, x_{J'})}.
\end{align}
Herein, for equality (3a) we make explicit use of the definition~\eqref{CS_KJ}, and of the NOHE structure in Eq.~\eqref{NOHE_def}. Specifically, from Eq.~\eqref{CS_KJ} we see that $\overline{CS}(J'|J)$ is an operation where qubits $\left\{2^{J'}-1,\dots, 2\left(2^{J'}-1\right)\right\}$ act as controls; from Eq.~\eqref{NOHE_def} we see that among these qubits, we can only have a $\ket{1}$ state anywhere if $x_{J'}=1$; thus, we infer the first part of the equality, that if $x_{J'}=0$ the target qubits are unaffected. For the second part of the equality, again from Eq.~\eqref{NOHE_def} we see that if instead $x_{J'}=1$, then among the control qubits of $\overline{CS}(J'|J)$, we have one, and only one $\ket{1}$ state at position $2^{J'-1}-1 + \mu_{J'}(x_0,\dots,x_{J'-1})$; moreover, from Eq.~\eqref{CS_KJ}, we see that this corresponds exactly to the one control, whose first target is the pointer of $\ket{\bm{ohe}(x_0,\dots, x_{J'-1})}$: this therefore shifts the pointer by $2^{J'}$ positions, resulting in step (3a) above. Finally, for step (3b) we simply rewrite the standard OHE. Together, these equalities thus result in step (3) of Eqs.~\eqref{decomposition_align1}. The proof of Eqs.~\eqref{decomposition_align1} is thus completed by noting that in step (4) we just rewrite the standard OHE as before. In summary, this therefore proves the inductive step, and thus Claim 2 in Eq.~\eqref{claim 2}. \\
\indent We now use this result to prove Claim 1 in Eq.~\eqref{claim 1}. For this, we use the form in Eq.~\eqref{rewriting 1}, and we first note that the first part of the decomposition, i.e. the swap gates, has the effect of reordering the input as
\begin{equation}
    \Bigg[ \prod_{K=2}^{\log N-1} S_{K,2^K-1} \Bigg] \ket{\bm{x}} \ket{0}^{\otimes N-\log N - 1} = \bigotimes_{J=0}^{\log N - 1}\bigg[\ket{x_J}\ket{0}^{2^J-1}\bigg], 
\end{equation}
where again by linearity we only focus on computational-basis states as inputs. Thus, our goal for proving Claim 1 in Eq.~\eqref{claim 1} is now reduced to showing that 
\begin{equation}\label{claim 1 second form}
    \Bigg[\prod_{J=1}^{\log N -1}U^{(J\rightarrow J+1)}\Bigg] \Bigg[\bigotimes_{J=0}^{\log N - 1}\ket{x_J}\ket{0}^{2^J-1}\Bigg] = \ket{\text{NOHE}(\bm{x})}. 
\end{equation}
For this, it is sufficient to make direct use of Claim 2 in Eq.~\eqref{claim 2}. Specifically, we can now prove that for any $2 \leq J'\leq \log N$ the following holds:
\begin{equation}
    \Bigg[\prod_{J=1}^{J' -1}U^{(J\rightarrow J+1)}\Bigg] \Bigg[\bigotimes_{J=0}^{\log N - 1}\ket{x_J}\ket{0}^{2^J-1}\Bigg] = \ket{\text{NOHE}(x_0,\dots,x_{J'-1})} \Bigg[\bigotimes_{J=J'}^{\log N - 1}\ket{x_J}\ket{0}^{2^J-1}\Bigg]. 
\end{equation}
Again, to see this we proceed by induction on $2\leq J'\leq \log N$. First, for $J'=2$ we can check that
\begin{align}
    & U^{(1\rightarrow 2)} \Bigg[\bigotimes_{J=0}^{\log N - 1}\ket{x_J}\ket{0}^{2^J-1}\Bigg] = \nonumber \\ 
    & = \Bigg[U^{(1\rightarrow 2)} \ket{\text{NOHE}(x_0)}\ket{x_1}\ket{0}\Bigg]\Bigg[\bigotimes_{J=2}^{\log N - 1}\ket{x_J}\ket{0}^{2^J-1}\Bigg] = \nonumber \\ 
    & = \ket{\text{NOHE}(x_0,x_1} \Bigg[\bigotimes_{J=2}^{\log N - 1}\ket{x_J}\ket{0}^{2^J-1}\Bigg],
\end{align}
where we made explicit use of Eq.~\eqref{claim 2} for the case $J=1$. Finally, for the inductive step $J'\rightarrow J'+1$, we can use the straightforward rewriting
\begin{equation}
    \prod_{J=1}^{J' }U^{(J\rightarrow J+1)}=U^{(J'\rightarrow J'+1)}\prod_{J=1}^{J' -1}U^{(J\rightarrow J+1)}
\end{equation}
to recognize that the following chain of equalities holds true: 
\begin{align}
    & U^{(J'\rightarrow J'+1)}\prod_{J=1}^{J' -1}U^{(J\rightarrow J+1)} \Bigg[\bigotimes_{J=0}^{\log N - 1}\ket{x_J}\ket{0}^{2^J-1}\Bigg] = \nonumber \\
    & \overset{(1)}{=} U^{(J'\rightarrow J'+1)} \ket{\text{NOHE}(x_0,\dots,x_{J'-1})} \Bigg[\bigotimes_{J=J'}^{\log N - 1}\ket{x_J}\ket{0}^{2^J-1}\Bigg] = \nonumber \\
    & \overset{(2)}{=} \Bigg[ U^{(J'\rightarrow J'+1)} \ket{\text{NOHE}(x_0,\dots,x_{J'-1})} \ket{x_{J'}}\ket{0}^{\otimes 2^{J'}-1} \Bigg] \Bigg[\bigotimes_{J=J'+1}^{\log N - 1}\ket{x_J}\ket{0}^{2^J-1}\Bigg] = \nonumber \\
    & \overset{(3)}{=} \ket{\text{NOHE}(x_0,\dots,x_{J'})} \Bigg[\bigotimes_{J=J'+1}^{\log N - 1}\ket{x_J}\ket{0}^{2^J-1}\Bigg].
\end{align}
Specifically: For equality (1) we explicitly use the inductive hypothesis; For (2) we simply recollected part of the product; For (3) we used Claim 2 in Eq.~\eqref{claim 2}. Altogether, this proves by induction that Eq.~\eqref{claim 1 second form}, thereby directly proving the desired Claim 1 in Eq.~\eqref{claim 1}. \\
\\
\noindent \textbf{Parallelization.} By naively counting from the decomposition in Eq.~\eqref{U_OHE_decomposition}, it seems that a (realistic) circuit implementing $U^{(N)}_{\text{NOHE}}$ requires $\mathcal{O}(\log ^2 N)$ logical steps. This is due to the following fact: even though all the gates in the inner product over $J$ commute, and thus \emph{might} be executed in parallel, nevertheless they are manifestly non local: for any fixed $K$, we have $2^K$ control qubits, each acting on $\log N - K - 2$ targets; thus, we cannot assume that all those gates are executed in parallel on a real device, even though they commute.\\
\indent Intuitively, this can also be understood by recalling the scheme introduced in Section~\ref{main_idea_NOHE}. Therein, for implementing $U_\text{NOHE}^{(2N)}$, we proceeded with `main steps' labeled by $t=1,2,\dots,\log N -1$. However, during step $t$ each qubit in sector $K_c=t$ was used as a control-qubit for $\log N - t$ Friedkin gates - those involving sectors $K=K_c+1,K_c+2,\dots, \log N$. Even though they commute, these cannot be realistically executed in parallel, as they share a common control qubit; thus, step $t$ involves $\log N -t$. It follows that the total runtime of the protocol for $U_\text{NOHE}^{(2N)}$ is proportional to $\sum_{t=0}^{\log N - 1} (\log N - t) = \log N(\log N+1)/2$. Consequently, for implementing $U_\text{NOHE}^{(N)}$ via the decomposition in Eq.~\eqref{U_OHE_decomposition} the runtime is $\log N(\log N -1)/2=\mathcal{O}(\log^2N)$.\\ 
\indent In contrast, as mentioned in the main text and in Methods, here we show that $U^{(N)}_{\text{NOHE}}$ can be parallelized to a circuit of depth $\mathcal{O}(\log N)$. Specifically, we now show that the decomposition in Eq.~\eqref{U_OHE_decomposition} can be rearranged as 
\begin{equation}\label{fast decomposition}
    U^{(N)}_{\text{NOHE}} = \Bigg[\prod_{T=1}^{T_f}\prod_{K=0}^{D_T} \overline{CS}(K_T - K|J_T + K)\Bigg] \Bigg[ \prod_{K=2}^{\log N-1} S_{K,2^K-1} \Bigg],
\end{equation}
where the final step is $T_f=2\log N -3$ and we set the indexes $K_T = \left \lfloor{\frac{T-1}{2}}\right\rfloor$, $J_T = \text{min}\left\{T, \log N-1\right\}$ and $D_T = K_T - \text{max}\left\{0, T-\log N+1\right\}$. Importantly, for a fixed $1\leq T\leq T_f$ one always has
\begin{equation}
    \Big[\;\overline{CS}(K_T - K|J_T + K), \overline{CS}(K'_T - K'|J_T + K')\;\Big] = 0
\end{equation}
for $0\leq K,K' \leq D_T$; moreover, for a fixed $T$, the gates contained in the product over $K$ are such that any qubit (be it a control or a target) only appears once, so that the product over $K$ can be performed locally and in parallel for any $T$. From these two considerations, it follows that all the operations in the product over $K$ can be performed in parallel on a real processor; this therefore results in exactly $2\left(\log N-1\right)$ parallel circuit steps (also accounting for the initial swaps). \\
\indent To prove Eq.~\eqref{fast decomposition}, it is sufficient to recall the main intuition behind our protocol, and modify as follows. Specifically, below we modify the steps in Section~\ref{main_idea_NOHE} in such a way that the protocol eventually has duration $\mathcal{O}(\log N)$. Note that below we use the variable $t'$ to identify steps of runtime $\mathcal{O}(1)$. \\
\indent For steps $t'=0$ and $t'=1$, we proceed as in the previous scheme: we use the only one qubit of sector $K_c=0$ as the control of Friedkin gates in sectors $K=1$ and $K=2$ respectively. \\
\indent During step $t'=2$, we use again the qubit of sector $K_c=0$, now as control of a Friiedkin gate between the first two qubits in sector $K=3$; but crucially, in parallel we can also use the qubits in sector $K_c=1$ as controls of Friedkin gates in sector $K=2$. \\
\indent For step $t'=3$ we proceed similarly: the qubit of sector $K_c=0$ controls Friedkin gates in sector $K=4$, while in parallel the qubits of sector $K_c=1$ act as controls for Friedkin gates in sector $K=3$. \\
\indent For step $t'=4$, the qubits in sectors $K_c=0$ and $K_c=1$ control, respectively, Friedkin gates in sectors $K=5$ and $K=4$. But in parallel, now the qubits in sector $K_c=2$ can be used as controls for Friedkin gates in sector $K=3$. \\
\indent The main idea of our parallelized protocol proceeds by iterating this procedure: the main principle is that we start using qubits in sector $K$ as controls of Friedkin gates, as soon as they do not need to be involved as targets anymore. To calculate the new duration of the protocol, let $t'_K$ be the step at which qubits from sector $K$ start to be employed as controls; thus, for instance we have already seen that $t'_0=0$ and $t'_1=2$. First, note that at time $t'_K$ sector $K$ is interfaced with sector $K+1$. Then, one can see that $t'_{K+1} = t'_{K} + 2$. Indeed, at step $t'_K$ sector $K$ acts as control in Friedkin gates where sector $K+1$ is involved as target; at step $t'_K+1$, sector $K+1$ is ready to be employed as control, but is must wait because sector $K+2$ is now interfaced with sector $K$; finally, at step $t'_K + 2$ sector $K+2$ is ready to undergo Friedkin gates controlled by qubits in sector $K+1$. Thus, we get $t'_K = 2K$. For implementing $U_\text{NOHE}^{(2N)}$, we know that the largest sector which is employed as control is $K_c = \log N - 1$, which only controls Friedkin gates with qubits in sector $K=\log N$. Thus, the last step occurs exactly at $t'_{\log N - 1} = 2\log N - 2$;onsidering that we start counting from $t'=0$, it follows that the total number of steps is $2\log N - 1$. Thus, for implementing $U_\text{NOHE}^{(N)}$ the runtime is exactly $T_f = 2\log N - 3$ as claimed.

\section{Mathematical methods for the noise-resilience analysis}\label{math methods}

\noindent Here, we collect the mathematical methods used to prove several results in the noise-resilience analysis of Sections~\ref{SM_noise_analysis1} and~\ref{SM_noise_analysis2}. 

\subsection{Uniform distribution over the Hilbert space}\label{Math1}
\noindent We start by elaborating on the mathematical details concerning the uniform distribution over the (normalized states of the) Hilbert space.\\
\\
\noindent \textbf{Defining the distribution.} Conceptually, $\int_{\mathscr{H}} \mathrm{d}\psi\; f(\psi)$ represents an integral over the uniform distribution of normalized states in the Hilbert space $\mathscr{H}$, normalized such that $\int_{\mathscr{H}} \mathrm{d}\psi = 1$. To be more precise at the mathematical level, we find it useful to parameterize the wavefunction as
\begin{equation}
    \psi_{\bm{x}} = e^{i\phi_j} \sqrt{z_j} \;\;\;\;\; \text{for} \;\;\;\;\; j=1,\dots,N
\end{equation}
with $\phi_j\in\left[0,2\pi\right]$ and $z_j\in\left[0,+\infty\right]$ in general. The relation between $\bm{x}$ and the index $j$ can be e.g. simply $j=\mu(\bm{x})$. Then, the average of any function $f:\mathscr{H}\rightarrow \mathbb{R}$ takes the form
\begin{equation}
    \int_{\mathscr{H}}\!\! \mathrm{d}\psi \; f(\psi) = \frac{(N-1)!}{(2\pi)^N} \int_{\left[0,2\pi\right]^N} \mathrm{d}^N\phi\int_{\left[0,+\infty\right]^N}\mathrm{d}^Nz \; \delta( 1 - |\bm{z}|_1 ) \;f(\phi_1,\dots,\phi_N; z_1,\dots,z_N),
\end{equation}
where $\delta(x)$ is the conventional Dirac delta distribution, and $|\bm{z}|_1=\sum_i |z_i| $ is the $l^1$ norm. Note that, correctly, this is only meaningful for $N\geq 2$; indeed, if we had $N=1$, then the only one amplitude would be necessarily set to $1$ by the normalization requirement.\\
\indent In the following, it will be useful to factorize the distribution, separating the phase and squared-amplitude components, as 
\begin{equation}
    \mathrm{d}\psi = (2\pi)^{-N}p(z_1,\dots,z_N)\mathrm{d}^N\phi\;\mathrm{d}^Nz,
\end{equation}
with $p(z_1,\dots,z_N) = (N-1)! \;\delta( 1 - |\bm{z}|_1 )$ being the distribution over the amplitudes.   \\
\\
\noindent \textbf{Intermediate calculation.} For the sake of completeness, it is instructive to study the normalization of this distribution, which highlights how the counting factor $(N-1)!$ emerges naturally. To this end, we first introduce the following intermediate result, which will be of great usefullness also later on:
\begin{equation}
    f_n(x) = \int_0^{1-x}\mathrm{d}z_1\int_0^{1-(x+z_1)}\mathrm{d}z_2\int_0^{1-(x+z_1+z_3)}\mathrm{d}z_3\dots\int_0^{1-(x+z_1+\dots+z_{n-1})}\mathrm{d}z_n = \frac{(1-x)^n}{n!}. 
\end{equation}
This can be proved straightforwardly by induction. Indeed, one can easily certify that the claimed equality holds true e.g. for $n=1$ and $n=2$. Next, for the inductive step, we can write
\begin{align}
    f_{n+1}(x) & = \int_0^{1-x}\mathrm{d}z_1\int_0^{1-(x+z_1)}\mathrm{d}z_2\int_0^{1-(x+z_1+z_3)}\mathrm{d}z_3\dots\int_0^{1-(x+z_1+\dots+z_{n-1})}\mathrm{d}z_n\int_0^{1-(x+z_1+\dots+z_{n})}\mathrm{d}z_{n+1} = \nonumber \\
    & \overset{}{=}\; \int_0^{1-x}\mathrm{d}z_1 f_n(x+z_1) = \nonumber \\
    & \overset{(*)}{=}\; \frac{1}{n!} \int_0^{1-x}\mathrm{d}z_1 (1-x-z_1)^n = \frac{(1-x)^{n+1}}{(n+1)!};
\end{align}
in this chain of equalities, we used the inductive hypothesis for step (*).\\
\\
\noindent \textbf{Consistency check: normalization.} We can now check the correctness of our normalization by evaluating the integral for the constant function $f(\psi)=1$; using the intermediate result above, we obtain indeed
\begin{align}
    \int_{\mathscr{H}}\!\! \mathrm{d}\psi & = \frac{(N-1)!}{(2\pi)^N} \int_{\left[0,2\pi\right]^N} \mathrm{d}^N\phi\int_{\left[0,+\infty\right]^N}\mathrm{d}^Nz \; \delta( 1 - |\bm{z}|_1 ) = \nonumber \\
    & \overset{}{=} (N-1)! \int_{\left[0,+\infty\right]^N}\mathrm{d}^Nz \; \delta( 1 - |\bm{z}|_1 ) = \nonumber \\
    & \overset{}{=}  (N-1)! \int_0^1\mathrm{d}z_1 \dots \int_0^1\mathrm{d}z_N \; \delta \Big(1-(z_1+\dots+z_N)\Big) = \nonumber \\
    & \overset{(*)}{=}  (N-1)! \int_0^1\mathrm{d}z_1 \dots \int_0^1\mathrm{d}z_{N-1} \; \Theta\Big(1-(z_1+\dots+z_{N-1})\Big) = \nonumber \\
     & \overset{}{=}  (N-1)! \int_0^1\mathrm{d}z_1\int_0^{1-z_1}\mathrm{d}z_2\int_0^{1-(z_1+z_2)}\mathrm{d}z_3 \dots \int_0^{1-(z_1+\dots +z_{N-2})}\mathrm{d}z_{N-1} = \nonumber \\
      & \overset{}{=}  (N-1)! \; f_{N-1}(0) = (N-1)! \frac{1}{(N-1)!} = 1.
\end{align}
In this chain of equalities, the crucial step is (*), where we used the Dirac delta to write $z_N=1-(z_1+\dots+z_{N-1})$, and inserted the Heaviside function to ensure that with this substitution the condition $z_N\geq 0$ is preserved. 

\subsection{Bounding the erroneously queried portion of the wavefunction}\label{Math2}

\noindent Here, we prove the bound in Eq.~\eqref{eq66}. This is obtained straightforwardly from the following: 
\begin{align}
    \bigg|\langle \psi_D, A_k | \sum_{\bm{x}\notin\mathcal{G}_k}\psi_{\bm{x}}K_k\ket{\bm{x}} \bigg| & = \Bigg|  \sum_{\bm{y}\in\mathcal{G}_k}\sum_{\bm{x}\notin\mathcal{G}_k}\psi^*_{\bm{y}}\psi_{\bm{x}} \langle \bm{y}, D_{\bm{y}}, A_k|K_k|\bm{x}\rangle + \left[\sum_{\bm{y}\notin\mathcal{G}_k}\psi^*_{\bm{y}}\bra{\bm{y}, D_{\bm{y}}, A_k}\right]\left[\sum_{\bm{x}\notin\mathcal{G}_k}\psi_{\bm{x}}K_k\ket{\bm{x}}\right] \Bigg| = \nonumber \\
    & = \Bigg| \sqrt{p_k} \sum_{\bm{y}\in\mathcal{G}_k}\sum_{\bm{x}\notin\mathcal{G}_k}\psi^*_{\bm{y}}\psi_{\bm{x}} \langle\bm{y}|K_k^\dag K_k | \bm{x}\rangle + \left[\sum_{\bm{y}\notin\mathcal{G}_k}\psi^*_{\bm{y}}\bra{\bm{y}, D_{\bm{y}}, A_k}\right]\left[\sum_{\bm{x}\notin\mathcal{G}_k}\psi_{\bm{x}}K_k\ket{\bm{x}}\right] \Bigg|=\nonumber \\
    & = \Bigg|  \left[\sum_{\bm{y}\notin\mathcal{G}_k}\psi^*_{\bm{y}}\bra{\bm{y}, D_{\bm{y}}, A_k}\right]\left[\sum_{\bm{x}\notin\mathcal{G}_k}\psi_{\bm{x}}K_k\ket{\bm{x}}\right] \Bigg| \leq \nonumber \\
    & \overset{(*)}{\leq}\;\sqrt{ \sum_{\bm{y}\notin\mathcal{G}_k}|\psi_{\bm{y}}|^2 } \sqrt{ p_k \sum_{\bm{x}\notin\mathcal{G}_k}|\psi_{\bm{x}}|^2 } = \sqrt{p_k}\Big[ 1 - \sum_{\bm{x}\in\mathcal{G}_k}|\psi_{\bm{x}}|^2 \Big].
\end{align}
In step (*), we simply used the conventional Schwartz inequality.

\subsection{Bounding the conditional query fidelity}\label{Math3}

\noindent Here, we prove our bound to the conditional query fidelity given in Eq.~\eqref{eq67}. This is based on the following chain of inequalities: 
\begin{align}
    F_k(\psi|D) & = \frac{1}{p_k}\Bigg|\sqrt{p_k} \sum_{\bm{x}\in\mathcal{G}_k}|\psi_{\bm{x}}|^2 + \langle \psi_D, A_k | \sum_{\bm{x}\notin\mathcal{G}_k}\psi_{\bm{x}}K_k\ket{\bm{x}} \Bigg|^2 \geq \nonumber \\
    & \overset{(*)}{\geq}\;\frac{1}{p_k} \Bigg[  \sqrt{p_k} \sum_{\bm{x}\in\mathcal{G}_k}|\psi_{\bm{x}}|^2 -  \bigg|\langle \psi_D, A_k | \sum_{\bm{x}\notin\mathcal{G}_k}\psi_{\bm{x}}K_k\ket{\bm{x}} \bigg| \Bigg]^2 \geq \nonumber \\
    & \overset{(**)}{\geq}\; \frac{1}{p_k} \Bigg[  \sqrt{p_k} \sum_{\bm{x}\in\mathcal{G}_k}|\psi_{\bm{x}}|^2 -  \sqrt{p_k}\Big[ 1 - \sum_{\bm{x}\in\mathcal{G}_k}|\psi_{\bm{x}}|^2 \Big] \Bigg]^2 \Theta \Bigg( \sum_{\bm{x}\in\mathcal{G}_k}|\psi_{\bm{x}}|^2 - \frac{1}{2} \Bigg) = \nonumber \\
    & = \Bigg[ 2\sum_{\bm{x}\in\mathcal{G}_k}|\psi_{\bm{x}}|^2 - 1 \Bigg]^2 \Theta \Bigg( \sum_{\bm{x}\in\mathcal{G}_k}|\psi_{\bm{x}}|^2 - \frac{1}{2} \Bigg). 
\end{align}
We now analyze the two crucial steps $(*)$ and $(**)$ one by one.
\begin{itemize}
    \item \textbf{Step $(*)$}

    For this, we simply used the trivial fact that (if $a\in\mathbb{R}$ and $b\in\mathbb{C}$), it is always true that $|a+b|^2\geq (a-|b|)^2$. The inequality follows from defining $a = \sum_{\bm{x}\in\mathcal{G}_k}|\psi_{\bm{x}}|^2$ and $b = \langle \psi_D, A_k | \sum_{\bm{x}\notin\mathcal{G}_k}\psi_{\bm{x}}K_k\ket{\bm{x}}/\sqrt{p_k} $.

    \item \textbf{Step $(**)$}

    For this, more care is necessary. For concreteness, let $\lambda = 1 - \sum_{\bm{x}\in\mathcal{G}_k}|\psi_{\bm{x}}|^2 = 1-a$; we are interested in lower bounding the quantity $(a-|b|)^2$, by exploiting the fact that $|b|\leq \lambda \leq 1$. In this setting, a meaningful bound can be given as $(a-|b|)^2\geq (a-\lambda)^2$; importantly, for $\lambda = 1-a$ as here, this only holds when either $a\geq 1/3$. Indeed, let $f(|b|)=(a-|b|)^2-(a-\lambda)^2$. Then, it is easy to see that $f(|b|)\geq o$ if either $|b|\leq a-|a-\lambda|$, or $|b|\geq a+|a-\lambda|$; however, since $a+|a-\lambda|\geq \lambda$, we can already eliminate this option. To further evaluate, let us consider the two cases $a\geq \lambda$ and $a\leq 1/2$ separately. (i) If $a\geq \lambda$, $|b|\leq a-|a-\lambda|=\lambda$, thus we get $f(|b|)\geq 0$ straightforwardly. (ii) If else $a\leq \lambda$, we obtain $|b|\leq 2a-\lambda$; however, since in this case $2a-\lambda \leq \lambda$, this is not guaranteed by our hypothesis $|b|\leq \lambda$. In summary, from $\lambda = 1-a\geq |b|$, we get $f(|b|)\geq 0$ whenever $a\geq 1/2$. We thus use the Heaviside function to specify that $(a-|b|)^2\geq (2a-1)$ if $a\geq 1/2$, and just $a-|b|)^2\geq 0$ otherwise.    
\end{itemize}

\subsection{Statistics of the first $n$ amplitudes}\label{Math4}

\noindent Here, we are interested in the following problem: for $n\leq N$, what is the statistics of the first $n$ squared amplitudes $(z_1,\dots,z_n)$ of a randomly chosen state $\ket\psi \in\mathscr{H}$ in the Hilbert space? Note that only for simplicity here we consider `the first' $n$ squared amplitudes, but (since $\ket\psi$ is chosen randomly with uniform distribution) the problem is of course invariant on which $n$ squared amplitudes we actually consider. In summary, we are interested in calculating the marginal probability distribution
\begin{align}
    p(z_1,z_2,\dots,z_n) & = \frac{1}{(2\pi)^N} \int_{[0,2\pi]^N} \mathrm{d}^N\phi  \int_{[0,+\infty]^{N-n}} \mathrm{d}z_{n+1} \mathrm{d}z_{n+2}\dots \mathrm{d}z_{N} \; p(z_1,\dots,z_N) = \nonumber \\
    & = (N-1)!\int_0^1 \mathrm{d}z_{n+1}\dots \int_0^1 \mathrm{d}z_{N} \; \delta( 1 - |\bm{z}|_1 ),
\end{align}
where for the last line we simply evaluated the integral over the phases and wrote $p(z_1,\dots,z_N)$ explicitly. To calculate this marginal distribution, it is convenient to immediately introduce $s=z_1+\dots z_n$, which is the only term including the non-mute variables appearing in the integrand; we then evaluate the integral as
\begin{align}
    p(z_1,\dots,z_n) & = (N-1)! \Theta(1-s) \int_0^{1-s}\mathrm{d}z_{n+1}\int_0^{1-(s+z_{n+1})}\mathrm{d}z_{n+2}\dots \int_0^{1-(s+z_{n+1}+\dots z_{N-2})}\mathrm{d}z_{N-1} = \nonumber \\
    & = (N-1)! \;  f_{N-n-1} (s) = \nonumber \\
    & = \frac{(N-1)!}{(N-n-1)!}\Theta(1-s) (1-s)^{N-n-1} = n! \binom{N-1}{n} \Theta(1-s) (1-s)^{N-n-1}, 
\end{align}
yielding the marginal distribution
\begin{equation}
    p(z_1,\dots,z_n) = n! \binom{N-1}{n} \Theta(1-s) (1-s)^{N-n-1}.
\end{equation}
To understand this formula, note that it is only meaningful when $N\geq 2$ and $N\geq n \geq 1$. Indeed, for $N=1$ the only squared amplitude would be deterministically set to $z=1$ by the normalization requirement. Moreover, one can straightforwardly check that the limit case $n=N$ provides the correct result. It is also instructive to consider e.g. $n=1$, corresponding to the marginal distribution of one single squared amplitude $z_i$, for which we get
\begin{equation}
    p(z_i) = (N-1) \Theta(1-z_i) (1-z_i)^{N-2},
\end{equation}
which again stresses that $N\geq 2$. This allows for instance to calculate the momenta
\begin{align}
    \langle z_i^m \rangle & = \int d z_i\; p(z_i) z_i^m = \nonumber \\
    & = (N-1)\int_0^1 \mathrm{d}z (1-z)^{N-2} z^m = \nonumber \\
    & = (N-1) \; \mathcal{B}( m+1, N-1 )  = \nonumber \\
    & = (N-1)\frac{m!(N-2)}{(N+m-1)!} =  \frac{1}{\binom{N+m-1}{m}},    
\end{align}
where we inserted the well-known Beta function
\begin{equation}
    \mathcal{B}(a,b) = \int_0^1 \mathrm{d} x x^{a-1} (1-x)^{b-1}
\end{equation}
and we used the fact that when $(a,b)$ are positive integers, one has
\begin{equation}
    \mathcal{B}(a,b) = \frac{(a-1)!(b-1)!}{(a+b-1)!}.
\end{equation}
To check correctness, we note that the first moment reads $\langle z_i\rangle = 1/N$, as one could already infer from the beginning. Moreover, the variance of $z_i$ reads $\langle z_i^2\rangle = 2/N(N+1)$. 

\subsection{Distribution of the overlap}\label{Math5}

\noindent Let $\mathbb{P}$ be a projector onto a subspace of $\mathscr{H}$ of dimension $n\leq N$; that is, $\mathbb{P}$ is hermitian ($\mathbb{P}=\mathbb{P}^\dag$) and idempotent ($\mathbb{P}=\mathbb{P}^2$), and has trace $\text{Tr}(\mathbb{P})=n$. The overlap of an arbitrary state $\ket\psi \in\mathscr{H}$ with the space identified by $\mathbb{P}$ is given by $\langle \psi |\mathbb{P}|\psi\rangle$. Here, we treat this overlap as a statistical variable, and we are therefore interested in evaluating its momenta
\begin{equation}
    \Big\langle \langle \psi |\mathbb{P}|\psi\rangle^m  \Big\rangle = \int_\mathscr{H} \mathrm{d}\psi\; \langle \psi |\mathbb{P}|\psi\rangle ^m. 
\end{equation}
We can start by choosing a basis on $\mathscr{H}$ such that the subspace defined by $\mathbb{P}$ is spanned by the first $n$ terms, i.e.
\begin{equation}
    \mathbb{P} = \sum_{i=1}^n\ket{i}\bra{i}.
\end{equation}
Then, we get
\begin{equation}
    \langle \psi |\mathbb{P}|\psi\rangle = \sum_{i=1}^n |\langle i |\psi\rangle|^2=\sum_{i=1}^n z_i \equiv s(z_1,\dots, z_n),
\end{equation}
where the coordinates $z_i\geq 0$ are defined as previously explained. In the following, we will drop the dependence of $s=\langle\psi|\mathbb{P}|\psi\rangle$ on $\bm{z}$ whenever it is not necessary. With this redefinition, it is straightforward to see that
\begin{equation}
     \Big\langle \langle \psi |\mathbb{P}|\psi\rangle^m  \Big\rangle =  \int_0^1 \mathrm{d}s \; P(s) s^m,
\end{equation}
where $P(s)$ is the marginal distribution of the overlap, defined as 
\begin{align}
    P(s) & = (N-1)! \int_0^1 \mathrm{d}z_{1}\dots \int_0^1 \mathrm{d}z_{n} \;  \delta \Big(s - (z_1+\dots + z_n) \Big) \int_0^1 \mathrm{d}z_{n+1}\dots \int_0^1 \mathrm{d}z_{N} \; \delta( 1 - |\bm{z}|_1 ) = \nonumber \\
    & = n! \binom{N-1}{n}\Theta(1-s) (1-s)^{N-n-1} \int_{[0,+\infty]^n}\mathrm{d}z_1\dots \mathrm{d}z_n \; \delta\Big( s - (z_1+\dots + z_n) \Big) = \nonumber \\
    & = n \binom{N-1}{n} \Theta(1-s) (1-s)^{N-n-1}s^{n-1}.
\end{align} 
We now use this to evaluate the momenta:
\begin{align}
    \langle s^m \rangle & = \int_0^{+\infty} P(s) s^m = \nonumber \\
    & = n\binom{N-1}{n} \mathcal{B}(n+m, N-n) = \frac{\binom{n+m-1}{m}}{\binom{N+m-1}{m}}.
\end{align}
For instance, we note that, correctly, for $m=1$ we get $\langle s \rangle = n/N$, as could be deduced from the beginning. In general, we remark that the relation 
\begin{equation}
    \langle s^m \rangle = \binom{n+m-1}{m} \langle z_i^m \rangle
\end{equation}
holds between the momenta of one single squared amplitude, and the sum of $n$ of them. 

\subsection{Calculating some nontrivial averages}\label{Math6}

\noindent We are now interested in evaluating averages of the following form:
\begin{equation}
    J_m(\lambda) = \int_\mathscr{H} \mathrm{d}\psi\; \Theta\Big(\langle \psi |\mathbb{P}|\psi\rangle - \lambda\Big) \langle \psi |\mathbb{P}|\psi\rangle ^m,
\end{equation}
where again $\mathbb{P}$ is a projector onto a subspace of $\mathscr{H}$ of dimension $n\leq N$; that is, $\mathbb{P}$ is hermitian ($\mathbb{P}=\mathbb{P}^\dag$) and idempotent ($\mathbb{P}=\mathbb{P}^2$), and has trace $\text{Tr}(\mathbb{P})=n$. By simple considerations, it is straightforward to see that this can only depend on $n$ and $m$ and $N$. As above, we start by choosing a basis on $\mathscr{H}$ such that the subspace defined by $\mathbb{P}$ is spanned by the first $n$ terms, i.e. $\mathbb{P} = \sum_{i=1}^n\ket{i}\bra{i}$; then, we get
\begin{equation}
    \langle \psi |\mathbb{P}|\psi\rangle = \sum_{i=1}^n |\langle i |\psi\rangle|^2=\sum_{i=1}^n z_i \equiv s(z_1,\dots, z_n),
\end{equation}
where the coordinates $z_i\geq 0$ are defined as previously explained. In the following, we will drop the dependence of $s$ on $\bm{z}$ whenever it is not necessary. With this redefinition, it is straightforward to see that
\begin{equation}
    J_m(\lambda) = \int_0^1 \mathrm{d}s P(s) \Theta(s-\lambda) s^m,
\end{equation}
where we used the probability distribution over $s$ calculated previously. For obvious reasons, we only consider $\lambda \leq 1$; then, using the fact that $\Theta(x)=1-\Theta(-x)$, it is convenient to rewrite the integral above as
\begin{align}
    J_m(\lambda) & = \int_0^1\mathrm{d}sP(s)s^m - \int_0^1\mathrm{d}sP(s) \Theta(\lambda - s) s^m = \nonumber \\
    & = \langle s^m \rangle - n\binom{N-1}{n} \int_0^\lambda (1-s)^{N-n-1}s^{n+m-1} = \nonumber \\
    & = \langle s^m \rangle - n\binom{N-1}{n} \mathcal{B}(\lambda; n+m, N-n),
\end{align}
where we introduced the well-known incomplete Beta function $\mathcal{B}(\lambda; a, b) = \int_0^\lambda \mathrm{d}x x^{a-1}(1-x)^{b-1}$.

\bibliography{biblio.bib}